%% file: merged.tex
\pdfoutput=1
\documentclass[12pt,letterpaper]{article}

\usepackage[T1]{fontenc}
\usepackage[utf8]{inputenc}
\usepackage{amsmath,amsthm}
\usepackage{newtxtext,newtxmath}
\usepackage[margin=1in]{geometry}
\usepackage{graphicx}
\graphicspath{{figures/}}
\usepackage{mathtools}
\usepackage{booktabs}
\usepackage{multirow}
\usepackage{makecell}
\usepackage[dvipsnames,table]{xcolor}
\usepackage[export]{adjustbox}
\usepackage{pifont}
\usepackage{extarrows}
\usepackage{listings}
\usepackage{algorithm}
\makeatletter
\@ifundefined{algorithmic}{}{%

}
\makeatother
\usepackage{algpseudocode}
\usepackage{url}
\usepackage{setspace}
\usepackage{lineno}          %
\usepackage{scicite}         %
\newcommand{\citep}[1]{\cite{#1}}%
\usepackage{caption}
\usepackage{cleveref}

\Crefname{figure}{Fig.}{Figs.}
\crefname{figure}{Fig.}{Figs.}
\Crefname{table}{Table}{Tables}
\crefname{table}{Table}{Tables}
\crefname{equation}{Eq.}{Eqs.}
\Crefname{equation}{Eq.}{Eqs.}
\creflabelformat{equation}{#2#1#3}%
\crefname{section}{section}{sections}
\Crefname{section}{Section}{Sections}

\theoremstyle{plain}
\newtheorem{theorem}{Theorem}
\newtheorem{proposition}[theorem]{Proposition}
\newtheorem{lemma}[theorem]{Lemma}

\newtheorem{assumption}[theorem]{Assumption}
\theoremstyle{definition}
\newtheorem{definition}{Definition}
\theoremstyle{remark}

\definecolor{best}{gray}{0.85}      %
\definecolor{second}{gray}{0.93}    %

\newcommand{\siEdos}{section~S1.1}
\newcommand{\siPhdos}{section~S1.2}
\newcommand{\siCppr}{section~S2}
\newcommand{\siAlgo}{section~S3}
\newcommand{\siHyper}{section~S4}
\newcommand{\siOracle}{section~S4.3}
\newcommand{\siBase}{section~S4.4}
\newcommand{\siEval}{section~S4.5}
\newcommand{\siComp}{section~S4.6}
\newcommand{\siRes}{section~S5}
\newcommand{\siEle}{section~S5.1}
\newcommand{\siViz}{section~S5.2}
\newcommand{\siRobust}{section~S5.3}
\newcommand{\siMetastable}{section~S5.4}
\newcommand{\siAbl}{section~S5.5}
\newcommand{\siDisc}{section~S5.6}
\newcommand{\siCenter}{section~S5.7}
\newcommand{\siVH}{section~S5.8}
\newcommand{\siScaling}{section~S5.9}
\newcommand{\siTabAbl}{table~S5}
\newcommand{\siTabCenter}{table~S6}
\newcommand{\siTabStable}{table~S7}

\let\origfigure\figure
\let\endorigfigure\endfigure
\renewenvironment{figure}[1][htbp]{\origfigure[htbp]}{\endorigfigure}
\let\origtable\table
\let\endorigtable\endtable
\renewenvironment{table}[1][htbp]{\origtable[htbp]}{\endorigtable}
\makeatletter
\expandafter\let\csname origtable*\expandafter\endcsname\csname table*\endcsname
\expandafter\let\csname endorigtable*\expandafter\endcsname\csname endtable*\endcsname
\renewenvironment{table*}[1][htbp]{\csname origtable*\endcsname[htbp]}{\csname endorigtable*\endcsname}
\makeatother

\begin{document}

\begin{center}
{\LARGE\bfseries DeepDOSReasoner: Physics-grounded reasoning for density-of-states prediction and materials discovery}\\[1.5em]

{\large
Yingheng Wang$^{1\ast}$,
Francesco Ricci$^{2,3\ast}$,
Zhilong Wang$^{4,5,6}$,
Tao Yu$^{7}$,
Junwen Bai$^{1,8}$,
Xiangyun Lei$^{9,10}$,
Shufeng Kong$^{1}$,
Fengqi You$^{4,5,6}$,
John M. Gregoire$^{11,12}$,
Carla P. Gomes$^{1,4\ast}$
}\\[1.2em]

{\small\itshape
$^{1}$Department of Computer Science, Cornell University, Ithaca, NY, USA.\\
$^{2}$Institute of Condensed Matter and Nanosciences, Universit\'e catholique de Louvain, Louvain-la-Neuve, Belgium.\\
$^{3}$Matgenix SRL, Charleroi, Belgium.\\
$^{4}$Cornell University AI for Science Institute, Cornell University, Ithaca, NY, USA.\\
$^{5}$College of Engineering, Cornell University, Ithaca, NY, USA.\\
$^{6}$Robert Frederick Smith School of Chemical and Biomolecular Engineering, Cornell University, Ithaca, NY, USA.\\
$^{7}$School of Information and Communication Technology, Griffith University, Queensland, Australia\\
$^{8}$Current address: Google DeepMind, Mountain View, CA, USA.\\
$^{9}$Toyota Research Institute, Los Altos, CA, USA.\\
$^{10}$Current address: Ursa Simulation, Atlanta, GA, USA.\\
$^{11}$California Institute of Technology, Pasadena, CA, USA.\\
$^{12}$Current address: Autonomous Science Platform, Lila Sciences, Cambridge, MA, USA.
}\\[1em]

{\small $^{\ast}$Corresponding author. Email: yingheng@cs.cornell.edu (Y.W.); francesco.ricci@uclouvain.be (F.R.); gomes@cs.cornell.edu (C.P.G.)}
\end{center}

\vspace{1em}

\noindent\textbf{%
Predicting a material's behavior from its atomic structure is a longstanding goal of the physical sciences. The electronic and phonon densities of states (eDOS and phDOS) govern key electronic, optical, vibrational, and thermodynamic properties, yet their calculation using density functional theory (DFT) is costly: about 60\% of Materials Project entries lack   eDOS and over 80\% lack phDOS. Existing crystal-to-spectrum models rely on pooled structural representations and pointwise regression, which blur spectral features and violate total-state conservation. Here we introduce DeepDOSReasoner, a physics-grounded framework that decodes spectra through atom-to-energy source attention and refines the normalized DOS through learned continuity-equation transport, enforcing nonnegativity and total-state conservation by construction. DeepDOSReasoner achieves the highest prediction accuracy across all evaluated metrics for both eDOS and phDOS, recovers element-resolved contributions without supervision, and most reliably identifies rare valence-band (VB) gaps. In a prospective screen of 8,106 uncharted materials validated by new DFT calculations, its predictions remain the most accurate among the methods evaluated, with 25 of its top 50 candidates confirmed as VB-gap materials. Three case studies spanning substituted compounds, photovoltaic chalcogenides, and a high-entropy alloy demonstrate transfer across materials classes. Ablations confirm that both stages are necessary, while the model retains comparable accuracy when trained on only 25\% of the data. Predicting spectra in milliseconds enables exhaustive screening; we release predictions filling the missing eDOS for 91,405 and phDOS for 127,768 Materials Project entries. DeepDOSReasoner provides a blueprint for integrating machine learning with physical reasoning toward high-throughput spectral screening and autonomous materials discovery.
}

\vspace{1em}

\noindent\textbf{Teaser:} Physics-grounded machine learning predicts electronic and phonon spectra to accelerate materials discovery.

\newpage

\input{sections/introduction}

\input{sections/results}

\input{sections/discussion}

\input{sections/methods}

\singlespacing
\bibliography{sn-bibliography}
\bibliographystyle{Science}

\newpage

\section{Acknowledgments}

\noindent\textbf{Funding:}
This project was partially supported by Schmidt Sciences programs, an AI2050 Senior Fellowship and an Eric and Wendy Schmidt AI in Science Postdoctoral Fellowship, the National Science Foundation (NSF), and the Air Force Office of Scientific Research (AFOSR; FA9550-23-1-0322).

\medskip
\noindent\textbf{Author contributions:}
C.P.G. and J.M.G. (while at Caltech) conceived the project and supervised the work. Y.W. designed and implemented DeepDOSReasoner under the guidance of C.P.G. and J.M.G. J.M.G., F.R., Z.W., and X.L. designed the use cases. Y.W., T.Y., J.B., and S.K. conducted the experiments. F.R. and Z.W. performed the DFT calculations. Y.W., F.R., Z.W., F.Y., J.M.G., and C.P.G. interpreted the results. Y.W., F.R., Z.W., and C.P.G. wrote the manuscript, with contributions from all authors.

\medskip
\noindent\textbf{Competing interests:}
The authors declare that they have no competing interests.

\medskip
\noindent\textbf{Data and materials availability:}
All data needed to evaluate the conclusions in the paper are present in the paper and/or the Supplementary Materials.
The eDOS benchmark and stable-only subset used in this work are derived from raw eDOS data publicly available through the Materials Project (\url{https://materialsproject.org}).
The processed Materials Project identifier lists and the fixed training/validation/test splits defining the eDOS benchmark, its stable-only subset, and the phDOS benchmark are included in the code repository (\url{https://github.com/gomes-lab/deep-dos-reasoner}).
The DeepDOSReasoner-predicted spectra that fill the missing eDOS for $91{,}405$ and the missing phDOS for $127{,}768$ materials in the Materials Project are available at \url{https://huggingface.co/datasets/yingheng/mp-predicted-edos-phdos}.
The new DFT-computed reference spectra generated for the prospective-discovery evaluation and the case studies will be deposited in a public long-term repository upon publication.
An interactive demo is available at \url{https://gomes-lab.github.io/deepdosreasoner-demo/}.
The source code for DeepDOSReasoner, including training and evaluation scripts, is available at \url{https://github.com/gomes-lab/deep-dos-reasoner}.

\clearpage
\singlespacing

\setcounter{secnumdepth}{3}
\setcounter{section}{0}
\setcounter{subsection}{0}
\setcounter{figure}{0}
\setcounter{table}{0}
\setcounter{equation}{0}
\setcounter{theorem}{0}
\setcounter{definition}{0}
\setcounter{example}{0}
\setcounter{remark}{0}
\renewcommand{\thesection}{S\arabic{section}}
\renewcommand{\thesubsection}{S\arabic{section}.\arabic{subsection}}
\renewcommand{\thesubsubsection}{S\arabic{section}.\arabic{subsection}.\arabic{subsubsection}}
\renewcommand{\thefigure}{S\arabic{figure}}
\renewcommand{\thetable}{S\arabic{table}}
\renewcommand{\theequation}{S\arabic{equation}}
\renewcommand{\thealgorithm}{S\arabic{algorithm}}

\begin{center}
{\Large\bfseries Supplementary Materials for}\\[1em]
{\LARGE\bfseries DeepDOSReasoner: Physics-grounded reasoning for density-of-states prediction and materials discovery}\\[1.5em]

{\large
Yingheng Wang$^{\ast}$,
Francesco Ricci$^{\ast}$,
Zhilong Wang,
Tao Yu,
Junwen Bai,\\
Xiangyun Lei,
Shufeng Kong,
Fengqi You,
John M. Gregoire,
Carla P. Gomes$^{\ast}$
}\\[1.5em]

{\small $^{\ast}$Corresponding author. Email: yingheng@cs.cornell.edu (Y.W.); francesco.ricci@uclouvain.be (F.R.); gomes@cs.cornell.edu (C.P.G.)}
\end{center}

\input{appendix/intro_edos}

\input{appendix/formal_method}

\input{appendix/algorithm}

\input{appendix/details}

\input{appendix/results}

\end{document}

%% file: sections/introduction.tex
\section{INTRODUCTION}

The density of states (DOS) provides a compact bridge between a material's atomic structure and its physical responses~\citep{kittel2018introduction,hook2013solid}. The electronic DOS (eDOS) governs electronic and optical behavior, while the phonon DOS (phDOS) determines lattice vibrational and thermodynamic properties. Both are one-dimensional spectra defined over a bounded energy or frequency axis and constrained by strict total-state sum rules. Of the two, eDOS prediction is particularly demanding because many performance-relevant properties depend on sharp spectral features: the low-temperature electronic specific heat and Pauli susceptibility on the Fermi-level density of states, near-band-edge optical absorption on the steepness of the band edges, catalytic activity on the near-Fermi-energy states, and thermoelectric and transparent-conductor behavior on narrow gaps below but near the Fermi energy~\citep{ricciGappedMetalsThermoelectric2020a}. As a result, downstream discovery workflows for thermoelectrics, photovoltaics, electrocatalysts, superconductors, and transparent conductors depend not only on whether eDOS spectra can be predicted accurately on average, but also on whether their fine spectral structure is faithfully preserved at scale~\citep{lee2016computational}. Density functional theory (DFT)~\citep{hohenberg1964inhomogeneous,kohn1965self} remains the standard route for computing both eDOS and phDOS, but its per-material cost limits coverage. Among the 154{,}387 inorganic crystal entries in the Materials Project~\citep{Jain2013}, about 60\% lack a published eDOS and over 80\% lack a phDOS, constraining screening efforts that require full spectral signatures.

Machine learning offers a route to amortize this cost by learning direct mappings from crystal structure to spectrum. Crystal graph neural networks have become reliable surrogates for scalar targets such as formation energy, band gap, and elastic moduli~\citep{Xie2018,schutt2018schnet,Chen2019}, and such surrogates now drive materials discovery at scale~\citep{merchant2023scaling}, but predicting a full DOS spectrum is qualitatively harder. The target is not an unconstrained vector of bin values: it is a nonnegative spectral density whose validity depends on total-state conservation, and whose application-relevant content often lies in sharp, easily blurred features such as band edges, gaps below but near the Fermi energy (i.e., in the valence band) in metallic systems~\citep{ricciGappedMetalsThermoelectric2020a}, and van Hove singularities~\citep{van1953occurrence}. Existing eDOS prediction models~\citep{kaundinya2022prediction,Kong2022,lee2023density} fall short in three concrete ways. First, they often predict the band gap and the surrounding spectrum separately and reconcile them only post hoc, which can misalign band edges and introduce spurious in-gap density. Second, pointwise regression losses can yield low per-bin error while still shifting peaks, smoothing van Hove singularities, and drifting from total-state conservation, producing spectra that appear plausible but are physically inconsistent. Third, most models decode the spectrum from a single pooled crystal embedding, obscuring the element-resolved contributions that physically compose the total eDOS~\citep{mulliken1955electronic} and limiting interpretability.

Here we introduce \textbf{DeepDOSReasoner}, a physics-grounded framework for crystal-to-DOS prediction that addresses these shortcomings directly. In Stage~I, DeepDOSReasoner encodes the crystal as a graph using an edge-aware Transformer~\citep{velivckovic2017graph,vaswani2017attention} and decodes the eDOS bin by bin through atom-to-energy source attention. This design allows each energy bin to selectively aggregate atom- and element-resolved structural information, rather than reading from a single pooled representation. In Stage~II, DeepDOSReasoner treats the normalized DOS as a probability density and refines it through a learned, state-dependent velocity field under a continuity equation with no-flux boundary conditions~\citep{kundu2024fluid}, integrated with a differentiable upwind scheme~\citep{godunov1959finite}. By construction, this refinement preserves nonnegativity and total-state conservation throughout its trajectory. In practice, it sharpens band edges and van Hove singularities while resolving the band gap and surrounding spectrum jointly. Thus, DeepDOSReasoner embeds physical consistency directly into the model output rather than imposing it through soft penalties, contributing to a broader effort to couple data-driven learning with principled physical reasoning~\citep{wang2023scientific,jumper2021highly,chen2021automatingdrnet,wang2025scireasoner}.

On a Materials Project eDOS benchmark with band gaps retained~\citep{Kong2022,Jain2013}, DeepDOSReasoner advances the state of the art across all evaluated metrics ($R^2$, mean absolute error, mean squared error, and Wasserstein distance), outperforming every baseline simultaneously.
Beyond aggregate accuracy, it more faithfully preserves the sharp spectral features that govern downstream materials properties: it obtains the lowest peak-restricted mean absolute error on van Hove singularities near band edges, identifies valence-band (VB) gaps with the highest precision, recall, F1, and Matthews correlation coefficient (MCC) among all tested methods, and recovers element-resolved spectral contributions without explicit supervision.
To test generality beyond electronic spectra, we further evaluate DeepDOSReasoner on a phDOS dataset~\citep{Chen2021}: with minimal modification, it again achieves the best performance across all four evaluation metrics, demonstrating that our method transfers naturally across spectral domains. In fact, DeepDOSReasoner is the only method that leads on every reported metric across all of these tasks.

We next evaluate DeepDOSReasoner in prospective materials-discovery settings. For VB-gap discovery, we screen an $8{,}106$-material pool of nonmagnetic Materials Project entries lacking DFT eDOS labels, and obtain ground-truth labels for a $100$-material evaluation subset from new DFT calculations. DeepDOSReasoner outperforms all baselines simultaneously on precision, recall, F1, and MCC, and 25 of its 50 top-ranked candidates are confirmed as genuine VB-gap materials by the new DFT calculations. We further demonstrate the practical reach of eDOS prediction through three case studies spanning distinct functional-materials classes: (i) substituent-induced variants of two parent compounds, where the model captures chemically reasonable spectral shifts under substitution; (ii) I$_2$--II--IV--X$_4$ quaternary chalcogenide semiconductors relevant to thin-film photovoltaics~\citep{wang2021unsupervised}, where the model faithfully reproduces photovoltaic-relevant spectral features; and (iii) a CuPtFeCoNi high-entropy alloy~\citep{han2024highentropy}, where the model shows close agreement with the DFT eDOS, especially on a key electronic descriptor related to alloy catalytic activity.

Finally, ablation studies identify the structure-aware decoder and constraint-preserving physical reasoning as the two load-bearing components of the framework. Data-scaling experiments show that DeepDOSReasoner matches full-data baselines on several metrics using only $25\%$ of the training set, while stable-only controls confirm that its advantages persist when training is restricted to thermodynamically stable structures. Together, these results show that DeepDOSReasoner's gains arise from the tight coupling of neural structure-to-spectrum learning with physical reasoning grounded in first-principles constraints. Exploiting its millisecond-per-spectrum inference, we accompany this work with a community resource for high-throughput screening: predicted spectra filling the missing eDOS for $91{,}405$ materials and the missing phDOS for $127{,}768$ materials in the Materials Project.
More broadly, our results suggest that embedding physical reasoning directly into machine-learning architectures can improve both predictive accuracy and physical fidelity, offering a promising direction for scientific AI and high-throughput materials discovery.

%% file: sections/results.tex
\section{RESULTS}
We first summarize the DeepDOSReasoner architecture (\Cref{fig:main}) and then evaluate it on standard spectrum-prediction benchmarks, on VB-gap identification and discovery, on transfer to functional-materials classes, and through empirical analyses of the framework. Detailed methods are in Materials and Methods; hyperparameters, baseline descriptions, and metric definitions are in \siHyper{} of the Supplementary Materials.

\subsection{Overview of DeepDOSReasoner}\label{sec:results:overview}
\begin{figure*}[htbp]
    \centering
    \includegraphics[width=1\linewidth]{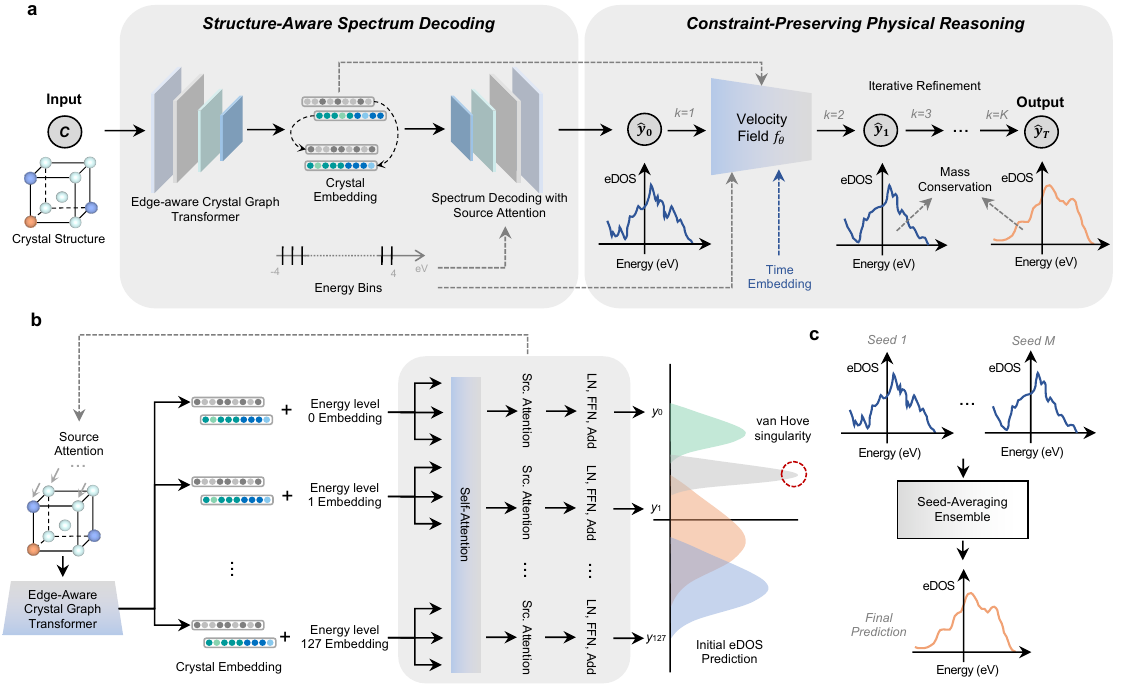}
    \caption{\textbf{DeepDOSReasoner architecture.} (\textbf{a})~End-to-end diagram of the two coupled stages: Stage~I (\emph{structure-aware spectrum decoding}) encodes the input crystal with an edge-aware crystal graph Transformer and decodes an initial eDOS sequence through source (atom-to-energy) attention over discrete energy bins; Stage~II (\emph{constraint-preserving physical reasoning}) refines the normalized eDOS through a learned transport field that moves spectral mass over energy while conserving total states, producing sharper band edges and peaks in the final prediction. (\textbf{b})~Architectural details of the source-attention decoder: each energy-bin query attends over the atom-level embeddings produced by the encoder through interleaved self-attention (across energy bins) and source attention (from energy bins to atoms), grounding each spectral coordinate in the atomic context that physically contributes at that energy. (\textbf{c})~Diagram of the seed-averaging ensemble used at inference: $M$ independently trained DeepDOSReasoner copies are run through the full two-stage pipeline, their normalized predictions are averaged (a convex combination that preserves nonnegativity and unit mass), and the averaged density is rescaled by the total mass $Z$ to recover the absolute DOS.}\label{fig:main}
\end{figure*}

DeepDOSReasoner maps a crystal structure to a full DOS spectrum through two coupled stages (\Cref{fig:main}). Stage~I encodes the crystal graph with an edge-aware graph Transformer and decodes the spectrum bin by bin through source (atom-to-energy) attention, so each energy bin aggregates the atom-level information most relevant to that spectral coordinate. Stage~II treats the normalized prediction as a probability density and refines it with a learned velocity field under a continuity equation with no-flux boundary conditions, sharpening band edges and peaks while conserving total spectral mass. Once trained, DeepDOSReasoner predicts a full DOS spectrum in ${\sim}10.5$~ms per crystal per model on a single GPU, versus ${\sim}10^2$ CPU-core-hours for a typical DFT eDOS calculation, a speedup of order $10^7$ after a one-time training cost of ${\sim}5$~GPU-hours per model (\siComp). This throughput makes exhaustive coverage of existing structure databases practical; as a companion resource, we release predicted spectra filling the missing eDOS for $91{,}405$ materials and the missing phDOS for $127{,}768$ materials in the Materials Project.
\Cref{fig:examples}(a) summarizes per-task performance against all baselines on the four benchmark tasks below.

\subsection{DOS prediction on standard benchmarks}\label{sec:results:standard}

\subsubsection{Electronic density of states prediction}\label{sec:results:edos}
We curate the eDOS benchmark from Materials Project structures~\citep{Jain2013} and DFT-computed eDOS spectra~\citep{Kong2022}, aligning each spectrum to its DFT Fermi level, broadening with a $\sigma=0.5$~eV Gaussian, and area-preservingly rebinning onto a uniform 128-bin grid over $[-4,4]$~eV; band gaps are retained explicitly rather than excised, so every method must predict both the gap and the surrounding spectrum on the same grid. The benchmark comprises $37{,}565$ crystals spanning stable and metastable entries, partitioned 80\%/10\%/10\% into training, validation, and test sets (test $n=3{,}756$; Materials and Methods). We evaluate against three baseline families: crystal-graph structure encoders (E3NN~\citep{geiger2022e3nn}, MEGNet~\citep{Chen2019}, GATGNN~\citep{louis2020graph}), interatomic-potential architectures adapted to spectrum output (NequIP~\citep{batzner2022nequip}, M3GNet~\citep{chen2022universal}, MACE~\citep{batatia2022mace}, EquiformerV2~\citep{liao2024equiformerv2}), and crystal-to-sequence predictors (Mat2Spec~\citep{Kong2022}, DOSTransformer~\citep{lee2023density}), using four complementary metrics: $R^2$, MAE, MSE, and Wasserstein distance (WD), the last two being most sensitive to sharp band edges and correct peak placement (definitions in \siEval).

DeepDOSReasoner achieves the best result across all four metrics (\Cref{tab:edos}). Relative to the strongest baseline on each metric, it improves $R^2$ from 0.645 to 0.691, lowers MAE from 2.605 to 2.567, reduces MSE from 44.008 to 38.142 ($-13.3\%$), and lowers WD from 0.223 to 0.191 ($-14.3\%$); against the median baseline the gains are larger still ($\Delta R^2{\approx}0.06$, $\Delta$WD${\approx}0.06$). The largest gains fall on MSE and WD, which most directly reward sharp band edges, accurate peak placement, and a faithful gap, exactly the features the structure-aware decoder and the mass-conserving reasoning stage are designed to recover.

These gains separate DeepDOSReasoner from all three families simultaneously, with relative $R^2$ improvements of $7$--$8\%$ over the best member of each family and $14$--$24\%$ lower WD. Unlike the baselines, which produce the spectrum by unconstrained pointwise regression and reconcile it with the band gap only post hoc, DeepDOSReasoner treats the spectrum as a constrained probability density: the $14.3\%$ WD reduction is the direct signature of the mass-conserving Stage~II, which can only transport existing mass along the energy axis and therefore cannot leak density into a gap or clip peaks below zero. The same mechanism preserves the sharp van Hove peaks near band edges, where DeepDOSReasoner attains the lowest peak-restricted MAE of any method ($15.95$ versus $33.47$ for EquiformerV2; see ``Empirical analyses of the framework''). Representative spectra against DFT are shown in \Cref{fig:examples}(b) and \siViz; across these examples the baselines err in three characteristic ways, broadening peaks, shifting them along the energy axis, or fabricating spurious oscillatory peaks, whereas DeepDOSReasoner more often places the dominant singularities at the correct energies with their sharp profiles intact. Error decompositions further confirm that the residual error is distributed broadly across material complexity, training-set similarity, subclass, and spectral composition rather than concentrated on any subset (\siRobust).

\begin{table*}[ht!]
\caption{\textbf{eDOS prediction with band gaps retained.} We report mean $\pm$ one standard deviation over five independent runs. Darkly shaded cells indicate the best results and lightly shaded cells indicate the second best.}
\label{tab:edos}
\begin{center}
\begin{sc}
\begin{tabular}{l*{4}{c}}
\toprule
Method
& $R^2 \uparrow$
& MAE $\downarrow$
& MSE $\downarrow$
& WD $\downarrow$ \\
\midrule

E3NN
& 0.615\scriptsize{$\pm$0.004} & 2.814\scriptsize{$\pm$0.001} & 47.034\scriptsize{$\pm$0.342} & 0.270\scriptsize{$\pm$0.001} \\

MEGNet
& 0.639\scriptsize{$\pm$0.004} & 2.711\scriptsize{$\pm$0.001} & 44.015\scriptsize{$\pm$0.185} & 0.251\scriptsize{$\pm$0.002} \\

GATGNN
& 0.618\scriptsize{$\pm$0.005} & 2.825\scriptsize{$\pm$0.002} & 47.115\scriptsize{$\pm$0.412} & 0.271\scriptsize{$\pm$0.002} \\

NequIP
& 0.615\scriptsize{$\pm$0.005} & 2.805\scriptsize{$\pm$0.001} & 46.932\scriptsize{$\pm$0.302} & 0.271\scriptsize{$\pm$0.003} \\

M3GNet
& 0.637\scriptsize{$\pm$0.003} & 2.702\scriptsize{$\pm$0.002} & 44.462\scriptsize{$\pm$0.312} & 0.250\scriptsize{$\pm$0.001} \\

MACE
& 0.575\scriptsize{$\pm$0.006} & 3.084\scriptsize{$\pm$0.003} & 46.027\scriptsize{$\pm$0.279} & 0.258\scriptsize{$\pm$0.002} \\

EquiformerV2
& \cellcolor{second}0.645\scriptsize{$\pm$0.002} & \cellcolor{second}2.605\scriptsize{$\pm$0.001} & 44.013\scriptsize{$\pm$0.615} & 0.230\scriptsize{$\pm$0.001} \\

\midrule

Mat2Spec
& 0.632\scriptsize{$\pm$0.001} & 2.786\scriptsize{$\pm$0.001} & 45.123\scriptsize{$\pm$0.605} & 0.234\scriptsize{$\pm$0.001} \\

DOSTransformer
& 0.642\scriptsize{$\pm$0.001} & 2.688\scriptsize{$\pm$0.003} & \cellcolor{second}44.008\scriptsize{$\pm$0.304} & \cellcolor{second}0.223\scriptsize{$\pm$0.001} \\

\midrule

DeepDOSReasoner
& \cellcolor{best}\textbf{0.691\scriptsize{$\pm$0.004}} & \cellcolor{best}\textbf{2.567\scriptsize{$\pm$0.002}} & \cellcolor{best}\textbf{38.142\scriptsize{$\pm$0.253}} & \cellcolor{best}\textbf{0.191\scriptsize{$\pm$0.005}} \\

\bottomrule
\end{tabular}
\end{sc}
\end{center}
\end{table*}

\subsubsection{Phonon density of states prediction}\label{sec:results:phdos}

To test whether the design transfers beyond electronic spectra, we benchmark DeepDOSReasoner on the phonon density of states (phDOS), which shares the eDOS's structure (a nonnegative one-dimensional density with a total-mass sum rule, $\int_0^{\infty} g(\omega)\,d\omega = 3N$ for a cell of $N$ atoms), but lives on the vibrational frequency axis. We apply the model without architectural change beyond replacing the energy axis with a frequency grid and substituting the analytic constant $Z=3N$ at inference, and evaluate on the held-out test set ($n=152$) of the phDOS dataset of Chen et al.~\citep{Chen2021} against the same baselines as for eDOS.

\begin{table*}[htbp]
\caption{\textbf{phDOS prediction performance on the original frequency scale ($n=152$ test materials).} We report mean $\pm$ one standard deviation over five independent runs. Darkly shaded cells indicate the best results and lightly shaded cells indicate the second best.}
\label{tab:phdos}
\begin{center}
\begin{sc}
\begin{tabular}{l*{4}{c}}
\toprule
Method
& $R^2 \uparrow$
& MAE $\downarrow$
& MSE $\downarrow$
& WD $\downarrow$ \\
\midrule

E3NN
& 0.506\scriptsize{$\pm$0.015} & 0.082\scriptsize{$\pm$0.003} & 0.026\scriptsize{$\pm$0.001} & 0.103\scriptsize{$\pm$0.004} \\

MEGNet
& 0.538\scriptsize{$\pm$0.014} & 0.080\scriptsize{$\pm$0.003} & 0.024\scriptsize{$\pm$0.001} & 0.094\scriptsize{$\pm$0.004} \\

GATGNN
& 0.270\scriptsize{$\pm$0.025} & 0.106\scriptsize{$\pm$0.002} & 0.038\scriptsize{$\pm$0.003} & 0.155\scriptsize{$\pm$0.008} \\

NequIP
& 0.327\scriptsize{$\pm$0.022} & 0.093\scriptsize{$\pm$0.004} & 0.035\scriptsize{$\pm$0.002} & 0.121\scriptsize{$\pm$0.006} \\

M3GNet
& 0.543\scriptsize{$\pm$0.012} & 0.075\scriptsize{$\pm$0.003} & 0.024\scriptsize{$\pm$0.001} & 0.075\scriptsize{$\pm$0.003} \\

MACE
& 0.013\scriptsize{$\pm$0.030} & 0.115\scriptsize{$\pm$0.004} & 0.052\scriptsize{$\pm$0.004} & 0.241\scriptsize{$\pm$0.012} \\

EquiformerV2
& 0.579\scriptsize{$\pm$0.010} & 0.074\scriptsize{$\pm$0.002} & 0.022\scriptsize{$\pm$0.001} & 0.073\scriptsize{$\pm$0.003} \\

\midrule

Mat2Spec
& 0.590\scriptsize{$\pm$0.010} & 0.072\scriptsize{$\pm$0.002} & 0.022\scriptsize{$\pm$0.001} & \cellcolor{second}0.069\scriptsize{$\pm$0.003} \\

DOSTransformer
& \cellcolor{second}0.598\scriptsize{$\pm$0.018} & \cellcolor{second}0.071\scriptsize{$\pm$0.001} & \cellcolor{second}0.020\scriptsize{$\pm$0.001} & 0.074\scriptsize{$\pm$0.003} \\

\midrule

DeepDOSReasoner
& \cellcolor{best}\textbf{0.688\scriptsize{$\pm$0.006}} & \cellcolor{best}\textbf{0.064\scriptsize{$\pm$0.002}} & \cellcolor{best}\textbf{0.016\scriptsize{$\pm$0.001}} & \cellcolor{best}\textbf{0.065\scriptsize{$\pm$0.002}} \\

\bottomrule
\end{tabular}
\end{sc}
\end{center}
\end{table*}

DeepDOSReasoner achieves the best phDOS prediction across all four metrics (\Cref{tab:phdos}). Relative to the strongest baseline per metric, it improves $R^2$ from 0.598 to 0.688, a $15.1\%$ relative gain, larger than the eDOS gain despite the smaller training set, lowers MAE from 0.071 to 0.064, reduces MSE from 0.020 to 0.016 ($-20\%$), and lowers WD from 0.069 to 0.065, separating from every baseline family by ${\geq}0.090$ in $R^2$.

\begin{figure*}[htbp]
    \centering
    \includegraphics[width=1\linewidth]{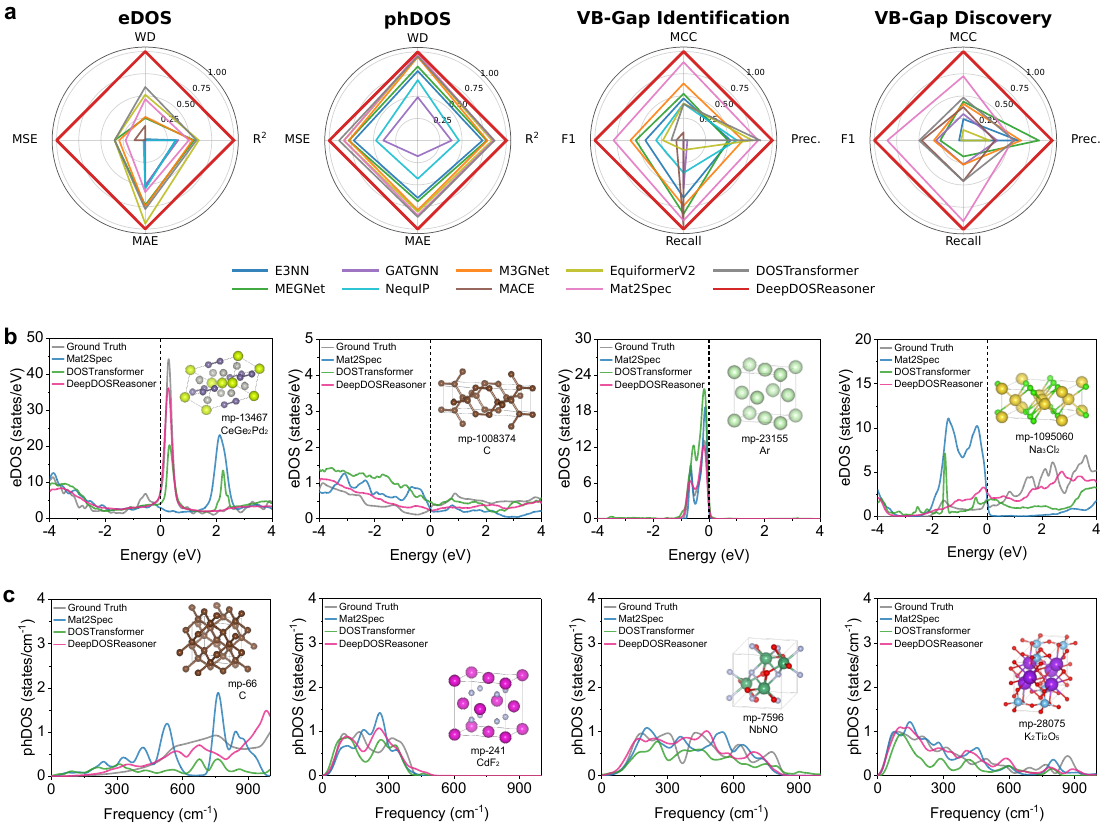}
    \caption{\textbf{Benchmark summary and representative spectrum-prediction examples.} (\textbf{a})~Per-task radar summary of relative performance across the four benchmark tasks, eDOS prediction, phDOS prediction, VB-gap identification, and prospective VB-gap discovery, with DeepDOSReasoner (red) shown against all baselines. For each task the four reported metrics are min-max normalized across methods so that the outer ring corresponds to the best value on every axis; DeepDOSReasoner reaches the best normalized value %
    on every axis of every task.
    (\textbf{b})~Representative eDOS predictions on materials from the primary eDOS benchmark: DeepDOSReasoner (pink) is shown alongside DFT ground truth (gray) and the two crystal-to-sequence baselines (Mat2Spec, DOSTransformer). (\textbf{c})~Representative phDOS predictions on materials from the held-out phDOS test set, with frequencies on the original cm$^{-1}$ axis. Additional eDOS and phDOS visualization examples are provided in \siViz.}\label{fig:examples}
\end{figure*}

This cross-physics transfer is essentially automatic: the continuity-equation form, no-flux boundary conditions, and upwind CDF integration are agnostic to the physical meaning of the spectral coordinate, so switching to phDOS requires only the frequency grid and the $3N$ sum rule, with no change to the velocity-field architecture, loss, or discretization. And the gain comes in a regime where the test set is only 152 materials and two baselines collapse outright (MACE $R^2{=}0.013$, GATGNN $R^2{=}0.270$). Representative predictions reproducing the dominant acoustic and optical structure are shown in \Cref{fig:examples}(c) and \siViz.

\subsection{Rare near-Fermi-energy feature identification and discovery}\label{sec:results:features}

\subsubsection{VB-gap identification}\label{sec:results:vb}
A stringent measure of eDOS quality is whether a predicted spectrum preserves rare near-Fermi-energy features. We focus on VB gaps, contiguous windows of suppressed density immediately below the Fermi level, which occur in only ${\sim}5\%$ of the test set but are linked to unconventional transport, thermoelectricity, and transparent conduction~\citep{ricciGappedMetalsThermoelectric2020a}. Adapting the criterion of Kong et al.~\citep{Kong2022}, we classify a material as VB-gapped from a four-condition rule on its predicted spectrum: the density must fall below $0.9$~state/eV across a contiguous window at least $1$~eV wide whose high-energy edge lies in $[-1.5,-0.25]$~eV (implementation details in \siEval). The rule is applied identically to every method so that differences reflect the predicted spectra rather than threshold tuning.

Because positives are rare (${\sim}5\%$), we report precision, recall, F1, and Matthews correlation coefficient (MCC), the last remaining informative under strong imbalance. DeepDOSReasoner achieves the best result on all four simultaneously, precision 0.904, recall 0.471, F1 0.619, MCC 0.640 (\Cref{tab:vb_summary}), improving precision by $8.5\%$ and F1 by $6.4\%$ over the next-best method. The runner-up on precision (DOSTransformer) and on recall (MACE) are different baselines, so DeepDOSReasoner is the only method that leads both ends of the precision--recall trade-off at once.

The baselines fall into the two failure modes this benchmark exposes: over-flagging (MACE and GATGNN, precision below 0.42) and over-conservatism (DOSTransformer: precision 0.833 but recall 0.296), both rooted in unconstrained regression that does not control spectral mass inside the near-Fermi-energy window. DeepDOSReasoner avoids both, exceeding DOSTransformer on precision \emph{and} recall, because the mass-conserving Stage~II transports mass out of genuinely depleted windows rather than regressing small values within them, while source attention identifies which windows are physically depleted (the ablation analysis below confirms the reasoning stage yields its largest gain on WD, the metric most tied to gap leakage).

\begin{table*}[htbp]
\caption{\textbf{VB-gap identification performance.} We report mean $\pm$ one standard deviation over five independent runs. Darkly shaded cells denote the best results and lightly shaded cells denote the second best.}
\label{tab:vb_summary}
\begin{center}
\begin{sc}
\begin{tabular}{l*{4}{c}}
\toprule
Method
& Prec. $\uparrow$ & Recall $\uparrow$ & F1 $\uparrow$ & MCC $\uparrow$ \\
\midrule

E3NN
& 0.661\scriptsize{$\pm$0.023} & 0.409\scriptsize{$\pm$0.024} & 0.499\scriptsize{$\pm$0.038} & 0.499\scriptsize{$\pm$0.021} \\

MEGNet
& 0.646\scriptsize{$\pm$0.014} & 0.441\scriptsize{$\pm$0.021} & 0.524\scriptsize{$\pm$0.049} & 0.513\scriptsize{$\pm$0.022} \\

GATGNN
& 0.411\scriptsize{$\pm$0.015} & 0.391\scriptsize{$\pm$0.035} & 0.399\scriptsize{$\pm$0.027} & 0.373\scriptsize{$\pm$0.012} \\

NequIP
& 0.698\scriptsize{$\pm$0.018} & 0.360\scriptsize{$\pm$0.015} & 0.471\scriptsize{$\pm$0.029} & 0.483\scriptsize{$\pm$0.013} \\

M3GNet
& 0.734\scriptsize{$\pm$0.011} & 0.425\scriptsize{$\pm$0.019} & 0.538\scriptsize{$\pm$0.034} & 0.544\scriptsize{$\pm$0.034} \\

MACE
& 0.389\scriptsize{$\pm$0.013} & \cellcolor{second}0.467\scriptsize{$\pm$0.022} & 0.418\scriptsize{$\pm$0.025} & 0.397\scriptsize{$\pm$0.029} \\

EquiformerV2
& 0.787\scriptsize{$\pm$0.008} & 0.315\scriptsize{$\pm$0.008} & 0.453\scriptsize{$\pm$0.034} & 0.484\scriptsize{$\pm$0.030} \\

\midrule

Mat2Spec
& 0.822\scriptsize{$\pm$0.014} & 0.454\scriptsize{$\pm$0.010} & \cellcolor{second}0.582\scriptsize{$\pm$0.025} & \cellcolor{second}0.608\scriptsize{$\pm$0.029} \\

DOSTransformer
& \cellcolor{second}0.833\scriptsize{$\pm$0.098} & 0.296\scriptsize{$\pm$0.059} & 0.431\scriptsize{$\pm$0.132} & 0.482\scriptsize{$\pm$0.128} \\

\midrule

DeepDOSReasoner
& \cellcolor{best}\textbf{0.904\scriptsize{$\pm$0.017}}
& \cellcolor{best}\textbf{0.471\scriptsize{$\pm$0.024}}
& \cellcolor{best}\textbf{0.619\scriptsize{$\pm$0.023}}
& \cellcolor{best}\textbf{0.640\scriptsize{$\pm$0.011}} \\

\bottomrule
\end{tabular}
\end{sc}
\end{center}
\end{table*}

\subsubsection{Prospective VB-gap discovery}\label{sec:results:vb_discovery}
To turn this into a discovery setting, we screen $8{,}106$ nonmagnetic Materials Project entries that lack a computed eDOS (filtered for proximity to the convex hull, fewer than five elements, and metallic character, following Kong et al.~\citep{Kong2022}). We rank the pool by a gap-likeness score, the minimum of the predicted eDOS over a fixed ${\sim}1$~eV window immediately below the valence-band maximum, so that the lowest scores mark spectra that dip closest to zero exactly where a VB gap's high-energy edge would sit. We assemble a 100-material evaluation set from the top-50 candidates of DeepDOSReasoner and the top-50 of Mat2Spec, the strongest in-distribution baseline (the two lists are disjoint), then compute ground-truth labels with new DFT calculations; this enrichment guarantees enough genuine positives to estimate recall and F1, and the labels score every method on identical candidates (protocol in \siDisc). Of the 100 candidates, 44 are genuine VB-gap materials.

\begin{figure*}[htbp]
    \centering
    \includegraphics[width=1\linewidth]{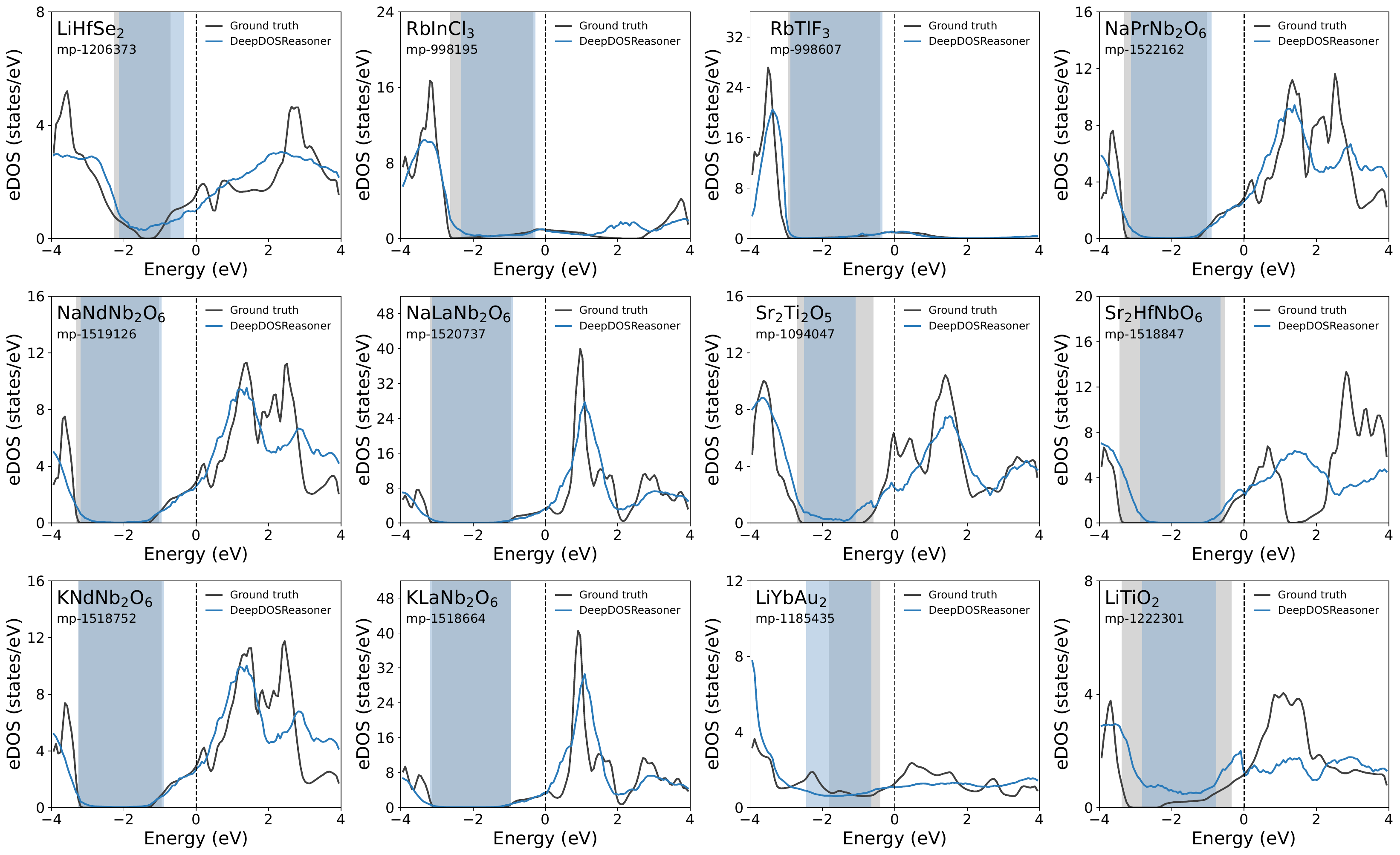}
    \caption{\textbf{Prospective VB-gap discovery visualization.} Predicted eDOS (blue) is shown alongside DFT ground truth (gray) for twelve representative candidates flagged by DeepDOSReasoner from the $8{,}106$-material discovery pool of nonmagnetic Materials Project entries lacking a published DFT eDOS. The shaded blue window in each panel marks the contiguous sub-Fermi-energy region that satisfies the VB-gap detection criterion; the gray shading indicates the corresponding window obtained from the DFT reference. The visualization complements the precision/recall/F1/MCC numbers in \Cref{tab:vb_discovery}, showing how DeepDOSReasoner places suppressed-density windows on previously unseen compositions while preserving the surrounding band-edge structure.}\label{fig:vb_discovery}
\end{figure*}

On this out-of-distribution set, DeepDOSReasoner attains the best precision (0.786), recall (0.500), F1 (0.611), and MCC (0.434), outperforming all nine baselines on every metric (\Cref{fig:examples}(a); \Cref{tab:vb_discovery}) and recovering 22 of the 44 confirmed VB-gap materials. Measured on each model's own ranked list, 25 of DeepDOSReasoner's 50 top-ranked candidates are DFT-confirmed VB-gap materials, versus 19 of 50 for Mat2Spec: a hit rate of $50\%$ against the ${\sim}5\%$ VB-gap prevalence of the labeled test set, i.e., a ${\sim}10\times$ enrichment. The confirmed discoveries cluster in chemically coherent families, including alkali--rare-earth niobates (e.g., NaLaNb$_2$O$_6$, KNdNb$_2$O$_6$), ternary halides (e.g., RbInCl$_3$, RbTlF$_3$), and chalcogenides (e.g., LiHfSe$_2$) (\Cref{fig:vb_discovery}), classes in which suppressed sub-Fermi-energy density is linked to transparent conduction and thermoelectric behavior~\citep{ricciGappedMetalsThermoelectric2020a}. The capability therefore does not collapse on a pool never seen in training and lacking any DFT reference; the MCC margin over the runner-up is in fact \emph{larger} out of distribution than in distribution (0.068 versus 0.032, an $18.6\%$ relative gain over Mat2Spec), consistent with the hard physical priors of Stage~II transferring beyond the training distribution. The per-baseline pattern reinforces this: the crystal-to-sequence models transfer most gracefully (Mat2Spec is the runner-up on recall, F1, and MCC), whereas the structure-encoder and interatomic-potential baselines compress into a narrow band (precision 0.65--0.77, recall 0.25--0.36), so stronger encoder inductive bias on the in-distribution benchmark does not translate into reliable discovery on unseen compositions. Representative flagged candidates with their DFT validations are shown in \Cref{fig:vb_discovery}; because the set is enriched from two models' top candidates rather than drawn uniformly, it supports a head-to-head comparison rather than an absolute precision estimate for the full pool.

\begin{table*}[htbp]
\caption{\textbf{Prospective VB-gap discovery.} Performance on 100 nonmagnetic Materials Project entries lacking a DFT eDOS, comprising the 50 top-ranked candidates of DeepDOSReasoner and the 50 top-ranked candidates of Mat2Spec under the gap-likeness screening score of \siDisc, drawn from the same $8{,}106$-material pool; the enrichment ensures enough DFT-validated VB-gap positives to estimate recall and F1. All methods are scored once against the same DFT-derived labels on this single fixed candidate set, so single values are reported. Darkly shaded cells denote the best results and lightly shaded cells denote the second best.}
\label{tab:vb_discovery}
\begin{center}
\begin{sc}
\begin{tabular}{l*{4}{c}}
\toprule
Method
& Prec. $\uparrow$ & Recall $\uparrow$ & F1 $\uparrow$ & MCC $\uparrow$ \\
\midrule

E3NN
& 0.733 & 0.250 & 0.373 & 0.248 \\

MEGNet
& \cellcolor{second}0.765 & 0.295 & 0.426 & 0.296 \\

GATGNN
& 0.700 & 0.318 & 0.438 & 0.262 \\

NequIP
& 0.647 & 0.250 & 0.361 & 0.189 \\

M3GNet
& 0.737 & 0.318 & 0.444 & 0.290 \\

MACE
& 0.696 & 0.364 & 0.478 & 0.281 \\

EquiformerV2
& 0.688 & 0.250 & 0.367 & 0.218 \\

\midrule
Mat2Spec
& 0.724 & \cellcolor{second}0.477 & \cellcolor{second}0.575 & \cellcolor{second}0.366 \\

DOSTransformer
& 0.727 & 0.364 & 0.485 & 0.307 \\

\midrule

DeepDOSReasoner
& \cellcolor{best}\textbf{0.786}
& \cellcolor{best}\textbf{0.500}
& \cellcolor{best}\textbf{0.611}
& \cellcolor{best}\textbf{0.434} \\

\bottomrule
\end{tabular}
\end{sc}
\end{center}
\end{table*}

\subsection{Transfer to functional-materials classes}\label{sec:results:applications}

\subsubsection{Substituent-induced spectral changes}\label{sec:results:doped}
A central use of an eDOS predictor is to survey how electronic structure responds to chemical substitution. We test two Materials Project parents that bracket contrasting regimes: CdI$_2$ (\texttt{mp-569994}), an ionic layered halide where Cl-for-I substitution should reshape the spectrum, and Dy$_4$Cu$_2$In$_6$ (\texttt{mp-1225405}), an intermetallic where isovalent Ga-for-In substitution should barely perturb it. Together they stress-test both failure modes at once: a parent-memorizing model would miss the first, an over-correcting model would spuriously perturb the second. We evaluate four substituted structures (CdCl$_{0.5}$I$_{1.5}$, CdICl, Dy$_4$Cu$_2$In$_5$Ga, Dy$_4$Cu$_2$In$_3$Ga$_3$), all absent from training, against DFT references.

\begin{figure*}[htbp]
    \centering
    \includegraphics[width=1\linewidth]{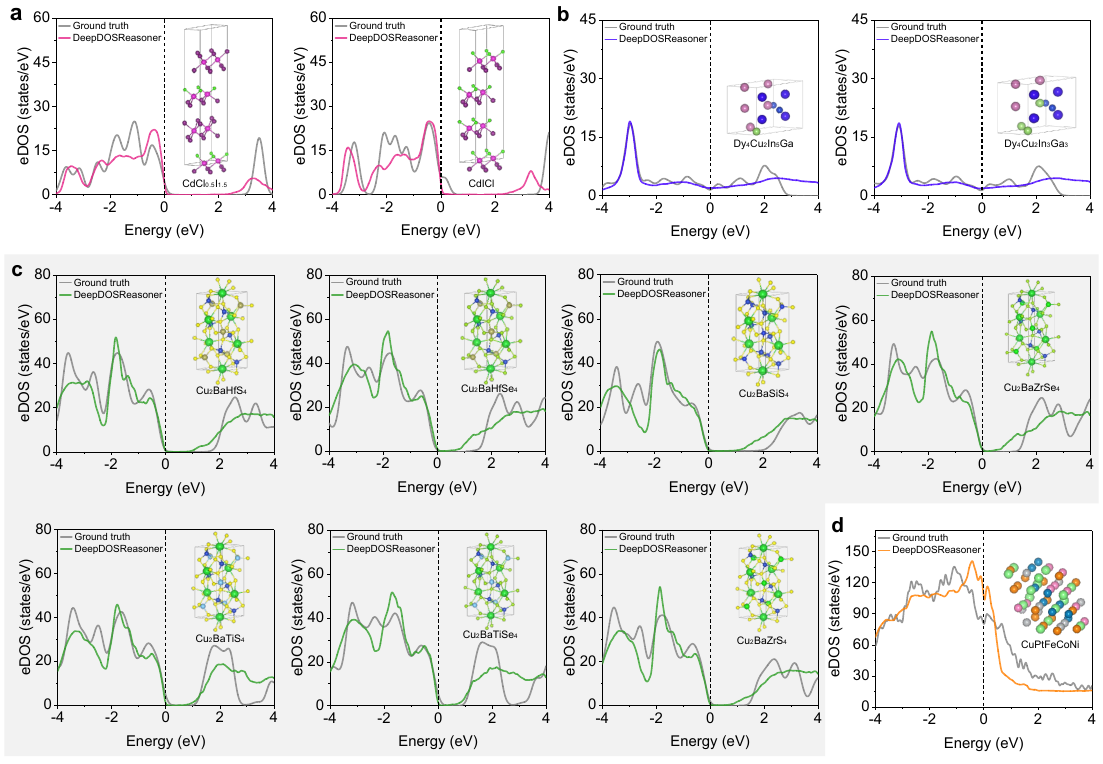}
    \caption{\textbf{Spectrum prediction case studies across material classes.} Predicted DeepDOSReasoner spectra (colored) are shown alongside DFT-derived references (gray). (\textbf{a})~Cl-for-I substitutional variants of CdI$_2$ (\texttt{mp-569994}): CdCl$_{0.5}$I$_{1.5}$ and CdICl. (\textbf{b})~Ga-for-In substitutional variants of Dy$_4$Cu$_2$In$_6$ (\texttt{mp-1225405}): Dy$_4$Cu$_2$In$_5$Ga and Dy$_4$Cu$_2$In$_3$Ga$_3$. (\textbf{c})~Seven I$_2$--II--IV--X$_4$ quaternary chalcogenide semiconductors of the form Cu$_2$Ba$M$X$_4$ ($M\in\{$Hf, Si, Ti, Zr$\}$, $X\in\{$S, Se$\}$): the presence of the band gap, the valence-band edge, and the dominant peaks are reproduced across the set. (\textbf{d})~CuPtFeCoNi high-entropy alloy: the model captures the gapless metallic spectrum and recovers the eDOS center $\bar\varepsilon$ with an absolute error of $0.20$~eV ($-1.16$~eV predicted versus $-0.97$~eV DFT). Panels (a) and (b) are discussed in the ``Substituent-induced spectral changes'' section, panel (c) in the ``Chalcogenide semiconductor case study'' section, and panel (d) in the ``High-entropy alloy case study'' section.}\label{fig:apps}
\end{figure*}

DeepDOSReasoner tracks the DFT trends in both regimes (\Cref{fig:apps}(a,b)). For the Cd--I--Cl variants, the predicted valence-peak amplitudes and band-edge positions shift with Cl content rather than transferring a fixed parent spectrum, consistent with the higher electronegativity of Cl deepening the anion-derived valence band. For the isovalent Dy$_4$Cu$_2$In$_{6-x}$Ga$_x$ variants, whose DFT spectra are nearly identical, the model returns the near-invariant spectrum, preserving the sharp Cu~3d feature near $-3$~eV, without manufacturing spurious differences from the altered input. In each panel the variant-to-variant differences in the predictions closely track those in the DFT reference, including their near-absence in panel~(b), so the model responds to substitution rather than interpolating to a generic halide or intermetallic profile, behavior well suited to high-throughput composition-variation workflows.

\subsubsection{Chalcogenide semiconductor case study}\label{sec:results:chalcogenide}

I$_2$--II--IV--X$_4$ quaternary chalcogenides, exemplified by Cu$_2$ZnSnS$_4$~\citep{zhu2017i2,sun2020structural,woods2020wide}, are earth-abundant thin-film photovoltaic candidates whose suitability depends sensitively on electronic structure~\citep{ju2017perovskite}: a Shockley--Queisser-relevant gap, sharp band edges, and the absence of mid-gap states. All three signatures can be read directly from the eDOS, making this family a natural application target and a stringent test of whether predicted spectra preserve photovoltaic-relevant detail.

We test DeepDOSReasoner on compositions from Wang et al.~\citep{wang2021unsupervised} without any finetuning; DFT references are computed as described in Materials and Methods.
Across seven representative Cu$_2$Ba$M$X$_4$ compositions ($M\in\{$Hf, Si, Ti, Zr$\}$, $X\in\{$S, Se$\}$), the model recovers the band gap as a contiguous low-DOS region, the sharp valence-band edge that sets the absorption onset, and the dominant peak structure on either side of the gap, tracking the composition-dependent shifts through the metal-$d$ and chalcogen-$p$ states (\Cref{fig:apps}(c)). The predicted gaps remain visually flat across all panels while the band edges stay steep. The very signatures that disqualify a candidate, namely mid-gap states and smeared edges, are precisely the artifacts that pointwise-regression baselines tend to produce.

\subsubsection{High-entropy alloy case study}
\label{sec:results:hea}

High-entropy alloys (HEAs) are another out-of-distribution stress test: they combine five elements at near-equimolar ratios in large disordered supercells and exhibit broad, gapless near-Fermi-energy spectra. We compare predicted and DFT-derived total eDOS for the CuPtFeCoNi alloy reported by Han et al.~\citep{han2024highentropy}. Among the publicly available HEA studies, this system provided the clearest validation case that simultaneously included a sufficiently specified atomic model and a DFT-derived total eDOS.
DeepDOSReasoner reproduces the alloy's gapless metallic character (no spurious in-gap features), the broad transition-metal-derived peak dominating the occupied side, and the gradual decay into the conduction band, without finetuning (\Cref{fig:apps}(d)).

As a quantitative check we evaluate the \emph{eDOS center} $\bar\varepsilon$, the first energy moment of the spectrum and a total-spectrum analogue of the d-band center that serves as a scalar catalytic-activity descriptor in the Hammer--N\o{}rskov framework~\citep{hammer2000theoretical}. Because $\bar\varepsilon$ is a normalized first moment, its accuracy depends on placing spectral mass correctly across the whole energy window, exactly what the mass-conserving stage keeps well-posed, since the refined spectrum is a true normalized density and $\bar\varepsilon$ a bona fide expectation rather than a ratio of two regression outputs. The predicted $\bar\varepsilon_{\mathrm{pred}}=-1.16$~eV differs from $\bar\varepsilon_{\mathrm{DFT}}=-0.97$~eV by only $0.20$~eV, modest against the ${\sim}1$~eV d-band dynamic range; we verified beforehand that DeepDOSReasoner predicts $\bar\varepsilon$ most accurately of all methods across the held-out test set (MAE $0.138$~eV, $R^2=0.921$; \siTabCenter, \siCenter), justifying its use here.

\subsection{Empirical analyses of the framework}\label{sec:results:analyses}

\begin{figure*}[htbp]
    \centering
    \includegraphics[width=1.0\linewidth]{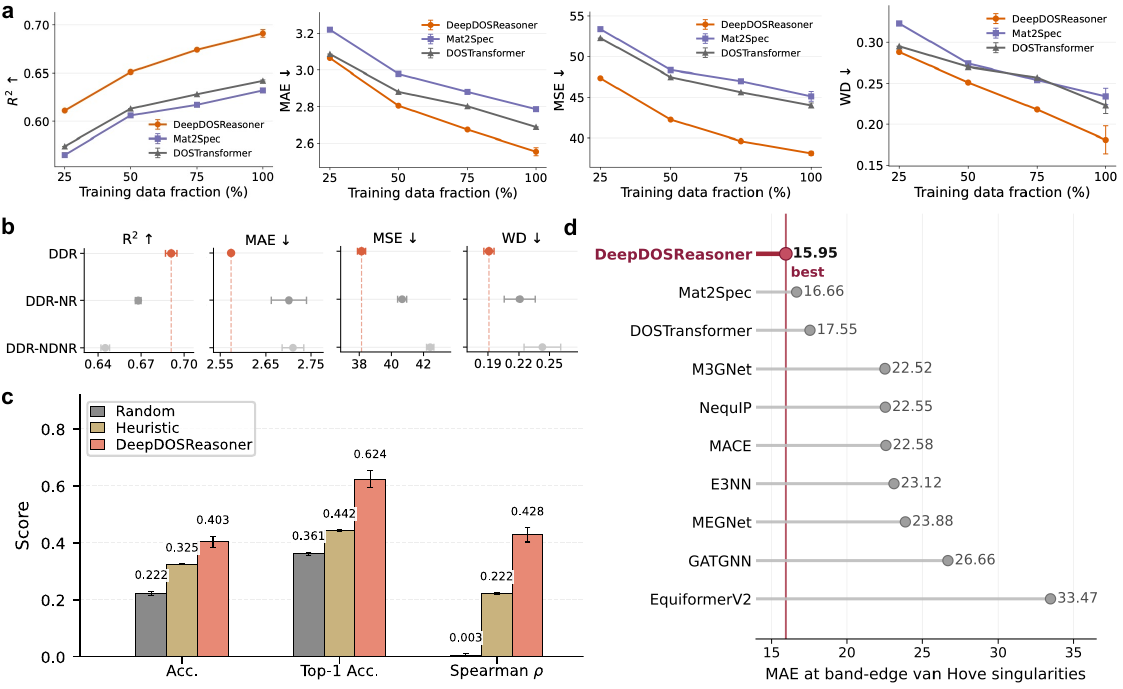}
    \caption{\textbf{Empirical analyses of DeepDOSReasoner.} (\textbf{a})~Data-scaling behavior on eDOS prediction as the training-set fraction grows from 25\% to 100\%, reported for $R^2$, MAE, MSE, and WD against the two crystal-to-sequence baselines (Mat2Spec, DOSTransformer). (\textbf{b})~Ablation study isolating the contribution of structure-aware spectrum decoding and constraint-preserving physical reasoning to overall eDOS performance: DDR is the full DeepDOSReasoner model, DDR-NR removes the constraint-preserving physical reasoning stage, and DDR-NDNR removes both the structure-aware spectrum decoding and the constraint-preserving physical reasoning (corresponding numbers in \siTabAbl). (\textbf{c})~Element-resolved contribution ranking on the eDOS test set: overall accuracy, top-1 accuracy, and Spearman $\rho$ of the element rankings recovered from source-attention weights (DeepDOSReasoner), compared against a random-ordering baseline and a composition-plus-oxidation-state chemistry heuristic (a trained-model-free baseline whose details are given in \siEle), all scored against DFT-derived element-resolved contributions. (\textbf{d})~Peak-restricted mean absolute error on band-edge van Hove singularities: DeepDOSReasoner attains the lowest value of any method evaluated on the same eDOS test set as \Cref{tab:edos}.}\label{fig:summary}
\end{figure*}

We characterize the framework through five analyses (\Cref{fig:summary}); full numbers and protocols are given in \siRes.

\paragraph{Ablation}
Both stages are load-bearing (\Cref{fig:summary}(b); \siAbl, \siTabAbl). Removing the constraint-preserving reasoning stage (DDR-NR) degrades every metric, most sharply WD ($+15.7\%$), the signature of misplaced spectral mass; additionally removing source attention (DDR-NDNR) drops the model to $R^2{=}0.645$, exactly the strongest-baseline level of \Cref{tab:edos}, while its WD degrades by $27.2\%$ overall, so adding each component in turn lifts the model from the top of the baseline envelope into a regime no baseline reaches. Reasoning is the larger contributor on the geometry-sensitive metrics (MAE, MSE, WD) and source attention on $R^2$, so the two components are complementary rather than redundant.

\paragraph{van Hove singularities}
On peak-restricted MAE over the band-edge van Hove singularities, DeepDOSReasoner is the most accurate method ($15.95$), improving $4.3\%$ over the next-best method (Mat2Spec, $16.66$) and $9.1\%$ over DOSTransformer ($17.55$) (\Cref{fig:summary}(d); \siVH). The ordering separates into three tiers: the two crystal-to-sequence baselines sit closest, the structure encoders and interatomic potentials cluster between $22$ and $27$, and EquiformerV2, the strongest baseline on aggregate $R^2$, becomes the \emph{worst} ($33.47$). High aggregate fit can thus be obtained by smearing peaks, a failure mode the mass-conserving refinement avoids by construction: under the continuity equation it can only move mass onto peaks, not thin them.

\paragraph{Emergent element resolution}
Although trained only on total spectra, the source-attention weights recover element-resolved contributions with no element-level supervision, reaching Spearman $\rho=0.428$ (vs.\ $0.003$ for random ordering) and top-1 accuracy $0.624$, and surpassing a composition-plus-oxidation-state chemistry heuristic that itself beats random ordering ($\rho=0.222$, top-1 $0.442$; \Cref{fig:summary}(c); \siEle). The decomposition is therefore an emergent property of structure-aware decoding rather than a fitted attribution, capturing structure-dependent contributions beyond what composition-only chemistry provides.

\paragraph{Data scaling} 
DeepDOSReasoner is the most sample-efficient method at every training fraction (\Cref{fig:summary}(a); \siScaling): its advantage over the baseline envelope is already established at $25\%$ of the training data and does not close as data grows, and at $25\%$ it already matches full-data baselines on several metrics, with the largest separation on MSE and WD.

\paragraph{Stable-only control} 
Restricting both training and test to the $14{,}055$ thermodynamically stable (on-hull) entries, DeepDOSReasoner remains best on all four metrics (\siTabStable; \siMetastable). Across methods, removing metastable structures degrades $R^2$, MAE, and MSE uniformly across architecturally diverse models, ruling out method-specific noise and identifying the metastable subset as genuinely useful supervision. Notably, DeepDOSReasoner's MAE lead over the strongest baseline \emph{grows} from $0.038$ to $0.116$ when metastable data are removed, the only metric whose inter-method gap widens, indicating that the model preserves per-bin accuracy under data scarcity better than the pooled-embedding baselines.

\paragraph{Generalization}
On the eDOS test set, DeepDOSReasoner's error is broadly distributed rather than concentrated on any regime (\siRobust). Prediction error grows only gently with material complexity (the number of distinct elements and unit-cell sites), and, most tellingly, test materials with few training neighbors of the same chemical system or space group retain comparably low error, indicating that the model generalizes beyond memorized training-set neighborhoods rather than interpolating among them. The residual error distributes broadly across material subclasses, with high-symmetry cubic structures the easiest, and correlates most strongly with the weight of intermediate-frequency spectral structure (Pearson $r$ up to $0.53$, versus $0.12$ for the smooth lowest-frequency band), the band-edge and critical-point regime that the mass-conserving stage is designed to sharpen, rather than with the smooth spectral envelope.

%% file: sections/discussion.tex
\section{DISCUSSION}

We introduced DeepDOSReasoner, a structure-aware, physics-grounded framework for DOS prediction that combines source-attentive spectrum decoding with mass-conserving physical reasoning over the spectrum. Across eDOS prediction,  
rare VB-gap identification,
prospective VB-gap discovery validated by new DFT calculations, 
and phDOS prediction without architectural change, DeepDOSReasoner consistently outperforms  existing learning-based DOS  methods, attaining the best result on every reported metric for each of these tasks. \textit{Notably, it is the only method that simultaneously improves spectrum fidelity, preservation of physically important spectral features, and downstream materials-discovery performance.}
Three application studies further demonstrate transfer beyond the primary benchmark distribution: chemically reasonable substituent-induced changes in two parent compounds whose substituted variants are absent from the training set, seven photovoltaic-relevant I$_2$--II--IV--X$_4$ chalcogenide compositions, and a CuPtFeCoNi high-entropy alloy whose predicted eDOS closely agrees with DFT.

A key conceptual contribution of this work is a demonstration of how to make physical conservation laws \emph{govern} a learned prediction rather than merely regularize it: physical consistency is most effective when built into the prediction process itself. Stage~II refines the normalized spectrum by transporting mass under a continuity equation, integrated by a differentiable, monotone CDF-upwind scheme, so nonnegativity and total-state conservation hold step by step for \emph{any} velocity field the network learns. Both training and inference are therefore confined to the manifold of physically valid spectra: conservation is a property of the model's output, not a penalty competing with the data-fit loss. This is what makes the reasoning stage essential rather than a marginal refinement:
removing it degrades the model toward the baseline envelope, most sharply on the geometry-sensitive metrics (see Results). It also explains why the gains concentrate on exactly the sharp, physically meaningful features that pointwise regression tends to wash out: band edges, sub-Fermi VB gaps, and band-edge van Hove singularities. Because the refinement can only move mass and never manufacture or delete it, an aggregate metric cannot be improved by smearing a peak or leaking density into a gap, and integrated descriptors such as the eDOS center remain bona fide expectations rather than ratios of unconstrained outputs. Source attention supplies the complementary ingredient, grounding each spectral coordinate in atom-level context; that the decoder recovers element-resolved contributions without ever seeing element labels is independent evidence that it learns the physically motivated decomposition of the total DOS rather than an opaque global representation.

These choices distinguish DeepDOSReasoner from prior crystal-to-spectrum work. Early models cast DOS prediction as pointwise regression on a single global crystal representation~\citep{fung2022physically,kaundinya2022prediction}, and recent ``crystal-to-sequence'' formulations decode the spectrum as an ordered sequence~\citep{lee2023density,Kong2022}; related sequence and regression approaches have been applied to other structured spectra, including X-ray absorption~\citep{carbone2020machine,kotobi2023integrating,lin2025universal}, mass spectra~\citep{nguyen2024advancing,goldman2024generating,youngfragnnet,wang2025neural}, and nuclear magnetic resonance~\citep{cobas2020nmr,yang2021predicting}. DeepDOSReasoner departs from all of these by grounding each energy bin in atom-level embeddings rather than a global embedding, by enforcing normalization and mass conservation by construction rather than through soft penalties, and by refining the spectrum through a conservation-preserving transport process rather than an unconstrained pointwise correction, moving physical constraints from the role of auxiliary regularizers to that of governing principles within the prediction process itself.

Three limitations qualify these conclusions and mark natural next steps. First, the current formulation assumes a fixed spectral grid; variable energy windows, adaptive discretizations, and material-dependent grids would require extending the velocity-field network to variable-length spectra, but would not change the underlying continuity-equation formulation. Second, accuracy is bounded by the fidelity of the supervision: the model is trained on PBE-level spectra and inherits their systematic errors, so predicted gaps should be read as qualitative indicators, and quantitative screening of properties such as absorption strength or carrier lifetimes would require higher-level reference data and complementary calculations. Third, the model predicts a single spin-unpolarized total DOS; extending the decoder to spin-resolved channels is conceptually straightforward and would broaden coverage of magnetic materials.

Beyond DOS, DeepDOSReasoner offers a reusable blueprint for learning structured scientific spectra governed by conservation laws and geometric constraints. Because the decoder is agnostic to the physical meaning of the spectral axis and the refinement requires only a conserved quantity expressible as a continuity equation, the same design should extend to other spectroscopies, such as X-ray absorption, mass spectra, and nuclear magnetic resonance, where pointwise regression routinely violates the underlying physics and where attention from atomic context to spectral coordinates carries physical meaning. As an immediate step in this direction, we release DeepDOSReasoner predictions filling the missing eDOS for $91{,}405$ materials and the missing phDOS for $127{,}768$ materials in the Materials Project; predictions for materials far from the respective training distributions, such as magnetic systems for eDOS, should be interpreted with commensurate care. More broadly, the work shows how data-driven learning and physical reasoning can be unified so that constraints derived from first principles shape predictions directly rather than through penalties or post-processing. We expect this direction to be productive wherever predictions must be both quantitatively accurate and physically faithful, and wherever the downstream payoff is materials discovery rather than spectrum reconstruction alone.

%% file: sections/methods.tex
\section{MATERIALS AND METHODS}
DeepDOSReasoner is organized around three design decisions that distinguish it from prior crystal-to-spectrum models and that together address the three failure modes identified in the Introduction. \emph{First, the decoder grounds every energy bin in atom-level embeddings via source (atom-to-energy) attention}, rather than reading the eDOS off a single pooled crystal representation. This preserves the element-projected structure that physically composes the total eDOS and yields interpretable contributions as a by-product of training on the total spectrum alone (see ``Structure-aware spectrum decoding''). \emph{Second, constraint-preserving reasoning through mass-conserving refinement is performed in probability space under a continuity equation, rather than as a pointwise correction in regression space}. This makes total-state conservation and nonnegativity properties of the model output rather than soft penalties added to a loss, so that the band gap and the surrounding spectrum are resolved jointly under a single conservation law (see ``Constraint-preserving physical reasoning''). \emph{Third, the velocity field that drives the refinement is trained end-to-end against the DFT-derived eDOS through a differentiable upwind integrator on the spectrum's cumulative distribution function (CDF)}. Conservation is enforced by the integrator itself, not by the training loss, so any velocity field the network learns yields a valid normalized density at every step; gradients of a final-output spectrum-matching loss can therefore be back-propagated through the full multi-step transport rollout without breaking the conservation guarantees (see ``Constraint-preserving physical reasoning''). The remainder of this section formalizes these choices in turn: the datasets; the problem formulation and physical constraints; the two stages; implementation, baselines, and evaluation; and dataset generation and the first-principles validation calculations.

\subsection{Datasets}\label{sec:methods:data}
\paragraph{eDOS benchmark.}
We curate an eDOS benchmark from Materials Project structures and DFT electronic-structure calculations~\citep{Jain2013,Kong2022}. Each raw DOS is (i) aligned to its DFT-computed Fermi level, (ii) broadened by Gaussian smearing with $\sigma=0.5$~eV, and (iii) area-preservingly rebinned onto a uniform 128-bin grid over $[-4,4]$~eV, so that integration of the discretized spectrum recovers the integral of the smoothed continuous spectrum. Unlike prior work that removes the energy interval between the valence-band maximum (VBM) and conduction-band minimum (CBM) for gapped materials, we retain the full eDOS so that band gaps are represented explicitly. The primary benchmark spans both thermodynamically stable and metastable Materials Project entries, with metastable entries dominating the dataset; it comprises $37{,}565$ crystals partitioned 80\%/10\%/10\% into training, validation, and test sets (test $n=3{,}756$), and we use this full set as our primary training and evaluation distribution. To analyze how metastable structures contribute to model performance, we additionally consider a stable-only subset comprising the structures with energy above hull equal to zero eV/atom, with the same preprocessing pipeline applied. A controlled comparison between the two protocols is reported in Results.

\paragraph{phDOS benchmark.}
For the phonon density of states experiments (see Results), we use the DFT-computed phDOS dataset released by Chen et al.~\citep{Chen2021}, which comprises $1{,}524$ crystalline materials drawn from the Materials Project with phonon spectra computed by density-functional perturbation theory~\citep{Petretto2018} as part of the high-throughput Materials Project phonon database. In the source release, the raw high-resolution DFPT phDOS is first smoothed with a Savitzky--Golay filter (window length 101, polynomial order 3) to attenuate finite-$q$-grid sampling artifacts while preserving the dominant peaks, and the smoothed spectrum over $[0,1000]$~cm$^{-1}$ is then interpolated onto a uniform 51-point frequency grid (spacing $20$~cm$^{-1}$). We adopt the same train/validation/test partition as the source release, so the held-out test set contains $n=152$ materials. At inference time, the absolute phDOS is recovered from our predicted normalized shape by rescaling to the analytic modes count $Z=3N$ (see ``Problem formulation and physical constraints'').

Full dataset curation and split details for both benchmarks are provided in \siHyper.

\subsection{Problem formulation and physical constraints}\label{sec:methods:pre}
\paragraph{Problem formulation.}
We represent an input crystal $c$ as a graph $G_c = (V, \mathcal{E})$, where
$V = \{u_i\}_{i=1}^N$ is the set of $N$ atomic nodes and each node
$u_i \in \mathbb{R}^{d_u}$ denotes an atom embedding (e.g., element type and other attributes).
The edge set $\mathcal{E} = \{(i,j)\}$ encodes interatomic relations; with each edge
$(i,j) \in \mathcal{E}$ we associate the scalar interatomic distance $r_{ij} \in \mathbb{R}$,
which is expanded into a $d_e$-dimensional edge embedding before entering the encoder (see ``Structure-aware spectrum decoding'').
In crystalline systems, atomic arrangements repeat periodically in three-dimensional space;
we construct $\mathcal{E}$ using a local neighborhood graph by limiting each atom to at most
$N_{\max}$ neighbors for computational efficiency, so that the node degree is bounded by $N_{\max}$.
Given a crystal $c$, our goal is to predict a target spectrum
$y \in \mathbb{R}^L$ (e.g., eDOS) defined on a fixed discretized grid
(energy $\varepsilon$ for eDOS), where $L$ is the spectrum length, i.e., the number of energy bins ($L=128$ for the eDOS benchmark).
The learning problem is to approximate a mapping $f : G_c \;\mapsto\; y$.

\paragraph{Physical constraints of DOS.}
A density of states $D(\varepsilon)$, electronic ($D(E)$, eDOS) or phononic ($g(\omega)$, phDOS), can be viewed as a (possibly unnormalized) density over a bounded spectral interval $\varepsilon\in[a,b]$.
Let $D(\varepsilon)\ge 0$ and define its normalized form as
$p(\varepsilon) := \frac{D(\varepsilon)}{\int_a^b D(\varepsilon')\,d\varepsilon'}.$
Physical validity requires \emph{nonnegativity}, $p(\varepsilon)\ge 0$ for all $\varepsilon\in[a,b]$, and \emph{total-state (mass) conservation},
$\int_a^b p(\varepsilon)\,d\varepsilon = 1$, equivalently $\int_a^b D(\varepsilon)\,d\varepsilon = Z$, where the total in-window mass $Z$ is a normalization constant separate from the spectral shape. The conservation law our refinement enforces is the conservation of this in-window mass: whatever value $Z$ takes for a given material, the refined spectrum integrates to it exactly. We stress that $Z$ is the integrated number of states \emph{within the finite analysis window} $[a,b]$, and equals a textbook full-range sum-rule constant only in the limit where the window captures the entire spectrum. The full-range totals are themselves analytic: for eDOS, $\int_{-\infty}^{\infty}D(E)\,dE$ is fixed by the number of included bands and the spin degeneracy, and for phDOS, $\int_0^{\infty} g(\omega)\,d\omega = 3N$ for a unit cell of $N$ atoms. The in-window mass $Z$, however, is the portion of this analytic total that falls inside the truncated window, and is therefore not fixed by composition alone: for eDOS over $[E_F-4,E_F+4]$~eV it depends on how much of the band structure lies near the Fermi level, and for phDOS over $[0,1000]$~cm$^{-1}$ it equals $3N$ only up to the (material-dependent) mode weight above the cutoff. For eDOS we therefore approximate $Z=N_{\mathrm{win}}$ from the crystal structure and the analytic full-range total with a trained oracle (see ``Structure-aware spectrum decoding''; \siOracle); for phDOS, where the $[0,1000]$~cm$^{-1}$ window captures all $3N$ modes for the large majority of materials, we use the analytic $Z=3N$ directly. These constraints must be preserved throughout the physical reasoning trajectory (e.g., during iterative refinement). In addition, structured spectral features such as sharp band edges/gaps and van Hove singularities must be preserved.
This motivates formulating DOS prediction as a physical reasoning problem rather than unconstrained pointwise regression. Additional background on eDOS is provided in \siEdos, and the analogous background for phDOS is provided in \siPhdos.

\subsection{Structure-aware spectrum decoding}\label{sec:methods:seq}
We introduce a structure-aware atom-to-energy spectrum decoding framework with hierarchical attention: (i) an edge-aware crystal-graph Transformer encodes atomic structure via gated, edge-modulated attention, and (ii) a structure-aware decoder predicts over energy bins with source (atom-to-energy) attention, preserving physically meaningful element-projected contributions.

\paragraph{Structure encoding with edge-aware attention.}
We encode crystal structures using a crystal graph Transformer~\citep{vaswani2017attention} with edge-aware graph attention~\citep{velivckovic2017graph}, explicitly modeling both atomic and interatomic (edge) features.

Let $h_i^l \in \mathbb{R}^{d_h}$ denote the hidden representation of node $i$ in layer $l$, and $h_{e_{ij}} \in \mathbb{R}^{d_e}$ the embedding of the edge feature between nodes $i$ and $j$. Each layer projects the node embeddings into query, key, and value spaces:
\begin{equation}
    q_i^l = W_q^l h_i^l + b_q^l, \quad
    k_i^l = W_k^l h_i^l + b_k^l, \quad
    v_i^l = W_v^l h_i^l + b_v^l,
\end{equation}
where $W_q^l, W_k^l, W_v^l \in \mathbb{R}^{d \times d_h}$ and $b_q^l, b_k^l, b_v^l \in \mathbb{R}^{d}$ are layer-specific learnable parameters, with $d=d_h$ so that the layers can be stacked.

Similarly, the edge embedding is linearly transformed into the same latent space, $g_{ij} = W_e h_{e_{ij}} + b_e$,
where $W_e \in \mathbb{R}^{d \times d_e}$ and $b_e \in \mathbb{R}^{d}$ are shared across all layers.

The normalized attention coefficient between node $i$ and its neighbor $j$ is then computed as
\begin{equation}
    \alpha_{ij}^l =
    \frac{\exp\left(\frac{(q_i^l)^\top (k_j^l + g_{ij})}{\sqrt{d}}\right)}
    {\sum_{p \in N(i)} \exp\left(\frac{(q_i^l)^\top (k_p^l + g_{ip})}{\sqrt{d}}\right)},
\end{equation}
where $N(i)$ denotes the set of neighboring nodes of $i$ and $\sqrt{d}$ is the standard scaling factor.

The aggregated representation of node $i$ is computed as a weighted sum of its neighbors' values, modulated by both attention scores and edge embeddings:
\begin{equation}
    \hat{h}_i^{l+1} = \sum_{j \in N(i)} \alpha_{ij}^l \cdot (v_j^l + g_{ij}).
\label{eq:gat}
\end{equation}
Both $v_j^l$ and $g_{ij}$ lie in $\mathbb{R}^d$ by construction (\Cref{eq:gat} is therefore dimensionally well-defined), so node features and edge features are combined in the same latent space without requiring an additional projection.
Compared to graph convolution~\citep{kipf2016semi,Xie2018}, this explicitly incorporates both pairwise similarity and interatomic features in message aggregation, thereby enhancing its expressiveness.

To improve training stability and support deeper stacks, we augment each layer with a \emph{gated residual} connection~\citep{he2016deep}. Let $\hat{h}_i^{l+1}$ denote the pre-residual output of layer $l$ (i.e., the representation after message passing/attention but before the residual update), and let $h_i^{l}$ be the input representation. We first project the residual branch to match dimensions, $r_i^{l} = W_r^{l} h_i^{l} + b_r^{l}$,
and then compute a gate that adaptively interpolates between the new update $\hat{h}_i^{l+1}$ and the residual signal $r_i^{l}$. Specifically, we set
\begin{equation}
\begin{aligned}
\beta_i^{l} &= \sigma\!\left(a^{l}\,[\, \hat{h}_i^{l+1} ~;~ r_i^{l} ~;~ \hat{h}_i^{l+1} - r_i^{l}\,]\right), \\
h_i^{l+1} &= \mathrm{LeakyReLU}\!\Big(f_{\mathrm{LN}}\!\big((1-\beta_i^{l}) \odot \hat{h}_i^{l+1} + \beta_i^{l} \odot r_i^{l}\big)\Big),
\end{aligned}
\end{equation}
where $\sigma(\cdot)$ is the sigmoid gate nonlinearity, $\mathrm{LeakyReLU}(\cdot)$ is the nonlinearity applied after normalization, $[\cdot;\cdot]$ denotes concatenation, $f_{\mathrm{LN}}$ is LayerNorm, and $\odot$ is element-wise multiplication. When $\beta_i^{l}$ is close to $0$, the layer trusts the newly computed update $\hat{h}_i^{l+1}$; when it is close to $1$, the model falls back to the residual branch, improving robustness early in training.

This edge-aware, gated attention design enables expressive modeling of the crystal structure by dynamically weighting interatomic interactions and adaptively controlling information flow across layers, while preserving physically meaningful element-projected contributions.

\paragraph{Spectrum decoding with source attention.}
Existing approaches~\citep{Kong2022,kaundinya2022prediction,lee2023density} typically compress a crystal into a single global embedding and decode the eDOS with a multilayer perceptron (MLP), which can obscure the element-projected contributions that physically constitute the spectrum. In contrast, our design is \emph{physics-informed}: since the eDOS can be expressed as an aggregation of element- or orbital-projected densities of states (pDOS), each energy bin should selectively attend to the atoms that contribute at that energy.

We introduce a structure-aware sequence decoding approach that models long-range correlations across energy bins with self-attention, while grounding each bin in atom-level embeddings through source (atom-to-energy) attention.
Specifically, at each decoding step $i$, the input is a concatenation of an energy bin embedding $s_i$ and the global crystal embedding $h_x$, forming the composite embedding $z_i = [\, s_i ~;~ h_x \,]$, where $s_i$ is a learned positional embedding of the energy-bin index $i$. The set $\{z_i\}_{i=1}^{L}$, one embedding per energy bin, initializes hidden states for the first self-attention layer.

This mechanism operationalizes the physically motivated view of eDOS as an aggregation of pDOS with element-projected contributions, while preserving fine-grained structure–energy dependencies. Moreover, by modeling the eDOS as a spectrum with contextual self-attention, our decoder captures smooth energy-wise correlations more effectively than static regression heads like MLPs, yielding predictions that are consistent with the underlying electronic structure.

Each decoder layer applies (i) self-attention over energy bins and (ii) source-attention to the encoder's final-layer node embeddings $\{h_j\}_{j=1}^{N}$ (here $j$ indexes atoms, while $i$ indexes energy bins):
\begin{align}
\mathrm{Att}(Q,K,V) &= \mathrm{softmax}\!\left(\frac{QK^\top}{\sqrt{d}}\right)V, \\
(Q,K,V) &=
\begin{cases}
(Q_{\mathrm{dec}},K_{\mathrm{dec}},V_{\mathrm{dec}}) & \text{(self)}\\
(Q_{\mathrm{dec}},K_{\mathrm{enc}},V_{\mathrm{enc}}) & \text{(source)}
\end{cases}
\end{align}
with $Q_{\mathrm{dec}}$ from decoder states and $(K_{\mathrm{enc}},V_{\mathrm{enc}})$ from $\{h_j\}$. The decoder outputs a DOS prediction $\hat{y}=\{\hat{y}_i\}_{i=1}^{L}$, which we normalize as $\tilde{\hat{y}}_i=\hat{y}_i/\sum_{m=1}^{L} \hat{y}_m$ and compare to the normalized ground truth $\tilde{y}_i=y_i/\sum_{m=1}^{L} y_m$. As noted above, the total integrated mass $Z=\int_a^b D(\varepsilon)\,d\varepsilon$, the integrated number of states within the analysis window ($N_{\mathrm{win}}$ for eDOS; the in-window modes mass for phDOS), sets only the absolute scale of the spectrum and is decoupled from its shape. We exploit this by learning only the spectral \emph{shape} and deferring the absolute scale to the total-state sum rule. The shape is supervised with the Kullback--Leibler (KL) divergence,
\begin{equation}
\mathcal{L}_{\mathrm{stage-1}}=D_{\mathrm{KL}}(\tilde{y}\,\|\,\tilde{\hat{y}})
=\sum_{i=1}^{L}\tilde{y}_i\log\frac{\tilde{y}_i}{\tilde{\hat{y}}_i},
\label{eq:kl_loss}
\end{equation}
which treats the normalized DOS as a distribution and emphasizes correct placement of spectral mass (peaks/band edges) rather than per-bin error alone. At inference, the absolute DOS is recovered by rescaling the normalized shape $\tilde{\hat{y}}$ by $Z$, the total integrated mass within the analysis window, so that the assigned in-window mass is reproduced exactly by construction. This in-window mass is not fixed by composition once the window is truncated. For eDOS we predict it with a lightweight machine-learning oracle that takes the crystal structure (encoded with CrystalFormer~\citep{wang2024conformal}) together with the analytic full-range total $\int_{-\infty}^{\infty}D(E)\,dE$ (fixed by the number of included bands and the spin degeneracy) and outputs the in-window count $N_{\mathrm{win}}=\int_{E_F-4}^{E_F+4}D(E)\,dE$ (in practice by regressing the size-normalized ratio of in-window to full-range states and rescaling; \siOracle). Regressing the intensive ratio rather than the extensive count is what keeps the problem well-conditioned: $N_{\mathrm{win}}$ spans more than three orders of magnitude across the benchmark, whereas the ratio is tightly concentrated. On held-out materials the oracle recovers the ratio to $R^2=0.921$ and the absolute count to $R^2=0.982$, with a mean absolute percentage error of $5.9\%$; this ${\sim}6\%$ scale uncertainty is decoupled from the spectral shape and leaves the conservation guarantees below intact, since rescaling a normalized density by a positive scalar preserves nonnegativity and reproduces the assigned in-window mass exactly. For phDOS the $[0,1000]$~cm$^{-1}$ window captures all $3N$ modes for the large majority of materials (about $94\%$ of the dataset; the exceptions are light-atom or strongly bonded compounds whose high-frequency stretches extend beyond the cutoff), so we use the analytic modes count $Z=3N$ directly as the scale (\siOracle).

\subsection{Constraint-preserving physical reasoning}\label{sec:methods:reason}
Given an initial prediction of the normalized eDOS from structure-aware spectrum decoding, we perform a second-stage reasoning that explicitly enforces hard physical constraints. While the previous stage captures element-projected contributions and energy-wise correlations, its output can still exhibit residual artifacts such as blurred band edges, non-zero noise in band gaps, or misplaced spectral mass.

To correct such errors without violating physical constraints, we refine the prediction directly in probability space. We introduce a constraint-preserving physical reasoning procedure for eDOS, \emph{mass-conserving iterative refinement}, which treats the normalized eDOS as a probability density and updates it via continuous transport that preserves mass by construction. This formulation enables principled redistribution of spectral mass across energy bins, sharpening band gaps and peaks while maintaining normalization. We present the high-level formulation and algorithmic intuition here; detailed derivations and proofs are deferred to \siCppr.

Formally, we model this refinement as transport dynamics governed by a continuity equation. Let $p_0(x)$ and $p_T(x)$ denote the normalized eDOS densities induced by the initial prediction $\tilde{\hat{y}}$ and the ground truth $\tilde{y}$ (the piecewise-constant densities whose bin masses are $\tilde{\hat{y}}_i$ and $\tilde{y}_i$), defined over the energy domain $x \in \mathbb{R}$:
$
    \int p_0(x)\,dx = \int p_T(x)\,dx = 1.
$
Our goal is to learn a transport that continuously evolves $p_0$ to $p_T$ while the total probability mass is preserved throughout the refinement trajectory. This formulation naturally admits a continuous flow interpretation, where a velocity field $f_\theta(x,t)$ governs the evolution of the normalized eDOS over time.

\paragraph{Mass-conserving transport.}
The iterative refinement must \emph{preserve total spectral mass}: for the normalized eDOS density $p(x,t)$ on an energy domain $\Omega=[a,b]$, $\int_{\Omega} p(x,t)\,dx = 1$, $\forall\, t.$
A convenient way to guarantee this is to model refinement as \emph{pure transport} of density along the energy axis, so that mass can move but cannot be created or destroyed. Let $f_\theta(x,t)$ denote a learnable velocity field. We evolve $p$ by the one-dimensional continuity equation
\begin{equation}\label{eq:continuity}
\partial_t p(x,t) + \partial_x j(x,t) \;=\; 0, \qquad j(x,t)\coloneqq p(x,t)\,f_\theta(x,t).
\end{equation}
Integrating \Cref{eq:continuity} over $\Omega=[a,b]$ gives
\begin{equation}
\frac{d}{dt}\int_a^b p(x,t)\,dx
= -\int_a^b \partial_x j(x,t)\,dx
= -\big[j(b,t)-j(a,t)\big].
\end{equation}
Therefore, imposing \emph{no-flux} boundary conditions $j(a,t)=j(b,t)=0$ ensures $\frac{d}{dt}\int_a^b p(x,t)\,dx=0$, so that total mass is conserved throughout refinement. For $\Omega=\mathbb{R}$, it suffices to assume vanishing flux at infinity, $\lim_{x\to\pm\infty} j(x,t)=0$ (with integrable tails), so the boundary term disappears. An equivalent log-density (characteristics) form of \Cref{eq:continuity} is given in \siCppr.

\paragraph{CDF-based upwind integrator.}
We integrate \Cref{eq:continuity} from a Stage~I initial density $p_0$ to a refined density $\hat{p}_T$ over $t\in[0,T]$ through a learned, state-dependent velocity field $f_\theta(x,t;p,\mathcal{C})$, where the conditioning includes the current density $p$ at the integration step and a crystal context $\mathcal{C}$ produced by Stage~I. For numerical stability we evolve in the Eulerian (CDF) domain via $F_\theta(x,t)=\int_{-\infty}^{x} p_\theta(x',t)\,dx'$, which obeys the equivalent first-order PDE
\begin{equation}
\label{eq:cdf-pde}
    \partial_t F_\theta(x,t) + f_\theta\big(x,t;p_\theta(\cdot,t),\mathcal{C}\big)\,\partial_x F_\theta(x,t) = 0,\qquad F_\theta(x,0)=F_0(x),
\end{equation}
and recover the refined density as $\hat{p}_T(x):=p_\theta(x,T)=\partial_x F_\theta(x,T)$. This Eulerian formulation gives a deterministic procedure for transporting $p_0$ to a refined density $\hat{p}_T$ while strictly conserving the total probability mass.

\paragraph{Discretization.}
We discretize the energy domain $\Omega=[a,b]$ into $N_x$ uniform bins of width $\Delta x=(b-a)/N_x$, where $N_x=L$ is the spectrum length defined above ($N_x=128$ for eDOS), and split the refinement horizon $[0,T]$ into $K$ update steps with time step $\Delta t=T/K$. Let $F_i^k\approx F_\theta(x_i,t_k)$ denote the discrete CDF at the upper edge of bin $i$ at time $t_k=k\Delta t$, $k=0,1,\ldots,K$, with the boundary values held fixed at $F_0^k\equiv 0$ (no spectral mass below $a$) and $F_{N_x}^k\equiv 1$. We evolve \Cref{eq:cdf-pde} using forward Euler in time and a first-order upwind derivative in space:
\begin{equation}
F_i^{k+1}=F_i^k-\Delta t\,v_i^k\,D_x^{\mathrm{up}}F_i^k,\qquad v_i^k\;\coloneqq\;f_\theta\big(x_i,t_k;\,p^k,\mathcal{C}\big),
\label{eq:discrete-cdf-update}
\end{equation}
where the per-bin velocity at step $k$ is conditioned on the current normalized density $p^k$ recovered from $F^k$ by finite differences; no-flux boundary conditions are enforced by holding the boundary CDF values fixed at $F_0\equiv 0$ and $F_{N_x}\equiv 1$ (in the discrete update, $v_{N_x}^k=0$ holds the upper endpoint fixed, while the lower endpoint $F_0$ lies outside the update range), so that no mass crosses $a$ or $b$; and the updated CDF values are clamped to $[0,1]$ for numerical robustness. Under the Courant--Friedrichs--Lewy (CFL) condition $\max_{i,k}|v_i^k|\,\Delta t/\Delta x\le \tfrac12$, this monotone upwind scheme, together with the order-preserving clamp, keeps the CDF nondecreasing, so the recovered density is nonnegative and total mass is conserved up to floating-point rounding. After $K$ steps, we recover the refined density by finite differences,
\begin{equation}
\hat{p}_T(x_i)=\frac{F_i^{K}-F_{i-1}^{K}}{\Delta x},\quad i=1,\ldots,N_x,
\label{eq:density-recovery}
\end{equation}
which uses $F_0^{K}\equiv 0$ for the first bin and returns exactly $N_x$ density values. Because $F_0^{K}=0$ and $F_{N_x}^{K}=1$, the recovered density already satisfies $\sum_{i=1}^{N_x}\hat{p}_T(x_i)\Delta x=1$; a final renormalization is applied only to absorb rounding. The same $K$-step rollout is used at both training and inference.

\paragraph{Conservation guarantee.}
The constructions above give the refinement stage a closed-form validity statement: regardless of the learned velocity field, subject only to the CFL stability condition stated below, the refined eDOS is guaranteed to be a valid normalized density and to differ from the initial prediction only through pure transport of mass along the energy axis. We state this formally below.

\begin{proposition}[Constraint-preserving refinement]
\label{prop:constraint_preserving}
Let $f_\theta$ be any velocity field that satisfies the no-flux boundary condition $j(a,t)=j(b,t)=0$ for $j\coloneqq p_\theta f_\theta$. Let $p_\theta(\cdot,t)$ evolve from a normalized initial density $p_0$ under the continuity equation~\eqref{eq:continuity} (continuous case), or via the discrete upwind CDF update~\eqref{eq:discrete-cdf-update} under the CFL condition $\max_{i,k}|v_i^k|\,\Delta t/\Delta x\le \tfrac12$ (discrete case). Then the refined density $\hat{p}_T$ produced by the rollout satisfies:
\begin{enumerate}
    \item \emph{Nonnegativity:} $\hat{p}_T(x)\ge 0$ for all $x\in\Omega$.
    \item \emph{Total-state conservation:} $\int_\Omega \hat{p}_T(x)\,dx = \int_\Omega p_0(x)\,dx = 1$ (exactly in the continuous case, up to rounding in the discrete case).
    \item \emph{Pure transport:} $\hat{p}_T$ differs from $p_0$ only through redistribution of spectral mass along the energy axis; mass is neither created nor destroyed.
\end{enumerate}
\end{proposition}
\begin{proof}[Proof sketch]
In the continuous case, integrating~\eqref{eq:continuity} over $\Omega$ and applying the no-flux condition gives $\tfrac{d}{dt}\int_\Omega p_\theta(x,t)\,dx = 0$, which yields (2); the CDF reformulation~\eqref{eq:cdf-pde} together with the monotone advection operator preserves CDF monotonicity, which yields (1); and pure transport (3) follows directly from the conservative form of~\eqref{eq:continuity}. In the discrete case, the upwind update preserves CDF monotonicity under the stated CFL condition, the finite-difference recovery of $\hat{p}_T$ from $F$ inherits nonnegativity, and holding the boundary CDF values $F_0=0$ and $F_{N_x}=1$ fixed makes the discrete mass sum equal one exactly (up to rounding). Full proofs, including the discrete monotonicity and mass-conservation lemmas, are given in \siCppr.
\end{proof}

\paragraph{End-to-end training.}\label{sec:methods:mir-training}
The $K$-step upwind rollout~\eqref{eq:discrete-cdf-update}--\eqref{eq:density-recovery} is fully differentiable in $\theta$: the upwind difference, the boundary masking, the CDF update, and the finite-difference density recovery are elementary operations that admit gradients. We therefore train the velocity field $f_\theta$ end-to-end by back-propagating a spectrum-matching loss through the rollout. With the Stage~I trained model frozen and $p_T$ denoting the DFT-derived normalized target, the loss is the KL divergence
\begin{equation}
\mathcal{L}_{\mathrm{stage-2}}(\theta)
= D_{\mathrm{KL}}\!\big(p_T\,\big\|\,\hat{p}_T(\theta)\big)
= \sum_{i=1}^{N_x} p_T(x_i)\log\frac{p_T(x_i)}{\hat{p}_T(x_i;\theta)}\,\Delta x,
\label{eq:mir-kl-loss}
\end{equation}
in the same distributional form as the Stage~I objective~\eqref{eq:kl_loss}. Because nonnegativity, total-mass conservation, and pure-transport behavior are enforced by the integrator itself (\Cref{prop:constraint_preserving}), $\hat{p}_T(\theta)$ is a valid normalized density at every step of training, and the gradient signal acts purely on \emph{where} the velocity field redistributes spectral mass. Because the Stage~I \texttt{Softplus} output is strictly positive and the short, CFL-bounded transport does not drive any bin to exactly zero, $\hat{p}_T$ stays positive wherever $p_T$ is, so the forward KL divergence in~\Cref{eq:mir-kl-loss} remains finite. No soft conservation penalties are added to~\Cref{eq:mir-kl-loss}. Both stages of DeepDOSReasoner therefore preserve the eDOS constraints by construction rather than relying on soft penalties.

\paragraph{Amortized inference.}
The training set consists of pairs of normalized eDOS distributions $\{(p_0^{(n)}, p_T^{(n)})\}$, one per crystal structure. To enable generalization across crystals rather than fitting a separate transport for each, we adopt an \emph{amortized inference} strategy: a single shared neural network with parameters $\theta$ outputs the per-bin velocity $f_\theta(x_i,t_k;p^k,\mathcal{C})$ that is reused (i) across all training and test crystals via conditioning on the crystal context $\mathcal{C}$ (encoder global feature and decoder-side context produced by Stage~I), and (ii) across all $K$ rollout steps within a single trajectory via conditioning on the current normalized density $p^k$ and the time index $t_k$. Both axes of amortization are served by the same network call, so the rollout is governed by one set of weights rather than a step-specific or crystal-specific velocity field; the resulting transport operator generalizes both within and across crystals. The detailed parameterization of $f_\theta$ is described in \siHyper.

\paragraph{Seed-averaging ensemble.}
To reduce variance from stochastic training and improve final accuracy, we use a seed-averaging ensemble at inference time. We train $M$ independent copies of DeepDOSReasoner from different random seeds, run each through the full two-stage inference procedure, and average the resulting refined normalized densities $\hat{p}_T^{(m)}$ across $m=1,\ldots,M$ before rescaling by $Z$. Because nonnegativity and unit mass are preserved under convex combination, the averaged density remains a valid probability distribution and the conservation guarantee of \Cref{prop:constraint_preserving} carries over to the ensemble. The ensemble incurs an $M$-fold inference cost but trades that for consistently improved spectral metrics over any single seed.

\subsection{Implementation details}\label{sec:methods:impl}
We train DeepDOSReasoner in two stages separately: Stage~I trains the structure encoder and the structure-aware spectrum decoder, and Stage~II trains the velocity field for mass-conserving iterative refinement with the Stage~I parameters frozen. Two implementation choices make the conservation guarantees of \Cref{prop:constraint_preserving} hold for any network weights from the first optimizer step. First, the output velocities are clamped so that the per-bin Courant number satisfies $|v_i^k|\,\Delta t/\Delta x\le\tfrac12$, enforcing the CFL condition by construction rather than by penalty. Second, the velocity head is zero-initialized, so that at the start of training the refinement reduces exactly to the identity and transport is learned as an additive correction on top of this warm start, with the integrator's guarantees active throughout. At test time, the refinement is integrated forward to the final step $t=T$ and the resulting density is used as the prediction (rescaled by $Z$ to recover the absolute DOS). Full hyperparameters and training schedule are provided in \siHyper, the complete training and inference procedure is summarized as pseudocode in \siAlgo, and computational cost is reported in \siComp.

\subsection{Baselines}\label{sec:methods:base}
We benchmark against baselines from three families. The first family is crystal-graph structure encoders with regression heads: E3NN~\citep{geiger2022e3nn} enforces geometric equivariance via spherical harmonics and tensor products, MEGNet~\citep{Chen2019} augments message passing with state-level features and a set-to-set graph-level readout, and GATGNN~\citep{louis2020graph} combines graph attention with a global attention readout. The second family is interatomic-potential architectures adapted for spectrum prediction, in which the per-atom features originally used for force/energy prediction are pooled and decoded through a fixed-grid regression head: NequIP~\citep{batzner2022nequip}, M3GNet~\citep{chen2022universal}, MACE~\citep{batatia2022mace}, and EquiformerV2~\citep{liao2024equiformerv2}. The third family is crystal-to-sequence predictors that explicitly treat the eDOS as an ordered sequence over energy bins: Mat2Spec~\citep{Kong2022} couples a probabilistic-embedding generator with supervised contrastive learning for spectra, and DOSTransformer~\citep{lee2023density} introduces a prompt-guided multi-modal Transformer with cross-attention between structure features and energy-coordinate tokens. Full baseline descriptions are provided in \siBase.

\subsection{Evaluation metrics}\label{sec:methods:eval}
We evaluate performance with task-specific metrics: (i) for overall eDOS prediction, we report mean absolute error (MAE), mean squared error (MSE), coefficient of determination ($R^2$), and Wasserstein distance (WD), all computed against the DFT ground-truth eDOS; (ii) for peak-fidelity assessment, we report a peak-restricted MAE evaluated only at energy bins corresponding to van Hove singularities adjacent to the VBM and CBM, extracted from the DFT ground truth (defined in \siEval); (iii) for VB-gap identification (the four-criterion rule is defined in Results), we report precision, recall, F1 score, and Matthews correlation coefficient (MCC); and (iv) for ranking element-projected contributions, we report overall accuracy, top-1 accuracy, and Spearman $\rho$. Unless stated otherwise, benchmark metrics are reported as the mean $\pm$ one standard deviation over five independent runs; the prospective-discovery evaluation reports single values on its fixed candidate set. Full metric definitions and computation details are provided in \siEval.

\subsection{Data generation}\label{sec:methods:datagen}
Beyond the preprocessing pipeline already described in the ``Datasets'' section, we record here the dataset acquisition and partitioning that underpin both benchmarks. The eDOS spectra are inherited from the Materials Project snapshot used by Kong et al.~\citep{Kong2022} (version 2021-03-22), where each spectrum was originally tabulated by VASP on a fixed-length energy grid (\texttt{NEDOS}$=2001$) before our area-preserving rebinning onto the 128-bin grid. Only nonmagnetic Materials Project entries (both metallic and semiconducting/insulating) enter the benchmark, and the set is partitioned at the random 80\%/10\%/10\% train/validation/test ratio adopted by Kong et al.~\citep{Kong2022}, with the test partition held out throughout training and hyperparameter selection. The phDOS dataset of Chen et al.~\citep{Chen2021} is partitioned into $1{,}220$/$152$/$152$ train/validation/test crystals; we adopt the released partition without modification so that the held-out test set used for the phDOS benchmark coincides with the original benchmark and remains directly comparable to numbers reported on the same split. The electronic-structure descriptors used in the case studies (e.g., the eDOS center $\bar\varepsilon$) are obtained from the predicted spectrum through the corresponding integral functions implemented in \texttt{pymatgen}~\citep{Ong2013}, so that the predicted and DFT reference values are computed by the same numerical procedure.

The prospective-discovery pool is a separate, label-free dataset drawn from the same Materials Project snapshot but restricted to nonmagnetic entries that lack a published DFT eDOS. After applying the candidate-screening filters described in Results (within $0.5$~eV/atom of the convex hull, fewer than five elements, structural-relaxation band gap of zero), the resulting $8{,}106$ candidate structures form the universe over which DeepDOSReasoner is asked to flag VB-gap materials. Crystal structures from this pool enter the pipeline only at inference time; no eDOS labels from these entries are used during training, validation, or hyperparameter selection, so the discovery evaluation reported in \Cref{tab:vb_discovery} is genuinely out-of-distribution with respect to the model.

\subsection{DFT calculations}\label{sec:methods:dft}
To produce the DFT-derived eDOS references against which DeepDOSReasoner is validated in the case studies and the prospective-discovery setting of Results, we perform first-principles calculations with the Vienna Ab initio Simulation Package (VASP)~\citep{kresse1996efficient}. Ion--electron interactions are treated with the projector-augmented wave (PAW) method~\citep{blochl1994projector,kresse1999from}, and exchange--correlation is described by the Perdew--Burke--Ernzerhof (PBE) functional within the generalized gradient approximation (GGA). Structure relaxations use a plane-wave cutoff energy of $450$~eV and are considered converged when the total energy change between self-consistency iterations falls below $10^{-5}$~eV and the maximum residual atomic force lies below $0.05$~eV/\AA. Brillouin-zone integration uses Monkhorst--Pack $k$-meshes of $3\times3\times3$ for the I$_2$--II--IV--X$_4$ chalcogenides, $2\times2\times1$ for the CdCl$_x$I$_{2-x}$ substitutional variants of CdI$_2$ (\texttt{mp-569994}), and $2\times4\times2$ for the Dy$_4$Cu$_2$In$_{6-x}$Ga$_x$ substitutional variants of Dy$_4$Cu$_2$In$_6$ (\texttt{mp-1225405}); the prospective-discovery candidates are computed under the same VASP setup, with the $k$-mesh density matched per candidate to the Materials Project default so that the validation reference is generated under the same recipe used for the eDOS training data of Kong et al.~\citep{Kong2022}. The CuPtFeCoNi high-entropy-alloy supercell, whose atomic configuration follows Han et al.~\citep{han2024highentropy}, is relaxed and its eDOS computed under the identical PBE/PAW setup. Spin-polarized calculations were performed for the alloy and substituted (doped) systems, with metal-site magnetic moments initialized to the Materials Project default ($0.6\,\mu_B$); the high-entropy-alloy and Dy$_4$Cu$_2$In$_{6-x}$Ga$_x$ calculations additionally included a Hubbard $+U$ correction on the localized transition-metal $d$ and Dy $4f$ states, respectively. Because DeepDOSReasoner predicts a single spin-unpolarized eDOS, the spin-up DOS channel of these spin-polarized references was used when comparing the model predictions against DFT. The converged self-consistent total and projected DOS are extracted with the VASPKIT post-processing tool~\citep{wang2021vaspkit}. The resulting raw spectra are then carried through the same preprocessing pipeline described above (Fermi-level alignment, $\sigma=0.5$~eV Gaussian smearing, and area-preserving rebinning onto the 128-bin grid over $[-4,4]$~eV) so that the DFT and model spectra are directly comparable, and the VB-gap criteria defined in Results are evaluated to produce the binary labels used in \Cref{tab:vb_discovery}. The same four-criterion rule transfers without modification to conduction-band gap searches by mirroring the window across $E_F$ (a contiguous low-DOS interval above $E_F$ rather than below, with the corresponding low-energy edge in $[0.25,1.5]$~eV); we focus on VB gaps in this work because they expose the more stringent of the two failure modes for the spectral baselines (Results).

%% file: appendix/intro_edos.tex
\section{Density of States}
\label{app:dos}

\subsection{Electronic Density of States}
\label{app:edos}

The \emph{electronic density of states} (eDOS) is a fundamental quantity in solid-state physics that characterizes how electronic states are distributed in energy within a crystalline material~\citep{kittel2018introduction}. For a periodic system with single-particle energy eigenvalues $\{\varepsilon_n(\mathbf{k})\}$ indexed by band $n$ and crystal momentum $\mathbf{k}$, the eDOS is defined as
\begin{equation}
D(E)
\;=\;
\frac{1}{V_{\mathrm{BZ}}}
\sum_{n}
\int_{\mathrm{BZ}}
\delta\!\left(E - \varepsilon_n(\mathbf{k})\right)
\, d\mathbf{k},
\label{eq:edos_def}
\end{equation}
where the integral is taken over the Brillouin zone (BZ), $V_{\mathrm{BZ}}$ denotes its volume, and the index $n$ runs over bands and spin channels, so that spin degeneracy is counted in $D(E)$. Intuitively, $D(E)\,dE$ measures the number of available electronic states whose energies lie within the interval $[E, E+dE]$.

The eDOS provides a compact yet information-rich summary of a material’s electronic structure. Many physical properties, including electrical conductivity, optical absorption, heat capacity, and magnetic response, depend on integrals or moments of the eDOS weighted by appropriate occupation statistics~\citep{hook2013solid}. In particular, the behavior of the eDOS near the Fermi level governs low-temperature transport, while its global structure reflects orbital hybridization and bonding characteristics encoded by the crystal geometry.

A defining feature of the eDOS is the presence of \emph{band edges} and \emph{band gaps}. For semiconductors and insulators, the eDOS vanishes over a finite energy interval separating the valence and conduction bands. The detailed shape of the eDOS near the valence-band maximum (VBM) and conduction-band minimum (CBM) determines effective carrier masses, carrier densities, and optical transition strengths~\citep{lee2016computational}. Accurately capturing these band edges, rather than merely predicting a scalar band gap, is therefore essential for faithful electronic structure modeling.

In addition to the fundamental band gap between the valence and conduction bands, the electronic density of states may exhibit valence-band gaps (\emph{VB gaps}), defined as energy intervals with vanishing or strongly suppressed density of states within the valence-band manifold below the Fermi level~\citep{harrison1989electronic}. Such VB gaps arise from orbital, bonding–antibonding, crystal-field, or correlation-induced splittings, and reflect intrinsic electronic structure beyond the global band gap~\citep{cox1987electronic}. Unlike the VBM–CBM band gap, VB gaps do not separate occupied from unoccupied states, but instead encode internal organization of occupied electronic states and are therefore important for faithfully representing the fine structure of the valence-band eDOS.

Beyond band gaps and VB gaps, the eDOS often exhibits sharp peaks known as \emph{van Hove singularities}~\citep{van1953occurrence}. These arise from critical points in the band structure where the group velocity vanishes, i.e.,
$\nabla_{\mathbf{k}} \varepsilon_n(\mathbf{k}) = \mathbf{0}$.
Van Hove singularities encode the topology of the electronic bands and are associated with enhanced electronic responses, influencing phenomena such as superconductivity, magnetism, and electronic instabilities. Resolving the location and amplitude of these peaks is particularly challenging, as they are highly sensitive to long-range crystal symmetry and dispersion relations.

In practice, the eDOS is computed using density functional theory (DFT) by discretizing the Brillouin zone and approximating the Dirac delta function with a smoothing kernel, such as a Gaussian filter. The resulting eDOS is represented on a finite energy grid and often normalized so that
\begin{equation}
\int_{a}^{b} D(E)\, dE
\;=\;
N_{\mathrm{win}},
\label{eq:edos_mass}
\end{equation}
where $[a,b]$ denotes the energy window of interest and $N_{\mathrm{win}}$ is the total number of electronic states within that window. Conservation of this in-window spectral mass is the physical constraint that our predictive model preserves by construction. The full-energy integral is itself fixed analytically: $\int_{-\infty}^{\infty}D(E)\,dE$ equals the number of included bands times the spin degeneracy. The in-window count $N_{\mathrm{win}}$ over the finite window used in this work is the portion of that analytic total falling inside $[a,b]$ and is not fixed by composition alone. We therefore estimate the in-window electronic-states count from the crystal structure together with the analytic full-range total using a trained oracle (\Cref{app:oracle}).

Importantly, the total eDOS admits a decomposition into \emph{element-projected} or \emph{orbital-projected} densities of states (pDOS), which quantify the contribution of each chemical element or orbital to the electronic states at each energy. Such decompositions provide interpretability and link spectral features to local chemical environments, allowing prominent peaks in the eDOS to be traced back to specific elements or bonding motifs.

Taken together, the eDOS constitutes a \emph{structured scientific spectrum}: it is nonnegative over energy, obeys strict conservation laws, contains sharp discontinuities and singular features, and encodes multi-scale information arising from atomic composition, crystal geometry, and long-range periodicity. These properties make eDOS prediction a challenging learning problem that goes beyond unconstrained regression, motivating models that reason over energy-conditioned structure while preserving physical validity.

\subsection{Phonon Density of States}
\label{app:phdos}

The \emph{phonon density of states} (phDOS) characterizes how vibrational normal modes are distributed in frequency within a crystalline material~\citep{kittel2018introduction,Petretto2018}. For a periodic system with phonon dispersion relations $\{\omega_\nu(\mathbf{q})\}$ indexed by branch $\nu$ and crystal momentum $\mathbf{q}$, the phDOS is defined as
\begin{equation}
g(\omega)
\;=\;
\frac{1}{V_{\mathrm{BZ}}}
\sum_{\nu}
\int_{\mathrm{BZ}}
\delta\!\left(\omega - \omega_\nu(\mathbf{q})\right)
\, d\mathbf{q},
\label{eq:phdos_def}
\end{equation}
where the integral runs over the Brillouin zone (BZ) and the branch index $\nu$ enumerates the $3N$ phonon branches of a unit cell containing $N$ atoms (three acoustic branches and $3N-3$ optical branches). Intuitively, $g(\omega)\,d\omega$ measures the number of vibrational modes whose frequency lies within $[\omega, \omega + d\omega]$.

Like the eDOS, the phDOS provides a compact bridge between atomic structure and macroscopic response. Many thermodynamic and transport quantities, including the vibrational contribution to the specific heat, the vibrational entropy and free energy, the Debye--Waller factor, and the high-temperature phonon mean free path, are obtained as integrals of $g(\omega)$ weighted by Bose--Einstein statistics. The low-frequency tail of $g(\omega)$, dominated by the acoustic branches, controls the low-temperature heat capacity through the Debye $T^3$ law, while the high-frequency content reflects bond stiffness and atomic mass contrast and largely sets the Debye temperature.

A characteristic feature of the phDOS is the separation between \emph{acoustic} and \emph{optical} branches and the corresponding low- and high-frequency peak structures. The three acoustic branches start linearly from $\omega=0$ at $\mathbf{q}=\mathbf{0}$, with slope set by elastic constants and atomic masses, and dominate the low-frequency portion of the spectrum. The remaining $3N-3$ optical branches lie at higher frequencies and produce distinct peak structures whose positions are largely determined by bond strengths, masses, and local symmetry. The presence and width of any \emph{phonon band gap}, an interval of vanishing $g(\omega)$ separating acoustic-dominated and optical-dominated manifolds, encode mass contrast within the unit cell and modulate three-phonon scattering channels relevant to lattice thermal transport.

In analogy to the electronic case, the phDOS exhibits van Hove singularities at frequencies where the phonon group velocity vanishes ($\nabla_{\mathbf{q}}\omega_\nu(\mathbf{q})=\mathbf{0}$). These critical-point peaks reveal the topology of the phonon dispersion, amplify electron--phonon coupling at the corresponding frequencies, and are therefore relevant to superconducting transition temperatures within the Eliashberg framework. As in the eDOS case, resolving the location and amplitude of these peaks is challenging because they are sensitive to long-range force constants and dispersion topology.

The phDOS is conventionally computed by density-functional perturbation theory (DFPT)~\citep{Petretto2018}, in which interatomic force constants are obtained via linear response and dynamical matrices are diagonalized on a fine $\mathbf{q}$-mesh. The resulting phDOS is represented on a finite frequency grid with the Dirac delta function approximated by a smoothing kernel and is constrained by a strict sum rule
\begin{equation}
\int_{0}^{\omega_{\max}} g(\omega)\,d\omega \;=\; 3N,
\label{eq:phdos_mass}
\end{equation}
where $3N$ is the total number of vibrational modes per unit cell and $\omega_{\max}$ exceeds the maximum phonon frequency. Conservation of the total number of modes is a hard physical constraint that any predictive model of the phDOS must respect, in direct analogy to the total-state conservation of \Cref{eq:edos_mass} for the eDOS. When the phDOS is represented on a truncated grid whose upper limit does not exceed the maximum phonon frequency, as for the $[0,1000]$~cm$^{-1}$ grid used in this work, \Cref{eq:phdos_mass} holds only approximately: the in-window mass equals $3N$ minus the weight of modes lying above the cutoff. For the large majority of materials the spectrum has decayed to zero well before $1000$~cm$^{-1}$, so this residual vanishes, $\int_0^{1000}g(\omega)\,d\omega = 3N$ holds, and the absolute scale is recovered by rescaling with the analytic $3N$ directly; only for the small fraction of compounds with high-frequency optical modes above the cutoff (e.g., light-element or strongly bonded systems) does $3N$ slightly over-count the in-window mass (see the phDOS Benchmark paragraph in \Cref{app:hyper}).

The total phDOS admits a mode-resolved decomposition into \emph{atom-projected} phonon densities of states, which quantify the mass-weighted contribution of each atom (or species) to the vibrational modes at each frequency. Such decompositions link spectral features to specific atomic motions and bonding motifs and parallel the element-projected decomposition of the eDOS.

Taken together, the phDOS shares the same structural form as the eDOS: it is a nonnegative one-dimensional density over a bounded spectral axis with a strict total-mass sum rule, contains sharp acoustic-onset features and van Hove singularities, and admits an atom-resolved decomposition. The principal differences between the two are the spectral coordinate (vibrational frequency versus electronic energy), the sum-rule constant ($3N$ versus the electronic-states count $N_{\mathrm{win}}$), and the absence of a Fermi level (and consequently no analogue of valence-band gaps in the eDOS sense). These shared properties are what allow the DeepDOSReasoner formulation, structure-aware spectrum decoding combined with a conservation-preserving continuity-equation refinement, to apply to both spectra essentially unchanged, with only the analytic sum-rule constant substituted at inference time.

%% file: appendix/formal_method.tex
\section[Constraint-Preserving Physical Reasoning: Mass-Conserving Iterative Refinement]{Constraint-Preserving Physical Reasoning: \\Mass-Conserving Iterative Refinement}
\label{app:cppr}

This appendix provides a formal, closed-loop derivation of the constraint-preserving physical reasoning stage, i.e., mass-conserving iterative refinement, introduced in the main Methods section. Throughout, let the energy domain be $\Omega=[a,b]\subset\mathbb{R}$, and let $p(\cdot,t)$ denote a time-dependent normalized eDOS density.

\subsection{Transport Model and Conservation Laws}
\label{app:cppr:transport}

\begin{assumption}[Regularity and boundary condition]
Assume $p(\cdot,t)$ is integrable on $\Omega$ for each $t$ and $f_\theta(\cdot,t)$ is measurable with $j(x,t):=p(x,t)f_\theta(x,t)$ well-defined. On bounded $\Omega=[a,b]$, we impose the \emph{no-flux} boundary condition
\begin{equation}
j(a,t)=j(b,t)=0,\quad \forall t.
\label{eq:noflux}
\end{equation}
(For $\Omega=\mathbb{R}$, replace \Cref{eq:noflux} by $\lim_{x\to\pm\infty} j(x,t)=0$ with integrable tails.)
\end{assumption}

\begin{definition}[Continuity equation]
Given a velocity field $f_\theta:\Omega\times[0,T]\to\mathbb{R}$, we define refinement as \emph{pure transport}:
\begin{equation}
\partial_t p(x,t) + \partial_x j(x,t)=0,
\qquad
j(x,t):=p(x,t)f_\theta(x,t).
\label{eq:app-continuity}
\end{equation}
\end{definition}

\begin{lemma}[Mass conservation]
\label{lem:mass_conservation}
If $p$ satisfies \Cref{eq:app-continuity} on $\Omega=[a,b]$ and the no-flux condition \Cref{eq:noflux}, then
\begin{equation}
\int_a^b p(x,t)\,dx = \int_a^b p(x,0)\,dx,\quad \forall t.
\label{eq:mass_conserved}
\end{equation}
\end{lemma}
\begin{proof}
Integrate \Cref{eq:app-continuity} over $[a,b]$:
$
\frac{d}{dt}\int_a^b p(x,t)\,dx
= -\int_a^b \partial_x j(x,t)\,dx
= -(j(b,t)-j(a,t)).
$
Apply \Cref{eq:noflux}. \qedhere
\end{proof}

\begin{lemma}[Log-density dynamics]
\label{lem:log_density}
Assume $p(x,t)>0$ at $(x,t)$. Expanding \Cref{eq:app-continuity} yields
\begin{equation}
\partial_t \log p(x,t) = - f_\theta(x,t)\,\partial_x \log p(x,t) - \partial_x f_\theta(x,t).
\label{eq:log_density}
\end{equation}
Equivalently, along characteristics $x'(t)=f_\theta(x(t),t)$,
\begin{equation}
\frac{D}{Dt}\log p(x(t),t) := \partial_t\log p + f_\theta\,\partial_x\log p
= -\partial_x f_\theta.
\label{eq:material_log_density}
\end{equation}
\end{lemma}
\begin{proof}
By product rule, \Cref{eq:app-continuity} is
$
\partial_t p + f_\theta\,\partial_x p + p\,\partial_x f_\theta = 0.
$
Divide by $p>0$ to obtain \Cref{eq:log_density}. Substitute $x'(t)=f_\theta$ to get \Cref{eq:material_log_density}. \qedhere
\end{proof}

\subsection{Learning the Velocity Field via Spectrum Matching}
\label{app:cppr:learning}

We parameterize the velocity field as a neural network $f_\theta(x,t;p,\mathcal{C})$ conditioned on the current density $p(\cdot,t)$ and a crystal context $\mathcal{C}$ produced by Stage~I (state-dependent / Markovian transport). Training does not regress $f_\theta$ against an analytic reference field; instead, we back-propagate a spectrum-matching loss directly through the $K$-step discretized rollout defined by \Cref{eq:upwind}--\Cref{eq:density_from_cdf}.

Let $\hat{p}_T(\theta)$ denote the output of that rollout starting from a frozen Stage~I initial density $p_0$, and let $p_T$ be the DFT-derived normalized target. We train against the KL divergence
\begin{equation}
\mathcal{L}_{\mathrm{stage-2}}(\theta) \;=\; D_{\mathrm{KL}}\!\big(p_T\,\big\|\,\hat{p}_T(\theta)\big) \;=\; \sum_{i=1}^{N_x}p_T(x_i)\,\log\frac{p_T(x_i)}{\hat{p}_T(x_i;\theta)}\,\Delta x,
\label{eq:velocity_kl_loss}
\end{equation}
in the same distributional form used by the Stage~I decoder. The upwind CDF update~\eqref{eq:upwind}, the boundary masking, and the finite-difference density recovery~\eqref{eq:density_from_cdf} are all elementary, differentiable tensor operations, so $\partial \mathcal{L}_{\mathrm{stage-2}}/\partial \theta$ is obtained by standard back-propagation through the entire $K$-step rollout.

Crucially, the conservation guarantees of \Cref{lem:discrete_monotone,lem:discrete_mass} are properties of the integrator, not of the loss: under the CFL condition they hold for \emph{any} velocity field at every step of training. The KL gradient in \Cref{eq:velocity_kl_loss} therefore acts purely on \emph{where} $f_\theta$ moves mass along the energy axis, and the iterates $\hat{p}_T(\theta)$ remain valid normalized densities throughout training. No auxiliary conservation-penalty terms are needed.

\subsection{CDF Transport and Equivalence to Density Transport}
\label{app:cppr:cdf}

\begin{definition}[Cumulative distribution function]
Define the CDF of $p(\cdot,t)$ on $\mathbb{R}$ (or on $\Omega$ with appropriate extension)
\begin{equation}
F(x,t):=\int_{-\infty}^x p(u,t)\,du.
\label{eq:cdf_def}
\end{equation}
\end{definition}

\begin{lemma}[CDF advection PDE]
\label{lem:cdf_pde}
If $p$ satisfies \Cref{eq:app-continuity} with velocity $f_\theta$, then $F$ satisfies the first-order advection equation
\begin{equation}
\partial_t F(x,t) + f_\theta(x,t)\,\partial_x F(x,t)=0.
\label{eq:cdf_pde}
\end{equation}
Conversely, if $F$ solves \Cref{eq:cdf_pde} and is differentiable in $x$, then $p=\partial_x F$ solves \Cref{eq:app-continuity}.
\end{lemma}
\begin{proof}
Differentiate \Cref{eq:cdf_def} in $t$ and use \Cref{eq:app-continuity}:
$
\partial_t F(x,t)=\int_{-\infty}^x \partial_t p\,du
= -\int_{-\infty}^x \partial_u(p f_\theta)\,du
= -(p f_\theta)(x,t).
$
Since $\partial_x F=p$, we obtain $\partial_t F + f_\theta\,\partial_x F=0$.
Conversely, differentiating \Cref{eq:cdf_pde} w.r.t.\ $x$ yields \Cref{eq:app-continuity}. \qedhere
\end{proof}

\subsection{Discrete Upwind Scheme and Discrete Validity}
\label{app:cppr:discretization}

Partition $\Omega=[a,b]$ into $N_x$ uniform bins of width $\Delta x=(b-a)/N_x$, indexed $i=1,\dots,N_x$, and let $\{t_k\}_{k=0}^{K}$ with step $\Delta t=T/K$. Let $F_i^k\approx F(x_i,t_k)$ denote the discrete CDF at the upper edge of bin $i$, with the fixed boundary values $F_0^k\equiv 0$ and $F_{N_x}^k\equiv 1$, and let $v_i^k:=f_\theta(x_i,t_k;p^k,\mathcal{C})$ with $v_{N_x}^k=0$ (no-flux at $b$; the pinned lower endpoint $F_0^k\equiv 0$ lies outside the update range, so no flux crosses $a$).

\begin{definition}[Upwind update]
We update the CDF by a forward-Euler, first-order upwind scheme:
\begin{equation}
F_i^{k+1} = F_i^k - \Delta t\, v_i^k\, D_x^{\mathrm{up}}F_i^k,
\label{eq:upwind}
\end{equation}
where
\[
D_x^{\mathrm{up}}F_i^k=
\begin{cases}
\frac{F_i^k-F_{i-1}^k}{\Delta x}, & v_i^k\ge 0,\\[6pt]
\frac{F_{i+1}^k-F_i^k}{\Delta x}, & v_i^k< 0.
\end{cases}
\]
\end{definition}
The update is applied for $i=1,\dots,N_x$ with $v_{N_x}^k=0$ (no-flux); the upper boundary value $F_{N_x}^k\equiv 1$ stays fixed because $v_{N_x}^k=0$, while the lower boundary value $F_0^k\equiv 0$ lies outside the update range; the case $v_i^k<0$ never references an index beyond $F_{N_x}^k$.

\begin{lemma}[Discrete monotonicity and nonnegativity]
\label{lem:discrete_monotone}
Assume the Courant--Friedrichs--Lewy (CFL) condition $\max_i |v_i^k|\frac{\Delta t}{\Delta x}\le \tfrac12$ for all $k$. If $F^k$ is nondecreasing in $i$ (i.e., $F_{i}^k\ge F_{i-1}^k$), then $F^{k+1}$ is also nondecreasing. Consequently, the recovered density
\begin{equation}
p_i^{k}:=\frac{F_i^{k}-F_{i-1}^{k}}{\Delta x}\ge 0
\label{eq:density_from_cdf}
\end{equation}
is nonnegative for all $i=1,\dots,N_x$ (using $F_0^k\equiv 0$).
\end{lemma}
\begin{proof}
Write $\lambda_i^k:=|v_i^k|\,\Delta t/\Delta x\in[0,\tfrac12]$. By the upwind rule \Cref{eq:upwind}, each interior update is a convex combination of two neighboring CDF values,
\[
F_i^{k+1}=
\begin{cases}
(1-\lambda_i^k)\,F_i^k+\lambda_i^k\,F_{i-1}^k, & v_i^k\ge 0,\\[4pt]
(1-\lambda_i^k)\,F_i^k+\lambda_i^k\,F_{i+1}^k, & v_i^k<0,
\end{cases}
\]
while the boundary bins are inert, $F_0^{k+1}=F_0^k\equiv 0$ and $F_{N_x}^{k+1}=F_{N_x}^k\equiv 1$ (the lower endpoint $F_0^k$ lies outside the update range, and $v_{N_x}^k=0$ holds the upper endpoint fixed). Each update is nondecreasing in every $F_j^k$ it depends on, as all coefficients lie in $[0,1]$. To check that the ordering is preserved, fix an interior interface and compare $F_i^{k+1}$ with $F_{i-1}^{k+1}$ under the hypothesis $F_j^k\ge F_{j-1}^k$ for all $j$. For the same-sign configuration $v_{i-1}^k,v_i^k\ge 0$,
\[
F_i^{k+1}-F_{i-1}^{k+1}=(1-\lambda_i^k)\big(F_i^k-F_{i-1}^k\big)+\lambda_{i-1}^k\big(F_{i-1}^k-F_{i-2}^k\big)\ge 0,
\]
and the cases $v_{i-1}^k,v_i^k<0$ and the \emph{converging} case $v_{i-1}^k\ge 0> v_i^k$ are nonnegative by the same kind of decomposition, all under $\lambda\le 1$ alone. The only delicate case is the \emph{diverging} interface $v_{i-1}^k<0\le v_i^k$, where
\[
F_i^{k+1}-F_{i-1}^{k+1}=\big(F_i^k-F_{i-1}^k\big)\big(1-\lambda_i^k-\lambda_{i-1}^k\big)\ge 0,
\]
which is nonnegative precisely because $\lambda_i^k+\lambda_{i-1}^k\le 1$---exactly what the bound $\lambda\le\tfrac12$ guarantees. (The looser bound $\lambda\le 1$ does \emph{not} suffice: taking $\lambda_{i-1}^k=\lambda_i^k=1$ at a diverging interface makes this increment negative, producing a negative recovered density.) Hence $F^{k+1}$ is nondecreasing, and \Cref{eq:density_from_cdf} is a forward difference of a nondecreasing sequence with $F_0^k\equiv 0$, so $p_i^k\ge 0$. \qedhere
\end{proof}

\begin{lemma}[Discrete mass conservation]
\label{lem:discrete_mass}
Let $p_i^k$ be defined by \Cref{eq:density_from_cdf}. With the fixed boundary values $F_0^k\equiv 0$ and $F_{N_x}^k\equiv 1$,
\[
\sum_{i=1}^{N_x} p_i^k\,\Delta x = F_{N_x}^k - F_0^k = 1
\]
for all $k$, i.e., total mass is conserved exactly (up to floating-point rounding, which a final renormalization absorbs). This holds for \emph{any} velocity field, independently of the CFL condition, since it uses only the fixed endpoints.
\end{lemma}
\begin{proof}
By telescoping,
$
\sum_{i=1}^{N_x} p_i^k\,\Delta x
=\sum_{i=1}^{N_x}(F_i^k-F_{i-1}^k)=F_{N_x}^k-F_0^k=1-0,
$
using the fixed boundary values. \qedhere
\end{proof}

\subsection{End-to-End Validity Guarantee}
\label{app:cppr:closure}

\begin{proposition}[Constraint-preserving refinement]
\label{thm:constraint_preserving}
Assume the refinement evolves a valid initial CDF $F(\cdot,0)$ (nondecreasing with limits $0$ and $1$) under \Cref{lem:cdf_pde} in the continuous setting, or under \Cref{eq:upwind} with the CFL condition in the discrete setting. Then the resulting refined density $p(\cdot,t)=\partial_x F(\cdot,t)$ (or $p_i^k$ via \Cref{eq:density_from_cdf}) satisfies:
\begin{enumerate}
\item \textbf{Nonnegativity:} $p(\cdot,t)\ge 0$ (or $p_i^k\ge 0$);
\item \textbf{Mass conservation:} $\int_\Omega p(\cdot,t)\,dx=1$ (or $\sum_i p_i^k\Delta x= 1$);
\item \textbf{Pure transport:} refinement redistributes spectral mass without creation/destruction.
\end{enumerate}
\end{proposition}
\begin{proof}
Continuous case: by \Cref{lem:cdf_pde}, density transport is equivalent to CDF advection; \Cref{lem:mass_conservation} gives conservation, and CDF monotonicity implies nonnegativity. Discrete case: \Cref{lem:discrete_monotone} gives nonnegativity and \Cref{lem:discrete_mass} gives conservation. \qedhere
\end{proof}

\paragraph{Remark.}
These guarantees hold by construction at every step of the discretized rollout, independently of how $f_\theta$ is parameterized or trained: mass conservation (\Cref{lem:discrete_mass}) follows from the fixed boundary values alone and is therefore unconditional, while nonnegativity (\Cref{lem:discrete_monotone}) additionally requires the CFL condition $\max_i|v_i^k|\Delta t/\Delta x\le\tfrac12$, which we enforce by zero-initializing and clamping the velocity field (\Cref{app:hyper}). Back-propagating \Cref{eq:velocity_kl_loss} through the rollout therefore yields an end-to-end closed-loop refinement module in which the spectrum-matching loss is the only training signal and constraint preservation is delegated entirely to the integrator.

%% file: appendix/algorithm.tex
\section{Algorithm}
\label{app:algo}

This section summarizes the training and inference pipeline of DeepDOSReasoner, including (i) structure-aware spectrum decoding, (ii) constraint-preserving physical reasoning via mass-conserving iterative refinement, and (iii) test-time recovery of the absolute spectrum by rescaling the predicted normalized shape with the total in-window mass $Z$, which is supplied by a trained states-count oracle for eDOS and by the analytic modes count $3N$ directly for phDOS (\Cref{app:oracle}).

\paragraph{Notation.}
Let a crystal be represented as a graph $G=(V,E)$ with node features and edge features. Let the energy axis $[a,b]$ be discretized into $N_x$ uniform bins of width $\Delta x=(b-a)/N_x$, indexed $i=1,\dots,N_x$. The ground-truth (unnormalized) eDOS is $y\in\mathbb{R}_+^{N_x}$ and its normalized density is
$p_T = y / (\sum_i y_i\,\Delta x)$, so that $\sum_i p_{T,i}\,\Delta x = 1$.
The decoder predicts $\hat{y}_0\in\mathbb{R}_+^{N_x}$ and its normalized density is
$p_0 = \hat{y}_0 / (\sum_i \hat{y}_{0,i}\,\Delta x)$.
Each density is represented by its discrete CDF with the fixed boundary values $F_0\equiv 0$ and $F_{N_x}\equiv 1$; we write $F^k$ for the CDF at rollout step $k$, so $F^0$ is the CDF of $p_0$.
The refinement velocity network is $f_\theta(x,t;p,\mathcal{C})$, conditioned on the current density $p$ and a pooled crystal context $\mathcal{C}$ (encoder global feature and decoder-side context).
The decoder and the refinement act only on the normalized \emph{shape}; the absolute spectrum is recovered at inference as $\hat{y}=Z\,p$, where the total in-window mass $Z$ is set differently per spectrum. For eDOS, a separate states-count oracle $\mathrm{Oracle}_\psi$ (parameters $\psi$) regresses the intensive ratio $R=N_{\mathrm{win}}/N_{\mathrm{tot}}$ from the crystal graph $G$ and the analytic full-range total $N_{\mathrm{tot}}=\int_{-\infty}^{\infty}D\,dE$ (number of included bands times spin degeneracy), and the scale is recovered as $Z=N_{\mathrm{win}}=R\,N_{\mathrm{tot}}$; for phDOS the $[0,1000]$~cm$^{-1}$ window captures all modes for the large majority of materials, so the analytic modes count $Z=3N$ is used directly (\Cref{app:oracle}).

\subsection{Training and Inference Procedure}
\label{sec:pipeline:algo}

We show the training and inference pipelines of DeepDOSReasoner in \Cref{alg:dosreasoner}. We also provide some useful notes for practical implementation below.

\medskip
\noindent\rule{\textwidth}{0.8pt}
\begingroup
\footnotesize
\captionof{algorithm}{(DeepDOSReasoner) Training and Inference}
\label{alg:dosreasoner}
\vspace{-4pt}
\noindent\rule{\textwidth}{0.4pt}
\begin{algorithmic}[1]
\Require Training set $\mathcal{D}=\{(G^{(n)},y^{(n)})\}_{n=1}^{N_{\mathrm{data}}}$; $N_x$ energy bins $\{x_i\}_{i=1}^{N_x}$ (CDF boundaries $F_0\equiv0$, $F_{N_x}\equiv1$); refinement horizon $T$; number of update steps $K$; $\Delta t=T/K$
\Require Decoder parameters $\phi$; refinement parameters $\theta$; states-count oracle $\mathrm{Oracle}_\psi$ (eDOS scale only)

\vspace{2pt}
\Statex \texttt{(Training) Structure-aware Spectrum Decoding}
\For{each minibatch}
    \State Encode crystal with edge-aware crystal graph Transformer: $\{h_j\},\,h_x \gets \mathrm{Enc}_\phi(G)$ \Comment{$j$ indexes atoms}
    \State Decode eDOS with energy self-attention + source (atom-to-energy) attention:
    \State \hspace{1.1em} $\hat{y}_0 \gets \mathrm{Dec}_\phi(\{h_j\},h_x,\{x_i\})$
    \State Normalize: $p_T \gets y/(\sum_i y_i\,\Delta x)$, \quad $p_0 \gets \hat{y}_0/(\sum_i \hat{y}_{0,i}\,\Delta x)$
    \State Compute distributional reconstruction loss:
    \State \hspace{1.1em} $\mathcal{L}_{\mathrm{stage-1}}(\phi)\gets D_{\mathrm{KL}}(p_T\|p_0)$
    \State Update decoder parameters: $\phi \leftarrow \phi - \eta_\phi \nabla_\phi \mathcal{L}_{\mathrm{stage-1}}$
\EndFor

\vspace{2pt}
\Statex \texttt{(Training) Mass-Conserving Iterative Refinement}
\For{each minibatch $\{(G,y)\}\sim \mathcal{D}$}
    \State \textit{(Freeze $\phi$)} Compute initial prediction $\hat{y}_0$ via Stage~I; normalize to $p_0$
    \State Normalize ground truth to $p_T \gets y/(\sum_i y_i\,\Delta x)$; build pooled context $\mathcal{C}$ (encoder global feature and decoder-side context) from Stage~I
    \State Initialize CDF: $F^0_0\gets 0$, $F^0_i \gets \sum_{m=1}^{i} p_{0,m}\Delta x$ for $i=1,\dots,N_x$; set step size $\Delta t \gets T/K$
    \For{$k=0,1,\dots,K-1$} \Comment{differentiable $K$-step upwind rollout}
        \State Recover current density: $p^k_i \gets (F^k_i-F^k_{i-1})/\Delta x$ for $i=1,\dots,N_x$ (using $F^k_0\equiv0$; for conditioning)
        \State Evaluate per-bin velocity: $v_i^k \gets f_\theta(x_i,t_k;\,p^k,\mathcal{C})$; set $v_{N_x}^k=0$ (no-flux)
        \State Upwind CDF update: $F^{k+1}_i \gets F^k_i - \Delta t\,v_i^k\,D_x^{\mathrm{up}}F^k_i$, clipped to $[0,1]$
    \EndFor
    \State Recover final density: $\hat{p}_{T,i} \gets (F^K_i-F^K_{i-1})/\Delta x$ for $i=1,\dots,N_x$ (using $F^K_0\equiv0$); renormalize
    \State KL spectrum-matching loss: $\mathcal{L}_{\mathrm{stage-2}}(\theta)\gets D_{\mathrm{KL}}\!\big(p_T\,\big\|\,\hat{p}_T(\theta)\big)$
    \State Update refinement parameters: $\theta \leftarrow \theta - \eta_\theta \nabla_\theta \mathcal{L}_{\mathrm{stage-2}}$ \Comment{back-prop through rollout}
\EndFor

\vspace{2pt}
\Statex \texttt{(Inference) Decoding $\rightarrow$ Reasoning}
\Function{Infer}{$G$}
    \State $\{h_j\},\,h_x \gets \mathrm{Enc}_\phi(G)$; \quad $\hat{y}_0 \gets \mathrm{Dec}_\phi(\{h_j\},h_x,\{x_i\})$
    \State $p_0 \gets \hat{y}_0/(\sum_i \hat{y}_{0,i}\,\Delta x)$; \quad $F^0_0\gets 0,\ F^0_i \gets \sum_{m=1}^{i} p_{0,m}\Delta x$ \Comment{initial CDF}
    \State Pooled context $\mathcal{C}$ (encoder global feature and decoder-side context) from Stage~I
    \For{$k=0,1,\dots,K-1$} \Comment{same upwind rollout as training, $\Delta t=T/K$}
        \State Recover current density: $p^k_i \gets (F^k_i-F^k_{i-1})/\Delta x$ for $i=1,\dots,N_x$ (using $F^k_0\equiv0$; for conditioning)
        \State Evaluate per-bin velocity: $v_i^k \gets f_\theta(x_i,t_k;\,p^k,\mathcal{C})$; set $v_{N_x}^k=0$
        \State Upwind CDF update: $F^{k+1}_i \gets F^k_i - \Delta t\,v_i^k\,D_x^{\mathrm{up}}F^k_i$, clipped to $[0,1]$
    \EndFor
    \State Recover final density: $p^K_i \gets (F^K_i-F^K_{i-1})/\Delta x$ for $i=1,\dots,N_x$ (using $F^K_0\equiv0$); renormalize \Comment{normalized shape}
    \Statex \textit{(Inference) Absolute-scale recovery (states-count oracle for eDOS; analytic $3N$ for phDOS)}
    \If{eDOS}
        \State $N_{\mathrm{tot}} \gets \int_{-\infty}^{\infty}D\,dE$ \Comment{analytic: \# included bands $\times$ spin degeneracy}
        \State $R \gets \mathrm{Oracle}_\psi(G,\,N_{\mathrm{tot}})$;\quad $Z \gets R\cdot N_{\mathrm{tot}}$ \Comment{oracle regresses $R=N_{\mathrm{win}}/N_{\mathrm{tot}}$}
    \ElsIf{phDOS}
        \State $Z \gets 3N$ \Comment{analytic modes count used directly; no oracle}
    \EndIf
    \State Rescale to absolute DOS: $\hat{y} \gets Z\cdot p^K$
    \State \Return absolute DOS $\hat{y}$ (and normalized shape $p^K$)
\EndFunction
\end{algorithmic}
\noindent\rule{\textwidth}{0.4pt}
\endgroup
\medskip

\paragraph{Upwind Derivative.}
For $D_x^{\mathrm{up}}$, use
\[
D_x^{\mathrm{up}}(F)_i =
\begin{cases}
(F_i-F_{i-1})/\Delta x, & v_i \ge 0,\\
(F_{i+1}-F_i)/\Delta x, & v_i < 0,
\end{cases}
\]
with boundary handling consistent with no-flux (or fixed-endpoint) conditions.

\paragraph{Practical Stability.}
Enforce a CFL condition $\max_i |v_i|\,\Delta t/\Delta x \le \tfrac12$ (via smaller $\Delta t$ or clamping) to keep the CDF monotone and the density nonnegative; the looser bound $\le 1$ does not guarantee monotonicity where the velocity field changes sign (see \Cref{lem:discrete_monotone}).

%% file: appendix/details.tex
\section{Additional Experimental Setup Details}
\label{app:hyper}

\subsection{Data Curation Setup}

\paragraph{eDOS Benchmark.}
Our primary eDOS benchmark is curated from Materials Project structures and DFT electronic-structure calculations~\citep{Jain2013,Kong2022} and spans both thermodynamically stable and metastable entries; metastable entries dominate the dataset. For the controlled stability analysis reported in the main text, we additionally use a stable-only subset comprising the structures lying on the convex hull (energy above hull equal to zero). The same preprocessing pipeline is applied to both protocols.

The target is the electronic density of states (eDOS) ($D(E)$), which describes how densely electronic states populate each energy level ($E$). A central feature of realistic spectra is the band gap: for an insulator/semiconductor, there exists an energy interval with (near-)zero density separating the valence band (occupied states) from the conduction band (unoccupied states). The valence-band maximum (VBM) and conduction-band minimum (CBM) define the gap edges, with $E_g = E_{\mathrm{CBM}} - E_{\mathrm{VBM}} \geq 0$. Unlike previous work that removes the interval $[E_{\mathrm{VBM}}, E_{\mathrm{CBM}}]$ for gapped materials—effectively “splicing” the spectrum and sidestepping gap prediction—we retain the full eDOS so the model must learn both (i) to place the gap at the correct energy and (ii) to maintain the sharp spectral transition at the band edges. This design is closer to the physical object returned by electronic-structure calculations and makes evaluation more stringent, since small errors near $E_{\mathrm{VBM}}$ and $E_{\mathrm{CBM}}$ can blur the gap or introduce spurious in-gap states.

The preprocessing pipeline applied to each raw DOS consists of three steps. (i) \emph{Fermi-level alignment}: each spectrum is shifted along the energy axis so that the DFT-computed Fermi level $E_F$ lies at $E=0$, placing valence and conduction features on a common reference across materials regardless of their absolute Fermi energies. (ii) \emph{Gaussian smearing}: the aligned spectrum is convolved with a Gaussian kernel of width $\sigma=0.5$~eV, attenuating discretization noise and finite-$k$-point sampling artifacts while preserving physical band-edge sharpness and van Hove singularities sufficiently to evaluate fine spectral structure. (iii) \emph{Area-preserving rebinning}: the smoothed spectrum is rebinned onto a uniform 128-bin grid over $[-4,4]$~eV by integrating the smoothed continuous spectrum over each target bin and dividing by the bin width, so that the discretized output exactly conserves the integral of the smoothed spectrum (and, consequently, the integrated number of states up to the grid endpoints). Together, retaining the full spectrum and using a conservation-respecting discretization yields a challenging, physically faithful benchmark that emphasizes band-edge fidelity, gap sharpness, and the model’s ability to reproduce fine-grained spectral structure rather than only coarse integrated properties.

\paragraph{phDOS Benchmark.}
For the phonon density of states experiments, we adopt the DFT-computed phDOS dataset released by~\citep{Chen2021}, which collects DFPT-derived phonon spectra~\citep{Petretto2018} for $1{,}524$ crystalline materials drawn from the Materials Project as part of the high-throughput Materials Project phonon database. In the source release, the raw high-resolution DFPT phDOS is first smoothed with a Savitzky--Golay filter (window length 101, polynomial order 3) to attenuate finite-$\mathbf{q}$-grid sampling artifacts while preserving the dominant acoustic-onset, optical-peak, and van Hove features, and the smoothed spectrum over $[0,1000]$~cm$^{-1}$ is then interpolated onto a uniform 51-point frequency grid (spacing $20$~cm$^{-1}$). We adopt the train/validation/test partition supplied by the source release, so the held-out test set used in the main text contains $n=152$ materials, and we use the released spectra without further reprocessing: each phDOS is already on the common 51-point grid and is treated as a nonnegative one-dimensional density over the bounded frequency axis.

The target is the phDOS $g(\omega)$, which describes how densely vibrational normal modes populate each frequency $\omega$ (see \Cref{app:phdos} for background). Unlike the eDOS, the phDOS has no Fermi level and no analogue of the VBM--CBM band gap, but the same total-mass conservation law applies, in the form of the modal sum rule $\int g(\omega)\,d\omega = 3N$ for a unit cell with $N$ atoms (\Cref{eq:phdos_mass}). This rule holds over the full spectrum $[0,\infty)$; on the truncated $[0,1000]$~cm$^{-1}$ grid it equals $3N$ only when the window captures every mode. In practice the cutoff is high enough that the phDOS has decayed to zero before $1000$~cm$^{-1}$ for the large majority of materials---about $94\%$ of the dataset ($143/152$ test materials), for which $\int_0^{1000}g\,d\omega = 3N$ holds and rescaling by the analytic $3N$ is exact (the trapezoidal endpoint term vanishes because $g\approx 0$ at the edge). For the remaining ${\sim}6\%$ ($9/152$ test; $86/1{,}524$ overall)---light-atom or strongly bonded compounds containing O, N, P, H, Be, B, C, or F, whose high-frequency optical stretches (e.g., O--H, B--N, B--O) extend beyond $1000$~cm$^{-1}$---the window truncates part of the spectrum and the analytic $3N$ slightly over-counts the in-window mass. At inference time we therefore rescale our predicted normalized shape by the analytic modes count $Z=3N$ directly (no learned scale oracle is used for phDOS, unlike the eDOS case of \Cref{app:oracle}), following the same normalize-and-rescale protocol used for eDOS but with the analytic constant substituted. No architectural or training changes are made for the phDOS benchmark: both Stage~I (structure-aware spectrum decoding) and Stage~II (mass-conserving iterative refinement) operate on the 51-point frequency grid in place of the 128-bin energy grid, and the only inference-time difference relative to the eDOS pipeline is the choice of $Z$. The phDOS benchmark therefore acts as a cross-physics transfer test of the framework: it changes the spectral coordinate, the grid resolution, and the sum-rule constant, while keeping the underlying conservation law in continuity-equation form.

\subsection{Training Setup}
\paragraph{Edge-Aware Crystal Graph Transformer.}
Our \emph{Edge-Aware Crystal Graph Transformer} encoder comprises three stacked \texttt{TransformerConv} layers~\citep{Fey/Lenssen/2019} with hidden size 512. We set both the node embedding dimension \(d_h\) and edge embedding dimension \(d_e\) to 512, and use 4 attention heads per layer. Each layer wraps the edge-modulated attention update in a gated residual connection that adaptively interpolates between the new representation and the projected input via a learnable gate, with \texttt{LayerNorm} applied to the gated combination before the post-norm nonlinearity, exactly as described in the main Methods. To regularize the encoder, we apply node-feature dropout with rate \(p=0.2\) in each layer.

\paragraph{Structure-to-Spectrum Decoder.}
The \emph{Structure-to-Spectrum Decoder} predicts eDOS on an energy grid using a two-stage decoding design. The first decoder module attends from positional (energy-bin) queries to the crystal graph representations to produce an intermediate spectrum representation; the second module then performs an additional round of attention refinement, conditioned on the same crystal representations. Each module contains 6 Transformer layers (12 layers total). We use an embedding dimension of 512, feed-forward dimension 2048, and 8 attention heads per layer. We adopt a Pre-\texttt{LayerNorm} formulation, where layer normalization precedes the attention and FFN sublayers, which improves optimization stability relative to Post-LayerNorm. The final output is passed through \texttt{Softplus}~\citep{zheng2015improving} to enforce nonnegativity, and is then normalized to sum to one across energy bins, yielding an initial normalized eDOS that can be treated as a discrete probability distribution.

\paragraph{Mass-Conserving Iterative Refinement.}
Starting from the initial prediction, we apply a \emph{Mass-Conserving Iterative Refinement} module that improves the density via a learned transport (velocity) field while preserving total mass. The velocity-field network follows a simplified variant of the decoder architecture and keeps the same core hyperparameters (embedding size, FFN size, and number of heads) to model fine-grained transport dynamics. For refinement, we discretize the time horizon \(t \in [0,T]\) with \(T=1\) and \(K=5\) steps, giving \(\Delta t = T/K = 0.2\).

\paragraph{Velocity-Field Network.}
Concretely, the velocity field \(f_\theta\) is parameterized as a per-bin Transformer. For each of the $N_x$ spectral bins we form an input by concatenating four features: a learned positional embedding for the bin index, the current normalized density value \(p^k\) at that bin, a scalar time fraction \(t_k/T\) broadcast across bins, and a Stage~I crystal context of dimension \(2\,d_{\mathrm{model}}\) (encoder global feature concatenated with the decoder-side context) broadcast across bins. The concatenation is linearly projected back to \(d_{\mathrm{model}}\) through a single \texttt{Linear} layer followed by \texttt{LeakyReLU}, and the resulting sequence is processed by two stacked self-attention blocks whose embedding dim, FFN dim, head count, and per-block layer count match the Stage~I decoder (\Cref{tab:hyperparams}). A final \texttt{Linear} layer maps each per-bin hidden state to a scalar velocity, producing the per-bin velocities \(\{v_i^k\}_{i=1}^{N_x}\) that drive the upwind CDF update. The output velocities are clamped so that the per-bin Courant number satisfies \(|v_i^k|\,\Delta t/\Delta x\le\tfrac12\) (i.e., \(|v_i^k|\le\Delta x/(2\Delta t)\)); this enforces the CFL condition of \Cref{lem:discrete_monotone} by construction for any network weights, keeping every coefficient of the convex-combination upwind update nonnegative, so the discrete maximum principle holds and the recovered density stays nonnegative at every rollout step. The weight and bias of this output projection are zero-initialized, so at the start of training the velocity field is identically zero and the discrete upwind step reduces to the identity; refinement is then learned as an additive correction on top of this identity warm-start, which keeps the mass-conservation and nonnegativity guarantees of the integrator active from the first optimizer step.

\paragraph{Training Details.}
Training proceeds in two stages. In Stage~I, we jointly train the structure encoder and spectrum decoder for 160 epochs with batch size 128 using Adam~\citep{kinga2015method} with \(\beta=(0.9,0.98)\) and \(\epsilon=10^{-9}\), and a Transformer-style inverse-square-root learning-rate schedule~\citep{vaswani2017attention},
\begin{equation}
\mathrm{lr}(\mathrm{step}) = \mathrm{lr}_{\mathrm{base}}\,\cdot\,d_{\mathrm{model}}^{-1/2}\,\cdot\,\min\!\big(\mathrm{step}^{-1/2},\;\mathrm{step}\cdot\mathrm{warmup}^{-3/2}\big),
\label{eq:noam_lr}
\end{equation}
with \(d_{\mathrm{model}}=512\), \(\mathrm{warmup}=300\) steps, and \(\mathrm{lr}_{\mathrm{base}}=1.0\). The value \(\mathrm{lr}_{\mathrm{base}}=1.0\) reported in \Cref{tab:hyperparams} is the scaling multiplier on the Noam schedule rather than a literal Adam learning rate; the effective peak learning rate at \(\mathrm{step}=\mathrm{warmup}\) is \(\mathrm{lr}_{\mathrm{base}}\cdot d_{\mathrm{model}}^{-1/2}\cdot\mathrm{warmup}^{-1/2}\approx 2.55\times 10^{-3}\), consistent with standard Transformer training. After \(c_{\mathrm{epochs}}=121\) epochs (corresponding to \(c_{\mathrm{steps}}\) optimizer updates), the schedule is additionally divided by \(\mathrm{rate}_{\mathrm{decay}}^{\,\mathrm{step}/c_{\mathrm{steps}}}\) with \(\mathrm{rate}_{\mathrm{decay}}=4\), giving a slow decay tail in the late-training phase. In Stage~II, we freeze the pretrained encoder and decoder and train only the refinement (velocity-field) network for 20 epochs with batch size 128 using AdamW~\citep{loshchilov2017decoupled} (learning rate 0.01, weight decay 0.001), with a \texttt{Cosine} annealing learning-rate schedule and a \texttt{Linear} warmup over the first 30\% of training steps. All modules are implemented in \texttt{PyTorch} and trained on a single NVIDIA H100 GPU.

\begin{table}[ht!]
\centering
\caption{Hyperparameter settings for DeepDOSReasoner.}
\label{tab:hyperparams}
\footnotesize
\setlength{\tabcolsep}{4pt}
\resizebox{\textwidth}{!}{%
\begin{tabular}[t]{ll}
\toprule
\textbf{Component / Setting} & \textbf{Value} \\
\midrule
\multicolumn{2}{l}{\textbf{Edge-Aware Crystal Graph Transformer}} \\
\multicolumn{2}{l}{\textbf{(Encoder)}} \\
\midrule
Backbone operator & \texttt{TransformerConv}~\citep{Fey/Lenssen/2019} \\
\# encoder layers & 3 \\
Hidden size & 512 \\
Node embedding dim \(d_h\) & 512 \\
Edge embedding dim \(d_e\) & 512 \\
\# attention heads / layer & 4 \\
Residual connection & Gated (learnable gate) \\
Normalization & \texttt{LayerNorm} \\
Post-norm activation & \texttt{LeakyReLU} \\
Node feature dropout \(p\) & 0.2 \\
\midrule
\multicolumn{2}{l}{\textbf{Structure-to-Spectrum Decoder}} \\
\midrule
Decoder modules & 2 \\
Layers per module & 6 \\
Total decoder layers & 12 \\
Embedding dim & 512 \\
FFN dim & 2048 \\
\# attention heads / layer & 8 \\
LayerNorm placement & Pre-\texttt{LayerNorm} \\
Output activation & \texttt{Softplus}~\citep{zheng2015improving} \\
Output normalization & \texttt{SumNorm} \\
\midrule
\multicolumn{2}{l}{\textbf{Mass-Conserving Iterative Refinement}} \\
\multicolumn{2}{l}{\textbf{(MCIR)}} \\
\midrule
Velocity net embedding dim & 512 \\
Velocity net FFN dim & 2048 \\
Velocity net \# heads / layer & 8 \\
Time horizon \(T\) & 1 \\
\# refinement steps \(K\) & 5 \\
Step size \(\Delta t\) & 0.2 \\
\bottomrule
\end{tabular}\hspace{1.5em}%
\begin{tabular}[t]{ll}
\toprule
\textbf{Component / Setting} & \textbf{Value} \\
\midrule
\multicolumn{2}{l}{\textbf{Stage~I: Encoder + Decoder Training}} \\
\midrule
Epochs & 160 \\
Batch size & 128 \\
Optimizer & Adam~\citep{kinga2015method} \\
Adam \(\beta\) & (0.9, 0.98) \\
Adam \(\epsilon\) & \(10^{-9}\) \\
LR schedule & Inverse-sqrt (Noam) \\
 & \quad + decay tail \\
\(\mathrm{lr}_{\mathrm{base}}\) (schedule multiplier) & 1.0 \\
Effective peak learning rate & \(\approx 2.55\times 10^{-3}\) \\
\(d_{\mathrm{model}}\) (schedule) & 512 \\
Warmup steps & 300 \\
Decay onset \(c_{\mathrm{epochs}}\) & 121 \\
\(\mathrm{rate}_{\mathrm{decay}}\) & 4 \\
\midrule
\multicolumn{2}{l}{\textbf{Stage~II: MCIR Training}} \\
\multicolumn{2}{l}{\textbf{(Encoder/Decoder Frozen)}} \\
\midrule
Epochs & 20 \\
Batch size & 128 \\
Optimizer & AdamW~\citep{loshchilov2017decoupled} \\
Learning rate & 0.01 \\
Weight decay & 0.001 \\
LR schedule & \texttt{Cosine} annealing \\
 & \quad + \texttt{Linear} warmup \\
Warmup ratio & 30\% of total steps \\
\midrule
\multicolumn{2}{l}{\textbf{Implementation}} \\
\midrule
Framework & \texttt{PyTorch} \\
Hardware & 1 $\times$ NVIDIA H100 (80GB) \\
Seed-averaging ensemble size $M$ & 2 \\
\bottomrule
\end{tabular}}
\end{table}

\subsection{States-Count Oracle for Absolute eDOS Scale}\label{app:oracle}
DeepDOSReasoner predicts the normalized eDOS \emph{shape}; recovering the absolute spectrum requires the integrated number of states inside the analysis window, $Z=N_{\mathrm{win}}=\int_{E_F-4}^{E_F+4}D(E)\,dE$. The full-energy total is analytic---$\int_{-\infty}^{\infty}D(E)\,dE$ equals the number of included bands times the spin degeneracy---but the fraction of that total falling inside the finite $[-4,4]$~eV window depends on the material, so $N_{\mathrm{win}}$ is not fixed by composition alone. We estimate it with a lightweight \emph{states-count oracle}. For phDOS no such oracle is needed: the $[0,1000]$~cm$^{-1}$ window captures all $3N$ modes for the large majority of materials (\Cref{app:hyper}), so the analytic modes count $Z=3N$ is used directly as the scale.

\paragraph{Inputs and Target.}
The oracle encodes the crystal structure with a crystal graph Transformer (CrystalFormer~\citep{wang2024conformal}) and is additionally conditioned on the analytic full-range total $N_{\mathrm{tot}}=\int_{-\infty}^{\infty}D(E)\,dE$, which is fixed by the number of included bands and the spin degeneracy. For a fixed DFT recipe this total is set by the composition and the band-count convention, so it is available even for candidate materials lacking a computed eDOS (such as those in the prospective discovery pool). Because $N_{\mathrm{win}}$ is \emph{extensive}---it scales with cell size and bands count, spanning more than three orders of magnitude across the benchmark (\Cref{tab:oracle_stats})---regressing it directly is poorly conditioned. The oracle instead predicts the \emph{intensive} ratio $R\coloneqq N_{\mathrm{win}}/N_{\mathrm{tot}}$, the size-normalized fraction of the total states that fall in the window, and recovers the absolute scale as $N_{\mathrm{win}}=R\cdot N_{\mathrm{tot}}$. Targets are the in-window states counts integrated from the DFT reference spectra of the eDOS benchmark (\Cref{app:hyper}), with an outlier filter clipping $R$ at $1.0$. The ratio is tightly concentrated (mean $0.567$, interdecile range $0.368$--$0.805$; \Cref{tab:oracle_stats}).

\paragraph{Accuracy.}
\Cref{tab:oracle_acc} reports held-out accuracy. The oracle predicts the intensive ratio to $R^2=0.921$ (MAE $0.031$, MAPE $5.9\%$); after rescaling by the analytic full-range total, the recovered absolute count $N_{\mathrm{win}}$ attains $R^2=0.982$ (MAE $3.58$ states); this higher $R^2$ largely reflects the exact, high-dynamic-range $N_{\mathrm{tot}}$ factor, which carries most of the variance of $N_{\mathrm{win}}$, so the intensive-ratio $R^2=0.921$ and the scale-free MAPE of $5.9\%$ (identical for $R$ and $N_{\mathrm{win}}$) are the more informative measures of predictive skill. Accuracy is marginally higher for metals (gap${=}0$, $R^2=0.989$) than for gapped materials ($R^2=0.971$), consistent with the additional near-$E_F$ band-edge structure in the latter. This ${\sim}6\%$ scale error is small relative to the shape-fidelity gains of the two-stage model and leaves the conservation guarantees of \Cref{thm:constraint_preserving} intact: rescaling a normalized density by a positive scalar preserves nonnegativity and reproduces the assigned in-window mass exactly.

\begin{table}[ht]
\centering
\caption{Distribution of the absolute eDOS in-window states count $N_{\mathrm{win}}$ and the intensive ratio $R=N_{\mathrm{win}}/N_{\mathrm{tot}}$ across the eDOS benchmark, where $N_{\mathrm{tot}}=\int_{-\infty}^{\infty}D\,dE$ is the analytic full-range total (number of included bands times spin degeneracy). $N_{\mathrm{win}}$ is extensive (it scales with cell size and bands count); $R$ is the size-normalized quantity the oracle regresses. The upper tail of $R$ is clipped at $1.0$ by the outlier filter.}
\label{tab:oracle_stats}
\small
\setlength{\tabcolsep}{6pt}
\begin{tabular}{lccccccc}
\toprule
Quantity & Mean & Median & Std & Min & P10 & P90 & Max \\
\midrule
$N_{\mathrm{win}}$ & 65.7 & 51.0 & 51.6 & 0.46 & 18.4 & 130.5 & 740.3 \\
$R=N_{\mathrm{win}}/N_{\mathrm{tot}}$ & 0.567 & 0.535 & 0.186 & 0.060 & 0.368 & 0.805 & 1.00 \\
\bottomrule
\end{tabular}
\end{table}

\begin{table}[ht]
\centering
\caption{Held-out accuracy of the eDOS states-count oracle. The oracle regresses the intensive ratio $R=N_{\mathrm{win}}/N_{\mathrm{tot}}$ and recovers the absolute count $N_{\mathrm{win}}=R\cdot N_{\mathrm{tot}}$ by rescaling with the analytic full-range total $N_{\mathrm{tot}}$. The bottom rows stratify the recovered $N_{\mathrm{win}}$ by metallic (gap${=}0$) versus gapped materials. RMSE, root-mean-square error; MAPE, mean absolute percentage error.}
\label{tab:oracle_acc}
\small
\setlength{\tabcolsep}{8pt}
\begin{tabular}{lcccc}
\toprule
Target & MAE & RMSE & $R^2$ & MAPE \\
\midrule
$R$ & 0.0309 & 0.0515 & 0.921 & 5.9\% \\
$N_{\mathrm{win}}$ & 3.58 & 6.90 & 0.982 & 5.9\% \\
\midrule
\quad $N_{\mathrm{win}}$, metals (gap${=}0$) & 2.95 & 5.71 & 0.989 & 5.2\% \\
\quad $N_{\mathrm{win}}$, gapped & 4.37 & 8.14 & 0.971 & 6.9\% \\
\bottomrule
\end{tabular}
\end{table}

\subsection{Baselines}\label{app:base}
We benchmark against representative methods for learning from periodic crystal structures and for predicting structured spectra. The baselines fall into three families: (i) crystal-graph structure encoders with regression heads, which embed a crystal graph into a global representation and then regress the eDOS pointwise; (ii) interatomic-potential architectures repurposed for spectrum output, which produce node-level features used for force/energy prediction in their original setting and which we adapt with a regression head over the same fixed energy grid; and (iii) crystal-to-sequence predictors, which model the eDOS as an ordered sequence over energy bins.

\paragraph{Adapting Baselines to DOS Prediction.}
The structure-encoder and interatomic-potential baselines were originally developed for scalar or per-atom regression targets (e.g., formation energy, atomic energies and forces) rather than for full spectra. To adapt each of them to DOS prediction, we leave the published backbone architecture unchanged and attach a single regression head on top of its pooled, crystal-level embedding: the head is a multilayer perceptron that maps the pooled embedding to the $L$ values of the spectrum on the fixed grid used throughout this work ($L=128$ energy bins for eDOS over $[-4,4]$~eV; $L=51$ frequency bins for phDOS over $[0,1000]$~cm$^{-1}$). No other modification is made to the backbones. All such baselines are trained and evaluated on the same structures, preprocessing pipeline, and train/validation/test splits as DeepDOSReasoner, so that differences in the reported metrics reflect the choice of structural representation and decoding pathway rather than differences in data or grid. The crystal-to-sequence baselines (Mat2Spec, DOSTransformer) natively decode the spectrum over energy bins in their original formulations and therefore use their own spectrum decoders rather than an added regression head.

\paragraph{Structure Encoders with Regression Head.}
\begin{enumerate}
    \item \textbf{E3NN (E(3)-equivariant neural networks)}~\citep{geiger2022e3nn}. E3NN is a general framework for building neural networks that are equivariant to 3D Euclidean transformations (rotations/translations, and in some settings reflections), using spherical harmonics and tensor products to ensure features transform predictably under coordinate changes. This inductive bias is well matched to atomistic systems where absolute orientation should not affect predictions. In our benchmarking, E3NN represents the family of geometric equivariant message-passing / tensor-field models commonly used for 3D materials learning, which encode local atomic environments in an equivariant manner and then pool to a crystal-level representation before regression.
    \item \textbf{MEGNet (materials graph network with set-to-set readout)}~\citep{Chen2019}. MEGNet is a universal graph network for molecules and crystals that augments a CGCNN-style message-passing backbone with state-level features (e.g., composition or external conditions) and a set-to-set graph-level readout. The state inputs and learned graph pooling allow MEGNet to capture global features and aggregate node information into a robust crystal-level embedding, making it a strong representative of the ``encoder + global readout + regression head'' family in our comparison.
    \item \textbf{GATGNN (graph attention GNN with global attention)}~\citep{louis2020graph}. GATGNN is a crystal graph neural network designed for inorganic materials property prediction. It stacks multiple (augmented) graph-attention layers for message passing and includes a global attention mechanism to aggregate node features into a graph-level embedding. Compared to vanilla MPNNs, its core distinction is the attention-based weighting of neighbor contributions and the explicit global attention readout, aimed at improving expressivity and long-range information integration.
\end{enumerate}

\paragraph{Interatomic Potentials with Spectrum Head.}
\begin{enumerate}
    \item \textbf{NequIP (E(3)-equivariant interatomic potential)}~\citep{batzner2022nequip}. NequIP is an E(3)-equivariant message-passing interatomic potential that uses tensor-product message functions over irreducible representations of the rotation group. Originally designed for accurate, data-efficient energy and force prediction, we repurpose it for eDOS prediction by attaching a regression head over a fixed energy grid to the pooled equivariant features.
    \item \textbf{M3GNet (universal materials graph network with three-body terms)}~\citep{chen2022universal}. M3GNet is a universal interatomic potential that augments MEGNet-style two-body message passing with three-body geometric terms, enabling broad coverage of the periodic table for energy/force/stress prediction. We use the same pooled crystal representation with a spectrum-prediction head and train it under the same protocol as the other spectrum baselines.
    \item \textbf{MACE (higher-order equivariant message passing)}~\citep{batatia2022mace}. MACE constructs many-body equivariant features by hierarchically composing tensor products, achieving high accuracy on energy/force benchmarks. We initialize MACE from a pretrained interatomic-potential checkpoint and adapt it to eDOS prediction by pooling its node features into a crystal representation and regressing the spectrum from this pooled embedding.
    \item \textbf{EquiformerV2 (scalable equivariant Transformer)}~\citep{liao2024equiformerv2}. EquiformerV2 is a Transformer-style equivariant network that scales to higher-degree irreducible representations through efficient tensor-product operations and improved attention. In its native setting it is used for energy/force prediction; we adapt it for eDOS prediction by attaching a spectrum-regression head, and it is the strongest structure-encoder/interatomic-potential baseline on the primary eDOS and phDOS prediction benchmarks in our experiments.
\end{enumerate}

\paragraph{Crystal-to-Spectrum Predictors.}
\begin{enumerate}
    \item \textbf{Mat2Spec (probabilistic embedding with contrastive learning for spectra)}~\citep{Kong2022}. Mat2Spec is a materials-to-spectrum framework introduced for predicting spectral properties such as electronic/phonon DOS. It frames spectrum prediction through probabilistic embeddings and uses contrastive learning to learn representations that support accurate spectrum inference; the work demonstrates utility for tasks such as identifying gaps and screening candidates with follow-up ab initio validation. In our comparisons, Mat2Spec serves as a canonical baseline that treats DOS prediction as a dedicated spectrum-learning problem rather than a generic scalar property regression.
    \item \textbf{DOSTransformer (prompt-guided multi-modal Transformer)}~\citep{lee2023density}. DOSTransformer explicitly models the fact that DOS values depend on both the crystal structure and the energy coordinate. It integrates heterogeneous inputs from the material and from energy levels using a multi-modal Transformer with cross-attention/self-attention, and introduces crystal-system prompts to guide learning of interactions that are shared across (and within) crystal systems. This makes it a strong baseline for directly predicting the DOS sequence over energy bins, rather than predicting a single pooled target.
\end{enumerate}

\paragraph{Summary.}
Structure-encoder baselines (E3NN, MEGNet, GATGNN) primarily test whether improved crystal representations and inductive biases (equivariance, set-level readout, attention-based message passing) suffice for accurate eDOS regression after pooling. Interatomic-potential baselines (NequIP, M3GNet, MACE, EquiformerV2) ask whether atomistic representations originally trained for force/energy prediction transfer to spectrum output when re-attached to a fixed-grid regression head. Crystal-to-sequence baselines (Mat2Spec, DOSTransformer) more directly address the functional/sequence nature of eDOS, with architectures or objectives designed around predicting structured spectra across energy. All baselines are trained and evaluated on the same structures, preprocessing pipeline, and data splits as DeepDOSReasoner; differences in reported numbers therefore primarily reflect the choice of representation and decoding pathway rather than differences in data.

\subsection{Evaluation Metrics}\label{app:eval}
We evaluate performance using task-specific metrics tailored to (i) eDOS prediction, 
(ii) valence-band (VB) gap identification, and (iii) element-resolved contribution ranking.

\begin{itemize}
\item \textbf{eDOS prediction} (spectrum regression). 
Let \(\hat{y}(E)\) denote the predicted eDOS and \(y(E)\) the DFT ground truth, discretized on
\(L\) energy bins \(\{E_i\}_{i=1}^L\) with values \(\hat{y}_i\) and \(y_i\); the pointwise metrics below (MAE, MSE, \(R^2\)) are computed on the original (un-normalized) scale. We report
\begin{itemize}
    \item {Mean Absolute Error (MAE):}
    \[
    \mathrm{MAE} = \frac{1}{L} \sum_{i=1}^L |\hat{y}_i - y_i|.
    \]
    \item {Mean Squared Error (MSE):}
    \[
    \mathrm{MSE} = \frac{1}{L} \sum_{i=1}^L (\hat{y}_i - y_i)^2.
    \]
    \item {Coefficient of determination (\(R^2\)):}
    \[
    R^2 = 1 - \frac{\sum_i (\hat{y}_i - y_i)^2}{\sum_i (y_i - \bar{y})^2},
    \]
    where \(\bar{y}\) denotes the mean of the ground-truth spectrum.
    \item {Wasserstein distance (WD):} to capture distributional and shape differences between spectra,
    we first normalize the predicted and reference spectra to unit mass and linearly rescale the spectral axis to the unit interval, \(u=(E-a)/(b-a)\in[0,1]\). With \(F(u)\) and \(F^\star(u)\) denoting the cumulative distributions of the normalized predicted and reference spectra, each ranging in \([0,1]\), we compute the 1-D Wasserstein-1 distance
    \[
    \mathrm{WD} = \int_0^1 |F(u) - F^\star(u)|\, du,
    \]
    implemented on the discrete grid. WD is therefore dimensionless: a value of \(w\) corresponds to optimally transporting the unit spectral mass across a fraction \(w\) of the analysis window. WD is computed on the normalized spectra and the normalized axis so that the two cumulative distributions share unit total mass, whereas MAE, MSE, and \(R^2\) above use the original un-normalized scale.
    \item {Peak-restricted MAE (van Hove singularities near band edges):} to assess fidelity to the sharp critical-point features that dominate near-edge optical processes and effective carrier behavior, we evaluate MAE only at the energy bins corresponding to van Hove singularities adjacent to the VBM and CBM. The peak bins are extracted from the DFT ground-truth spectrum (never from any model prediction), as follows. For each test material we first locate, on the DFT-derived eDOS, the valence-band maximum (VBM) and conduction-band minimum (CBM) that bound the fundamental gap, taken as the highest occupied and lowest unoccupied energies at which the density rises above a small fraction of its surrounding magnitude (for gapless or metallic systems both edges coincide with the Fermi level at $E=0$). We then place a fixed-width energy window immediately adjacent to each edge, extending into the valence band below the VBM and into the conduction band above the CBM, and within each window detect the local maxima of the DFT eDOS by standard discrete peak finding. The dominant peak on each side, i.e., the van Hove singularity nearest that band edge, contributes its bin index to the peak-index set \(\mathcal{P}\). The same DFT-derived \(\mathcal{P}\) is then applied to every method, so that differences in \(\mathrm{MAE}_{\mathrm{peak}}\) reflect only how faithfully each predicted spectrum reproduces the amplitude and location of these band-edge singularities. The peak-restricted MAE is then
    \[
    \mathrm{MAE}_{\mathrm{peak}} \;=\; \frac{1}{|\mathcal{P}|}\sum_{i\in\mathcal{P}}|\hat{y}_i - y_i|,
    \]
    averaged across the test set. Because \(\mathcal{P}\) excludes the low-DOS regions where pointwise errors are small almost by construction, \(\mathrm{MAE}_{\mathrm{peak}}\) is a substantially more stringent test of peak placement and amplitude than the unrestricted MAE above.
\end{itemize}

\item \textbf{VB-gap identification} (per-material binary classification).
To evaluate whether a model correctly identifies materials that exhibit a valence-band gap, we cast the
task as a per-material binary classification problem: a material is labeled positive if its eDOS satisfies
the four-criterion VB-gap rule of the main text (density below $0.9$~state/eV across a contiguous window
at least $1$~eV wide whose high-energy edge lies in $[-1.5,-0.25]$~eV) and negative otherwise. Each model's
predicted labels, obtained by applying the same rule to its predicted spectrum, are compared against the
DFT-derived ground-truth labels. We report
\begin{itemize}
    \item {Precision:} \(\mathrm{Precision} = \frac{\mathrm{TP}}{\mathrm{TP} + \mathrm{FP}}\).
    \item {Recall:} \(\mathrm{Recall} = \frac{\mathrm{TP}}{\mathrm{TP} + \mathrm{FN}}\).
    \item {F1 score:}
    \[
    \mathrm{F1} = \frac{2 \cdot \mathrm{Precision} \cdot \mathrm{Recall}}
    {\mathrm{Precision} + \mathrm{Recall}}.
    \]
    \item {Matthews correlation coefficient (MCC):}
    \[
    \mathrm{MCC} =
    \frac{\mathrm{TP}\cdot\mathrm{TN} - \mathrm{FP}\cdot\mathrm{FN}}
    {\sqrt{(\mathrm{TP}+\mathrm{FP})(\mathrm{TP}+\mathrm{FN})
    (\mathrm{TN}+\mathrm{FP})(\mathrm{TN}+\mathrm{FN})}},
    \]
    which is robust to the strong class imbalance of this task, where positive (VB-gap) materials are only ${\sim}5\%$ of the set.
\end{itemize}

\item \textbf{Element-Resolved Contribution Ranking.}
For evaluating element-resolved (pDOS-style) contributions, we assess both discrete identification
and ranking consistency of element-level contributors, obtained by averaging the source-attention
weights over atoms of the same element within each crystal. We report
\begin{itemize}
    \item {Overall accuracy:} agreement between the full predicted and ground-truth element rankings, computed as the rank-biased overlap (RBO)~\citep{webber2010similarity}, a top-weighted ranking-similarity measure that ranges from $0$ (entirely discordant orderings) to $1$ (identical orderings) and weights agreement among the highest-ranked contributors most heavily.
    \item {Top-1 accuracy:} whether the highest-ranked predicted element matches the
    ground-truth top contributor.
    \item {Spearman's rank correlation (\(\rho\))~\citep{spearman1904proof}:}
    \[
    \rho = 1 - \frac{6 \sum_i (r_i - r_i^\star)^2}{n(n^2 - 1)},
    \]
    where \(r_i\) and \(r_i^\star\) denote the predicted and ground-truth ranks of element \(i\)
    and \(n\) is the number of distinct elements in the crystal,
    measuring how well the relative ordering of contributions is preserved. The rank correlation is evaluated on crystals with \(n\ge 2\) distinct elements (unary compositions, whose ranking is trivial, are excluded), and tied contribution scores are assigned their average rank; with ties handled this way, \(\rho\) is computed as the Pearson correlation between the predicted and ground-truth rank vectors, to which the closed form above reduces when no ties are present.
\end{itemize}
\end{itemize}

\subsection{Computational Cost}\label{app:comp_cost}
We assess the computational efficiency of our proposed framework on a server equipped with a single NVIDIA H100 GPU (80GB) and Intel Xeon Platinum 8462Y+. The implementation is based on PyTorch and PyTorch Geometric libraries. \Cref{tab:comp_cost} summarizes the computational budget for both Structure-Aware Spectrum Decoding and Mass-Conserving Iterative Refinement.

\begin{table}[ht]
\centering
\caption{Computational cost and hardware specifications, reported for both the primary (full) eDOS benchmark and the stable-only subset. The per-crystal encoding and decoding times are single-sample module latencies (forward and backward pass, one crystal at a time), reported to show how the budget splits between the encoder and the decoder. The per-epoch wall-clock times are measured under the training configuration of \Cref{tab:hyperparams} (batch size $128$), where crystals are processed in parallel on the GPU. The two sets of figures are therefore measured under different batching and do not multiply: the per-epoch time is an order of magnitude below the per-crystal latency times the number of training structures. The single-crystal inference latency quoted in the main text (${\sim}10.5$~ms, forward pass only) is likewise lower than the per-crystal training figures here.}
\label{tab:comp_cost}
\begin{small}
    \begin{sc}
        \begin{tabular}{p{4.5cm}cc}
        \toprule
        & \multicolumn{2}{c}{Protocol} \\
        \cmidrule(lr){2-3}
        Metric & Full set & Stable-only \\
        \midrule
        \multicolumn{3}{l}{\textbf{Hardware Specifications}} \\
        \midrule
        GPU Specification & \multicolumn{2}{c}{1 $\times$ NVIDIA H100 (80GB)} \\
        CPU Specification & \multicolumn{2}{c}{Intel Xeon Platinum 8462Y+} \\
        \midrule
        \multicolumn{3}{l}{\textbf{Stage~I: Encoder + Decoder Training}} \\
        \midrule
        Encoding Time / Crystal (ms) & 4.39 & 4.75 \\
        Decoding Time / Crystal (ms) & 8.64 & 10.93 \\
        Train Time / Epoch (min) & 1.43 & 0.82 \\  
        \midrule
        \multicolumn{3}{l}{\textbf{Stage~II: MCIR Training}} \\
        \midrule
        Train Time / Epoch (min) & 3.72 & 1.43 \\
        \bottomrule
        \end{tabular}
    \end{sc}
\end{small}
\end{table}

%% file: appendix/results.tex
\begin{figure}[ht]
    \centering
    \includegraphics[width=1\linewidth]{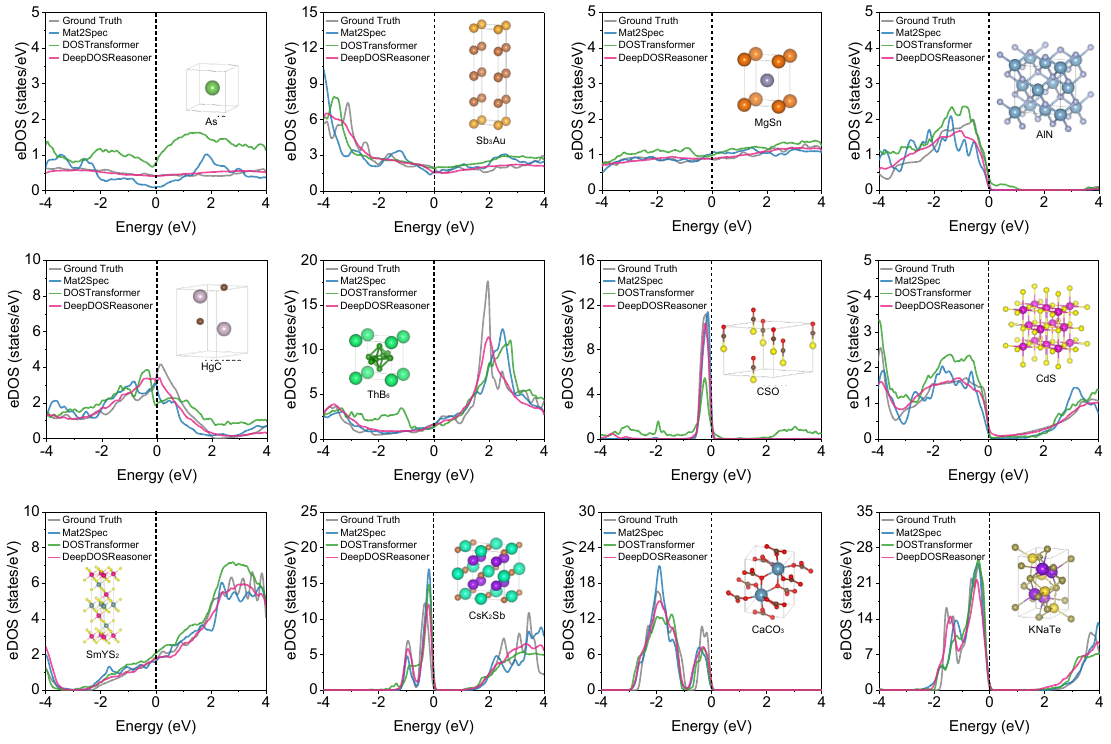}
    \caption{Additional eDOS visualization examples extending main-text Figure~2(b). Predicted DeepDOSReasoner spectra (pink) are shown alongside DFT ground truth (grey) and the two crystal-to-sequence baselines (Mat2Spec, DOSTransformer) for twelve representative materials drawn from the eDOS test set, with the corresponding crystal structure inset in each panel.}\label{fig:viz_edos_more}
\end{figure}

\begin{figure}[ht]
    \centering
    \includegraphics[width=1\linewidth]{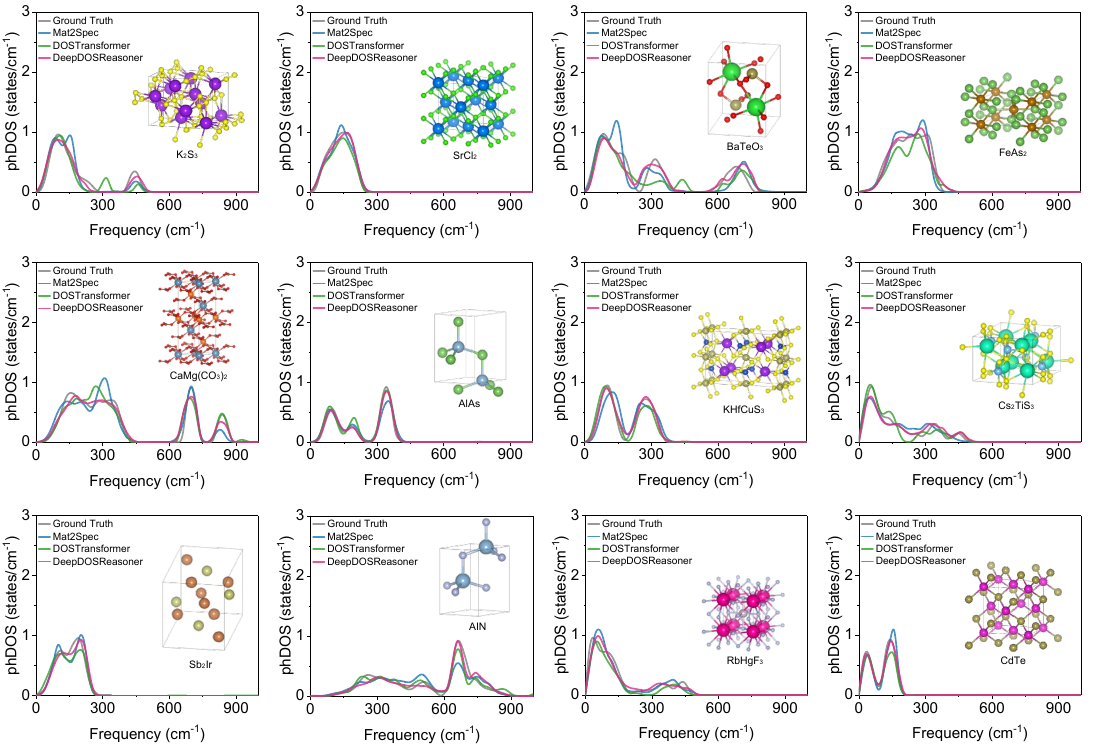}
    \caption{Additional phDOS visualization examples extending main-text Figure~2(c). Predicted DeepDOSReasoner phonon spectra (pink) are shown alongside DFT ground truth (grey) and the two crystal-to-sequence baselines (Mat2Spec, DOSTransformer) for twelve representative materials drawn from the held-out phDOS test set ($n=152$), with the corresponding crystal structure inset in each panel. Frequencies are reported on the original cm$^{-1}$ axis.}\label{fig:viz_phdos_more}
\end{figure}

\section{Additional Experimental Results}
\label{app:res}

\subsection{Element-Resolved Contributions of eDOS}\label{app:ele_con}
To probe DeepDOSReasoner’s ability to recover \emph{element-resolved} contributions from source attention, we average the source-attention weights over atoms of the same element within each crystal to obtain a per-element contribution score and rank the elements accordingly. We evaluate this ranking against the DFT-derived element-resolved contributions to the valence-band eDOS in the energy window $[-0.5,0]$~eV, reporting overall ranking accuracy, top-1 accuracy, and Spearman $\rho$ under two protocols (the full set and the stable-only subset); the full-set comparison is summarized in main-text Figure~5(c). Because the model is trained only on the total eDOS and never sees element-resolved labels, any agreement with the DFT decomposition is an emergent property of structure-aware decoding rather than a fitted attribution.

Across both protocols, {DeepDOSReasoner} consistently outperforms a random-ordering baseline. Under the full-set protocol, DeepDOSReasoner attains Acc.\ $=0.4026$ and Spearman $\rho=0.4283$, well above the corresponding random values (Acc.\ $=0.2215$, $\rho=0.0025$), indicating that the learned attention weights are aligned with ground-truth element-resolved contributions rather than producing chance orderings. Similar gains are observed under the stable-only protocol, where DeepDOSReasoner reaches Acc.\ $=0.4167$ and $\rho=0.5159$ versus the random baseline's Acc.\ $=0.2208$ and $\rho=0.0251$, reflecting faithful recovery of relative atomic importance.

Notably, the Top-1 accuracy under the stable-only protocol (0.7042 vs.\ 0.3875 for random ordering) suggests that the model reliably identifies the dominant atomic contributor to the valence-band eDOS. Since the model is never supervised with element-resolved labels, these results indicate that hierarchical source attention induces a meaningful internal decomposition of the total eDOS, implicitly capturing atom-level contributions as an emergent property of structure-aware decoding. This provides evidence that DeepDOSReasoner not only predicts accurate spectra, but also learns physically interpretable structure--spectrum relationships beyond its training objective.

\paragraph{Composition--Oxidation-State Chemistry Heuristic.}
As a stronger, trained-model-free reference than random ordering, we also evaluate a chemistry heuristic that ranks element contributions near $E_F$ from composition and a guessed oxidation state alone, with no learned parameters and no structural input. The heuristic encodes the established frontier-orbital picture of bonding in extended structures, in which the states near the Fermi level are often dominated by the energetically frontier valence shells of the constituent ions---most prominently the partially filled $d$/$f$ bands of open-shell transition metals and lanthanides---in the spirit of the Hammer--N{\o}rskov $d$-band framework~\citep{hammer1995why,hammer2000theoretical} and of the frontier-orbital analysis of solids~\citep{hoffmann1988solids}. Concretely, for each material we proceed in five steps:
\begin{enumerate}
    \item \emph{Composition.} The element set and per-element stoichiometric counts are read from the chemical formula alone.
    \item \emph{Oxidation-state assignment.} Each element is assigned a formal oxidation state using \texttt{pymatgen}'s composition-based guesser (\texttt{Composition.oxi\_state\_guesses()})~\citep{Ong2013}, which ranks charge-balanced assignments by oxidation-state statistics derived from the Inorganic Crystal Structure Database, rooted in bond-valence analysis~\citep{brown1985bond}; if no charge-balanced assignment is found, every element falls back to the neutral state (oxidation state $0$). Mixed-valence compositions yield fractional average oxidation states, which we round to the nearest integer before the configuration lookup.
    \item \emph{Valence configuration.} The resulting ion's valence configuration, i.e., the subshells outside its noble-gas core, is read off the ion's electronic structure as tabulated by \texttt{pymatgen}~\citep{Ong2013} (e.g., Fe$^{3+}\!\to 3d^5$, Er$^{3+}\!\to 4f^{11}$);\footnote{We use \texttt{pymatgen} $\geq$ 2024.6, in which \texttt{Species.electronic\_structure} accounts for the oxidation state.} for the rare ions whose configuration is not tabulated, we fall back to the neutral atom's configuration. This configuration serves only as a formal electron-counting proxy and is not intended to represent the actual orbital occupations in the solid.
    \item \emph{Frontier score.} The ion is scored by the frontier character of its valence subshells. Each orbital type carries a base weight reflecting how strongly it typically contributes to the density of states near $E_F$ ($w_d=w_f=1.0$, $w_p=0.5$, $w_s=0.15$). A partially filled subshell of occupancy $\mathrm{occ}$ and capacity $\mathrm{cap}$ contributes $w\bigl(1+\min(\mathrm{occ},\,\mathrm{cap}-\mathrm{occ})\bigr)$, peaking near half-filling, whereas a completely filled subshell contributes only a small residual $0.5\,w$; the ion's frontier score sums these contributions over its valence subshells. The weights were fixed a priori from qualitative chemical considerations and were not fit or selected using the test labels. The net effect matches chemical intuition: partially filled $d$/$f$ shells (open-shell transition metals and lanthanides) score high, filled $d^{10}$ ions (e.g., Cd$^{2+}$, Tl$^{3+}$) receive a small residual, partially filled $p$ shells score moderately, and closed-shell cations (e.g., Li$^{+}$, Mg$^{2+}$) and closed-shell anions (e.g., O$^{2-}$, $2p^6$) receive only small residual scores.
    \item \emph{Ranking.} Elements are ranked by their frontier scores, restricted to the element set of the DFT reference decomposition so that the predicted and ground-truth rankings cover identical items. Ties in the frontier score---common among closed-shell species---are broken uniformly at random, and we report metrics averaged over 100 random tie-breakings. The resulting ranking is evaluated with the same metrics as the source-attention rankings (\Cref{app:eval}).
\end{enumerate}
Under this protocol, the heuristic attains overall accuracy $0.325$, top-1 accuracy $0.442$, and Spearman $\rho=0.222$ on the full eDOS test set. This sits well above random ordering ($0.222$ / $0.361$ / $0.003$)---confirming that frontier-orbital chemistry alone already captures part of the near-$E_F$ element ordering---but well below DeepDOSReasoner ($0.403$ / $0.624$ / $0.428$), indicating that the source-attention decomposition learns structure-dependent contributions beyond what a composition-only chemical rule provides. The three methods are compared in main-text Figure~5(c).

\subsection{Visualization of Predicted Spectra}\label{app:viz}
This section provides qualitative evidence that complements the aggregate metrics in the main text. \Cref{fig:viz_edos_more} extends the eDOS examples of main-text Figure~2(b) to a larger panel of twelve representative materials, showing DeepDOSReasoner predictions alongside DFT ground truth and the two crystal-to-sequence baselines (Mat2Spec, DOSTransformer). \Cref{fig:viz_phdos_more} provides the analogous panel for phDOS, extending main-text Figure~2(c) across twelve materials drawn from the held-out phDOS test set. Across these visualizations, DeepDOSReasoner consistently better reproduces two physically salient aspects of the spectrum: (i) the band edges/band gap region for eDOS (and the acoustic onset/optical peak structure for phDOS) and (ii) sharp van Hove singularities.

First, in materials with a pronounced gap or steep band onsets, DeepDOSReasoner more faithfully captures the location and sharpness of the valence- and conduction-band edges, yielding a cleaner low-DOS region between them; visually, this appears as a closer match to the ground-truth curve near the onset/offset points, with less spurious ``leakage'' of density into the gap compared with competing methods that either smear the edges or introduce artificial shoulders in the forbidden region. Second, our curves better preserve the positions and relative magnitudes of prominent peaks, the van Hove singularities that encode critical-point structure in the band/phonon dispersion. In several panels, baseline models either broaden these peaks (over-smoothing), shift them along the energy/frequency axis (misalignment), or create extra oscillatory peaks (overfitting artifacts). In contrast, our predictions more often place the dominant singularities at the correct energies/frequencies and maintain their sharp, high-contrast profiles, leading to a closer qualitative match to the DFT lineshape in peak-dominated regions.

\begin{figure}[ht]
    \centering
    \includegraphics[width=1\linewidth]{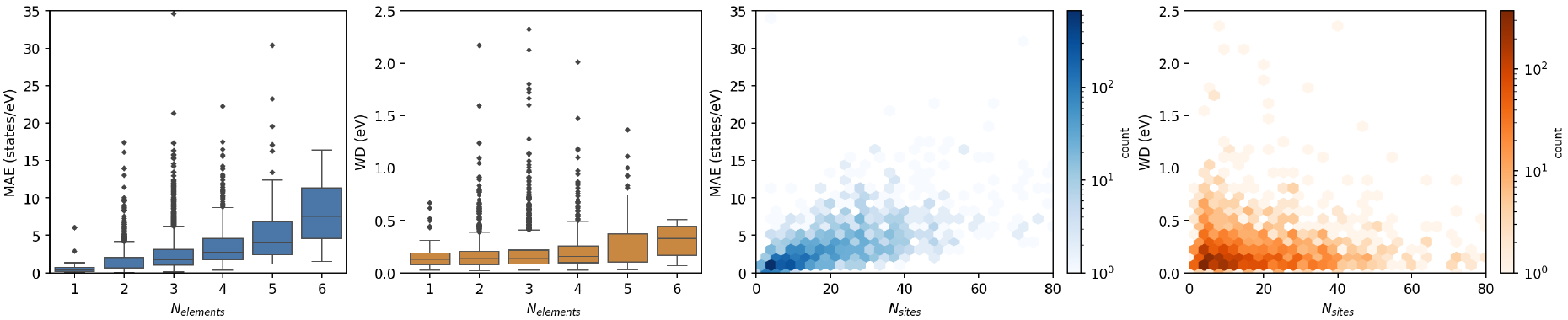}
    \caption{DeepDOSReasoner prediction error versus material complexity on the eDOS test set ($n=3{,}756$). The first two columns are boxplots of MAE and WD as a function of the number of distinct elements ($N_{\mathrm{elements}}$); the last two columns are hexbin density plots of MAE and WD against the number of atomic sites in the unit cell ($N_{\mathrm{sites}}$), with the colorbar reporting per-bin count on a log scale.}\label{fig:robust_complexity}
\end{figure}

\begin{figure}[ht]
    \centering
    \includegraphics[width=1\linewidth]{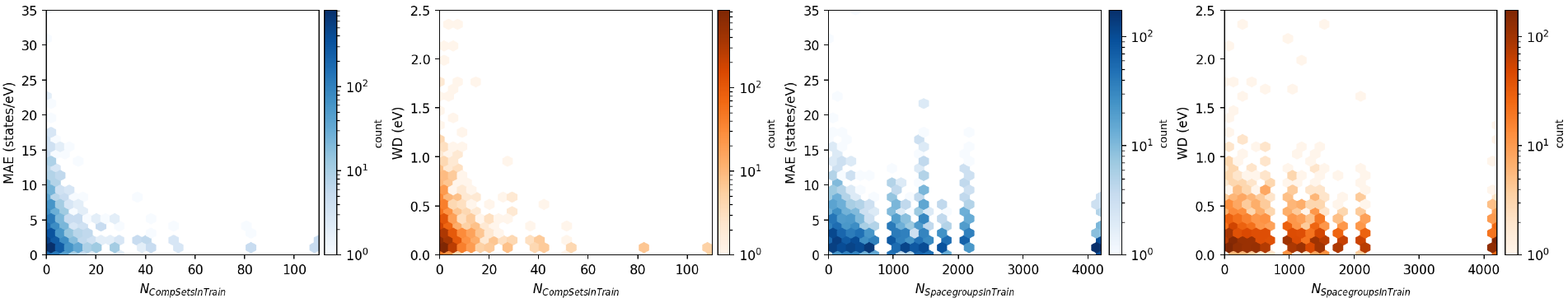}
    \caption{DeepDOSReasoner prediction error versus training-set similarity on the eDOS test set. The first two columns are hexbin density plots of MAE and WD against the number of training compositions sharing the test material's chemical system ($N_{\mathrm{CompSetsInTrain}}$); the last two columns are the analogous plots against the number of training entries sharing the test material's space group ($N_{\mathrm{SpacegroupsInTrain}}$). Lower-similarity bins retain a comparable error distribution, indicating that the model does not concentrate its accuracy on training-set neighborhoods.}\label{fig:robust_similarity}
\end{figure}

\begin{figure}[ht]
    \centering
    \includegraphics[width=1\linewidth]{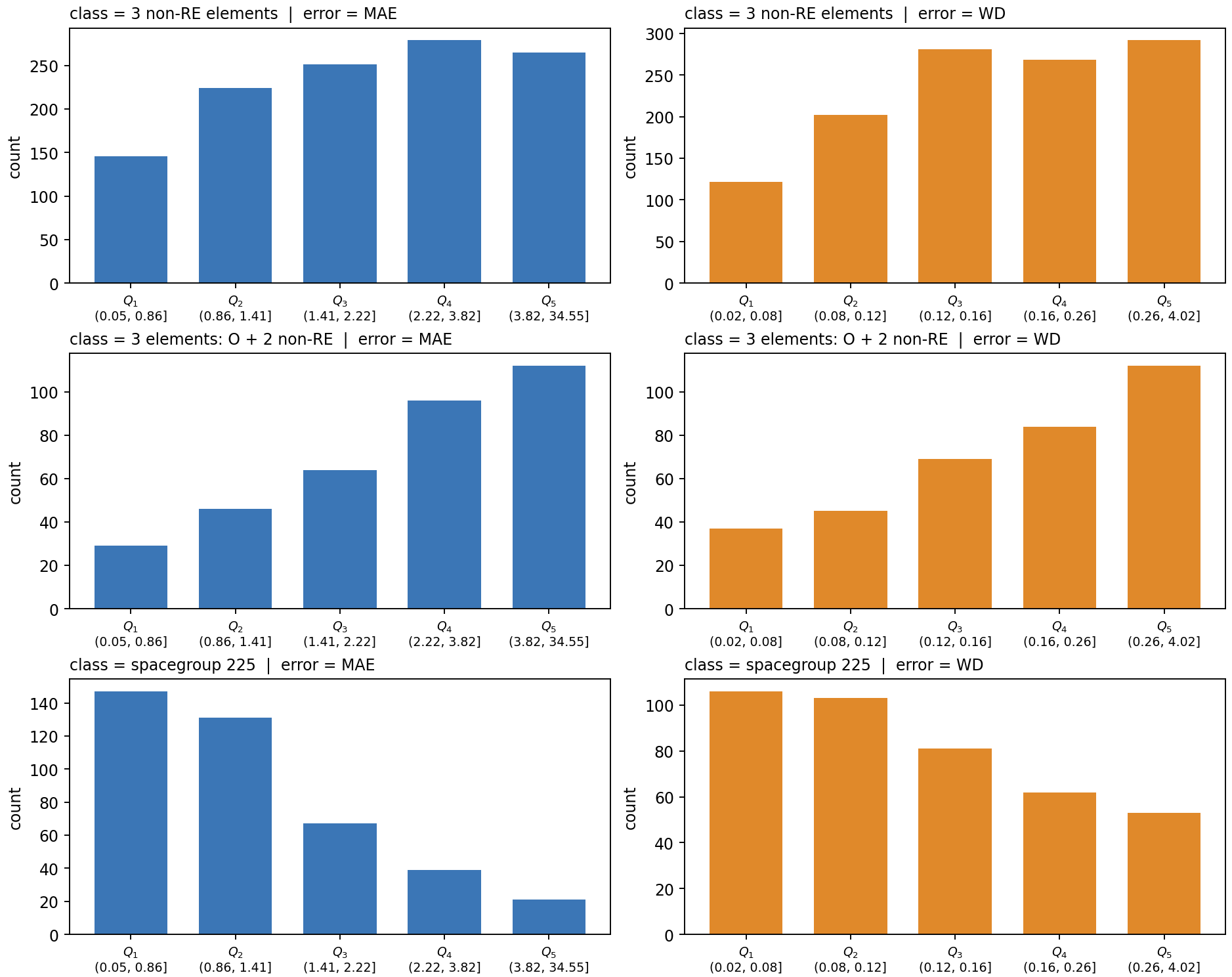}
    \caption{Distribution of DeepDOSReasoner test-set members across error quintiles for three material subclasses. Rows correspond to ternary compounds with three non-rare-earth elements (top), ternary compounds of the form O~+~two non-rare-earth elements (middle), and cubic structures with space group~$225$ (bottom). Columns show MAE quintiles (left, blue) and WD quintiles (right, orange); quintile edges are computed on the full eDOS test set and shown below each axis. A subclass that concentrates in $Q_1$--$Q_2$ is systematically easier than the test-set average.}\label{fig:robust_subclass}
\end{figure}

\begin{figure}[ht]
    \centering
    \includegraphics[width=1\linewidth]{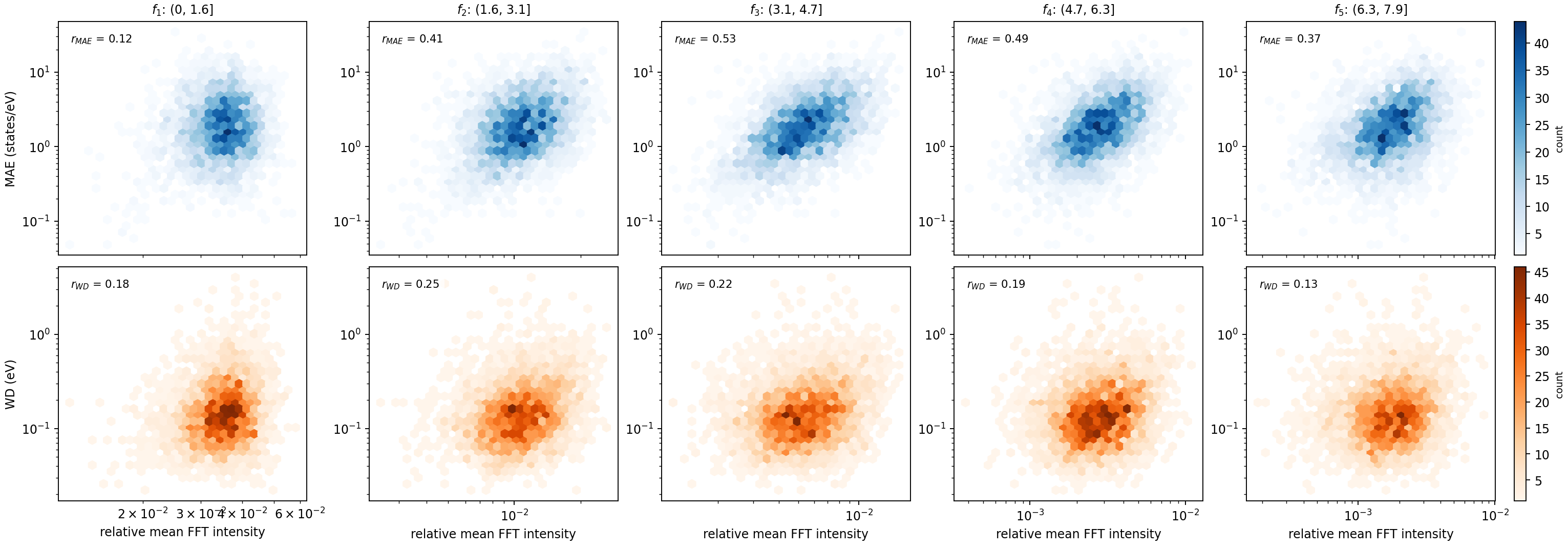}
    \caption{DeepDOSReasoner prediction error versus the relative Fourier composition of the DFT-derived eDOS. The frequency axis of the discrete Fourier transform of each target spectrum (on the 128-bin energy grid) is partitioned into five equal bands $f_1,\ldots,f_5$ from low frequency (left) to high frequency (right). For each band, we plot per-material MAE (top row, blue) and WD (bottom row, orange) against the relative mean FFT intensity in that band; the Pearson correlation between the relative-intensity proxy and the error metric ($r_{\mathrm{MAE}}$ for the top row, $r_{\mathrm{WD}}$ for the bottom row) is reported in each panel. The intermediate-frequency bands show the strongest positive correlation with MAE, the smooth lowest-frequency band the weakest, and WD correlations remain low in every band.}\label{fig:robust_fft}
\end{figure}

\subsection{Robustness and Error Decomposition}\label{app:robust}
To probe where DeepDOSReasoner's residual error concentrates and to test whether the model's accuracy is uniform across materials or driven by a particular subset of the test set, we report four complementary loss-decomposition analyses on the same $n=3{,}756$ eDOS test materials used in the eDOS evaluation of the main text. \Cref{fig:robust_complexity} examines the dependence on \emph{material complexity}, \Cref{fig:robust_similarity} on \emph{training-set similarity}, \Cref{fig:robust_subclass} on \emph{material subclass}, and \Cref{fig:robust_fft} on the \emph{Fourier composition of the eDOS}. Taken together, the four panels show that DeepDOSReasoner's residual error is not concentrated on a single regime: errors grow gently with material complexity, depend weakly on training-set similarity once a modest amount of in-distribution context is present, distribute broadly across material subclasses, and correlate most strongly with the relative weight of intermediate-frequency spectral structure rather than with the smooth spectral envelope.

\Cref{fig:robust_complexity} examines how DeepDOSReasoner's prediction error scales with material complexity, measured by the number of distinct elements ($N_{\mathrm{elements}}$, left two columns) and the number of atomic sites in the unit cell ($N_{\mathrm{sites}}$, right two columns). Both MAE and WD trend mildly upward with $N_{\mathrm{elements}}$ from $N_{\mathrm{elements}}=1$ to $N_{\mathrm{elements}}=6$, consistent with the larger combinatorial space of element-projected contributions that the decoder must resolve in compositionally complex materials, but the absolute spread within each $N_{\mathrm{elements}}$ bin is comparable to the across-bin trend, so material complexity is not the dominant driver of error. The hexbin plots against $N_{\mathrm{sites}}$ show a similar weak dependence and confirm that no narrow sites-count window concentrates the error.

\Cref{fig:robust_similarity} examines whether DeepDOSReasoner's prediction error depends on how similar a test material is to the training set, measured by the number of training compositions sharing its chemical system ($N_{\mathrm{CompSetsInTrain}}$) and the number of training entries sharing its space group ($N_{\mathrm{SpacegroupsInTrain}}$). MAE and WD decrease modestly as either similarity measure grows, but a substantial fraction of low-similarity test materials still attain low error, indicating that the model is not memorizing training-set neighborhoods.

\Cref{fig:robust_subclass} examines how prediction error distributes across three test-set material subclasses: ternary compounds composed of three non-rare-earth elements (top row), ternary compounds of the form O~+~two non-rare-earth elements (middle row), and cubic structures with space group~$225$ ($Fm\bar{3}m$, bottom row). For each subclass we partition the eDOS test set into quintiles of the corresponding error metric (MAE or WD), and report the count of subclass members in each quintile. A subclass whose count is uniform across quintiles contributes proportionally to the error distribution; a count that skews toward low quintiles ($Q_1$, $Q_2$) indicates that the subclass is systematically easier than average. DeepDOSReasoner shows the clearest such skew for the space-group-$225$ subclass, where over $60\%$ of subclass members fall in the two lowest MAE quintiles (and roughly half in the two lowest WD quintiles), consistent with the higher symmetry of cubic structures simplifying both the atomic-context inputs to the decoder and the resulting eDOS geometry. The two ternary subclasses are distributed more broadly across quintiles, with a modest concentration in the higher MAE quintiles for the O~+~two non-rare-earth case, indicating that oxygen-containing ternaries are slightly harder than the test-set average.

\Cref{fig:robust_fft} decomposes prediction error against the spectral composition of the target eDOS. For each test material we compute the discrete Fourier transform of the DFT-derived spectrum on the 128-bin grid, partition the resulting frequency axis into five equal bands $f_1,\ldots,f_5$, and report, for each band, the relative mean FFT intensity of that band (x-axis) against the per-material MAE (top row) or WD (bottom row). The Pearson correlation between the relative-intensity proxy and the error metric is reported in each panel ($r_{\mathrm{MAE}}$ and $r_{\mathrm{WD}}$). MAE correlates most strongly with the relative weight of the intermediate-frequency bands ($f_3$, $r_{\mathrm{MAE}}{=}0.53$; $f_4$, $r_{\mathrm{MAE}}{=}0.49$), more weakly with the sharpest band ($f_5$, $r_{\mathrm{MAE}}{=}0.37$), and only marginally with the smooth lowest-frequency band ($f_1$, $r_{\mathrm{MAE}}{=}0.12$): spectra dominated by a smooth envelope are comparatively easy pointwise, whereas spectra carrying substantial oscillatory peak-and-edge structure at intermediate frequencies are the hardest, exactly the band-edge and critical-point regime that the mass-conserving reasoning stage (see Methods of the main text) is designed to address. WD correlations are uniformly lower across bands ($\leq 0.25$), consistent with WD being sensitive to global mass placement rather than to amplitudes in any single frequency band.

\subsection{Per-Method Effects of Metastable Supervision}\label{app:metastable}
The main-text stable-only control restricts both training and test to the $14{,}055$ thermodynamically stable (on-hull) Materials Project entries, holding preprocessing, architecture, training, and evaluation fixed so that any change relative to the full set is attributable to the removal of metastable structures. This subsection reports the full per-method results and visualizes the same numerical comparison from two complementary angles.

\Cref{tab:edos_stable} reports per-method performance under the stable-only protocol. DeepDOSReasoner attains the best result on all four metrics ($R^2 = 0.674$, MAE $= 2.809$, MSE $= 41.476$, WD $= 0.197$); relative to the strongest stable-only baseline per metric, this is a $6.3\%$ $R^2$ improvement over EquiformerV2 ($0.634$), a $4.0\%$ MAE reduction (vs.\ EquiformerV2, $2.925$), a $10.7\%$ MSE reduction (vs.\ EquiformerV2, $46.445$), and a $10.0\%$ WD reduction (vs.\ MEGNet, $0.219$). The end-to-end ranking is largely preserved relative to the full set: DeepDOSReasoner first, EquiformerV2 the strongest structure-encoder/interatomic-potential baseline, M3GNet and MEGNet next, then DOSTransformer and Mat2Spec, E3NN/GATGNN/NequIP just behind, and MACE last, so the architectural advantages are not artifacts of training-set composition.

A per-method comparison against the full-set results (main-text Table~1) reveals an asymmetry across metrics. On $R^2$, MAE, and MSE --- the metrics tied to absolute spectral fidelity --- \emph{every} method degrades under stable-only training: $R^2$ drops by $0.6$--$6.4\%$ (MACE worst at $-6.4\%$, M3GNet most robust at $-0.6\%$, EquiformerV2 $-1.7\%$, DOSTransformer $-4.0\%$, DeepDOSReasoner $-2.5\%$); MAE rises by $8.5$--$14.3\%$ ($+9.4\%$ for DeepDOSReasoner, $+12.3\%$ for EquiformerV2, $+14.3\%$ for MACE); and MSE rises by $4.6$--$27.4\%$ (DeepDOSReasoner $+8.7\%$, DOSTransformer $+10.9\%$, MACE collapsing by $+27.4\%$). The unanimous direction of degradation across structurally diverse architectures rules out the interpretation that metastable entries inject method-specific noise; the metastable subset instead behaves as additional supervision for the structure-to-spectrum map, and the cost of removing it is largest for the architectures whose representations rely most on data abundance. Wasserstein distance behaves differently: five baselines improve under stable-only --- MEGNet ($0.251\to0.219$, $-12.7\%$), M3GNet ($0.250\to0.225$, $-10.0\%$), Mat2Spec ($0.234\to0.225$, $-3.8\%$), NequIP ($0.271\to0.263$, $-3.0\%$), and EquiformerV2 ($0.230\to0.226$, $-1.7\%$) --- while the other five worsen, consistent with stable-only test spectra being intrinsically simpler in global geometry; even within this mixed regime DeepDOSReasoner remains best on WD ($0.197$, a $10.0\%$ margin over MEGNet). Two patterns pull this together. MACE shows the largest absolute degradation on every metric (MSE $+27.4\%$, WD $+30.6\%$), consistent with its interatomic-potential pretraining transferring to spectrum regression only with abundant, chemically diverse data; and DeepDOSReasoner's MAE advantage over the second-best method \emph{grows} between protocols, from $0.038$ on the full set ($1.5\%$ relative) to $0.116$ on stable-only ($4.0\%$ relative) --- the only metric whose inter-method gap widens when metastable supervision is removed --- indicating that source-attentive decoding and mass-conserving refinement together preserve per-bin accuracy under data scarcity better than any pooled-embedding baseline. The element-resolved evaluation under the stable-only protocol (\Cref{app:ele_con}) shows that the emergent atom-to-energy decomposition also transfers to this regime, with rank-correlation gains over random ordering at least as large as in the full-set evaluation.

\begin{figure}[ht]
    \centering
    \includegraphics[width=0.95\linewidth]{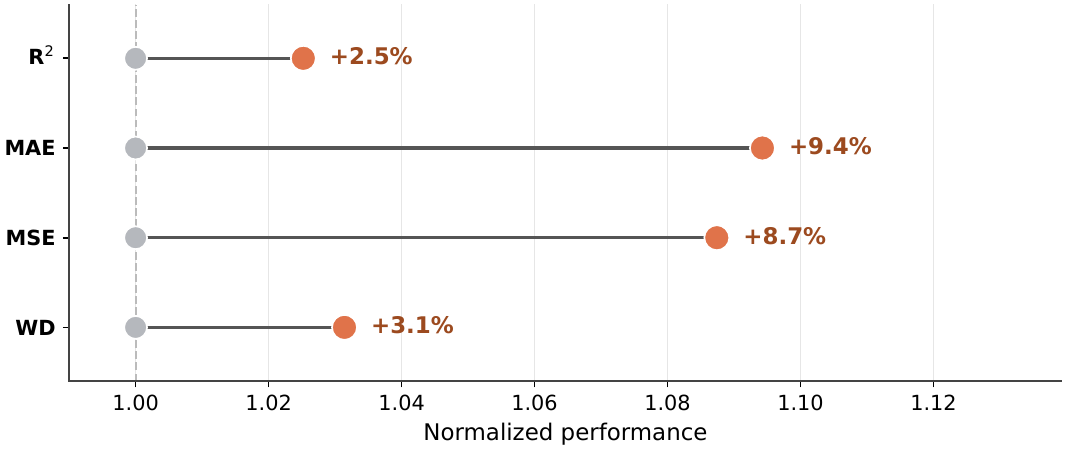}
    \caption{DeepDOSReasoner gains from training and evaluating with metastable structures, expressed as the relative improvement of full-set performance (orange) over the stable-only baseline (grey, normalized to~$1$) on each of the four eDOS metrics ($R^2$, MAE, MSE, WD; higher-is-better convention applied to each). MAE and MSE improve the most, consistent with the metastable subset acting as additional per-bin supervision.}\label{fig:metastable_boost}
\end{figure}

\Cref{fig:metastable_boost} isolates the effect of metastable training data on DeepDOSReasoner alone, showing the four metrics normalized to the stable-only protocol ($=1$): training and evaluating on the full set improves $R^2$ by $2.5\%$, MAE by $9.4\%$, MSE by $8.7\%$, and WD by $3.1\%$. The MAE and MSE gains are the largest, consistent with metastable structures providing a richer per-bin supervision signal that disproportionately benefits absolute spectral fidelity.

\begin{figure}[ht]
    \centering
    \includegraphics[width=0.95\linewidth]{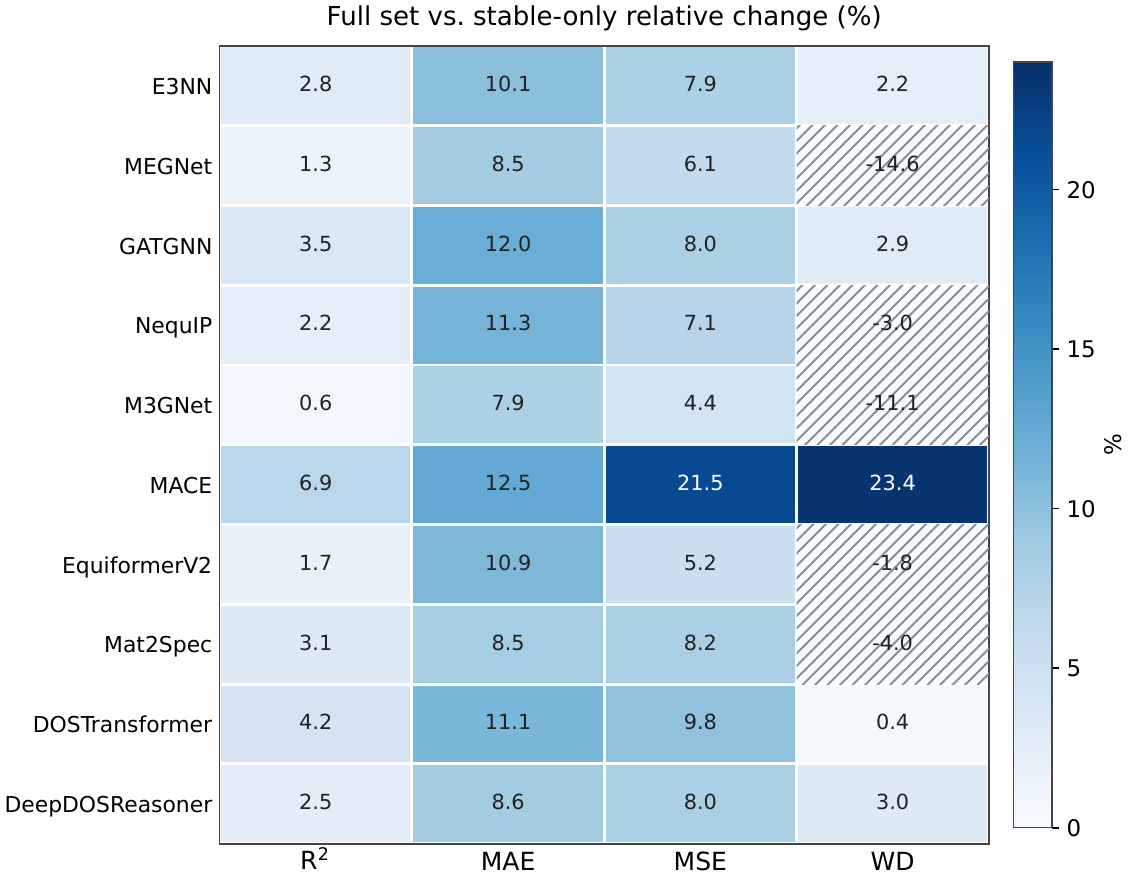}
    \caption{Per-method change in eDOS performance between the full-set and stable-only protocols. Each cell reports the relative advantage of the full-set protocol on the indicated metric, expressed as a percentage of the stable-only value: positive cells indicate that the method degrades when metastable training structures are removed, and hatched negative cells indicate that it improves under stable-only. The heatmap reproduces the numerical comparisons of \Cref{app:metastable} and complements the all-method ranking reported in \Cref{tab:edos_stable}.}\label{fig:metastable_heatmap}
\end{figure}

\Cref{fig:metastable_heatmap} extends the comparison to all ten methods evaluated in the main text. Each cell reports the relative advantage of the full-set protocol over the stable-only protocol on the indicated metric, expressed as a percentage of the stable-only value: positive cells mean the method performs better with metastable training data (i.e., degrades under stable-only), and hatched negative cells mean it improves under stable-only. Two patterns visible in the main text are reproduced here: (i)~all methods degrade on $R^2$/MAE/MSE under stable-only, with MACE degrading the most ($21.5\%$ on MSE and $23.4\%$ on WD in this normalization, corresponding to the $+27.4\%$ and $+30.6\%$ increases relative to the full-set values quoted in the main text); and (ii)~five baselines (M3GNet, MEGNet, Mat2Spec, EquiformerV2, NequIP) actually improve on WD under stable-only (hatched cells), while DeepDOSReasoner and four others worsen. The heatmap layout makes it visually immediate that DeepDOSReasoner is among the methods whose stable-only WD performance shifts in the more challenging direction, despite remaining the best WD performer in absolute terms (\Cref{tab:edos_stable}).

\subsection{Ablation Study}\label{app:abl}
\Cref{tab:abl} reports the full numerical results of the ablation study summarized in the main-text empirical-analyses section (and visualized in main-text Figure~5(b)). The full model (DDR) is compared to two ablated variants: DDR-NR removes the constraint-preserving physical reasoning stage and uses only the Stage~I structure-aware decoder, while DDR-NDNR additionally replaces the source-attention decoder with a global-embedding regression head. Both architectural components contribute across the four eDOS metrics, with the constraint-preserving refinement the larger contributor on MAE, MSE, and WD (most pronounced on WD) and the two components contributing about equally on $R^2$.

In detail, removing constraint-preserving physical reasoning (DDR-NR) degrades all four metrics: $R^2$ falls from $0.691$ to $0.668$ ($-3.3\%$), MAE rises from $2.567$ to $2.701$ ($+5.2\%$), MSE from $38.142$ to $40.668$ ($+6.6\%$), and WD from $0.191$ to $0.221$ ($+15.7\%$). The disproportionate WD degradation is consistent with the role of mass-conserving refinement: WD is most sensitive to misplaced spectral mass, and removing the continuity-equation refinement leaves the residual peak-misalignment and gap-leakage errors that Stage~II is designed to transport away. Further removing source-attentive decoding (DDR-NDNR) degrades every metric again, with $R^2$ falling to $0.645$, MAE to $2.710$, MSE to $42.415$ ($+11.2\%$ relative to the full model), and WD to $0.243$ ($+27.2\%$ relative to the full model). Comparing the two gaps, mass-conserving refinement (DDR\,$\to$\,DDR-NR) is the larger contributor on MAE, MSE, and WD --- accounting for essentially all of the MAE degradation and the majority of the MSE and WD degradation, with its largest single relative effect on WD ($+15.7\%$) --- while source-attentive decoding (DDR-NR\,$\to$\,DDR-NDNR) accounts for about half of the $R^2$ degradation and a substantial share of MSE and WD while leaving MAE almost unchanged, consistent with its role in supplying the structure-to-spectrum signal that the refinement stage then sharpens. Notably, even the most ablated variant (DDR-NDNR, $R^2{=}0.645$) ties the strongest baseline in main-text Table~1 (EquiformerV2, $R^2{=}0.645$), so adding each component in turn moves the model from the top of the baseline envelope into a regime that no baseline reaches.

\begin{table*}[htbp]
\caption{Ablation study on eDOS prediction. We report mean $\pm$ one standard deviation over five independent runs. Darkly shaded cells denote the best results and lightly shaded cells denote the second best. \textbf{Abbreviations:} DDR = DeepDOSReasoner; NR = without constraint-preserving physical reasoning; NDNR = without both structure-aware spectrum decoding and constraint-preserving physical reasoning.}
\label{tab:abl}
\begin{center}
\begin{sc}
\begin{tabular}{l*{4}{c}}
\toprule
Method
& $R^2 \uparrow$ & MAE $\downarrow$ & MSE $\downarrow$ & WD $\downarrow$ \\
\midrule

DDR
& \cellcolor{best}\textbf{0.691\scriptsize{$\pm$0.004}} & \cellcolor{best}\textbf{2.567\scriptsize{$\pm$0.002}} & \cellcolor{best}\textbf{38.142\scriptsize{$\pm$0.253}} & \cellcolor{best}\textbf{0.191\scriptsize{$\pm$0.005}} \\

DDR-NR
& \cellcolor{second}0.668\scriptsize{$\pm$0.002} & \cellcolor{second}2.701\scriptsize{$\pm$0.039} & \cellcolor{second}40.668\scriptsize{$\pm$0.282} & \cellcolor{second}0.221\scriptsize{$\pm$0.015} \\
DDR-NDNR
& 0.645\scriptsize{$\pm$0.003} & 2.710\scriptsize{$\pm$0.024} & 42.415\scriptsize{$\pm$0.227} & 0.243\scriptsize{$\pm$0.018} \\

\bottomrule
\end{tabular}
\end{sc}
\end{center}
\end{table*}

\subsection{Prospective VB-Gap Discovery: Candidate Screening}\label{app:vb_discovery}
This section details the candidate-screening protocol for the prospective VB-gap discovery experiment described in the main text, in which $100$ nonmagnetic Materials Project entries lacking a DFT eDOS---the $50$ top-ranked candidates of DeepDOSReasoner together with the $50$ top-ranked candidates of Mat2Spec from the same $8{,}106$-material pool, with no overlap between the two lists---are validated by new DFT calculations and used to score every method against identical labels (per-method results in main-text Table~4). Candidate screening is performed by ranking: each of the two models orders the full pool by a gap-likeness score, defined as the minimum of its predicted eDOS over a fixed ${\sim}1$~eV window immediately below the valence-band maximum---bins 47--64 of the 128-bin energy grid, spanning $-1.04$~eV to $+0.03$~eV, the region in which a VB gap's high-energy edge would sit. The lower this minimum, the closer the predicted spectrum dips to zero somewhere in the candidate gap region, i.e., the more gap-like the prediction; the pool is sorted in ascending order of this score, and each model contributes its $50$ lowest-scoring (most gap-like) entries to the evaluation set. This ranking score is used only to assemble the candidate set; all reported metrics are instead obtained by applying the four-criterion classification rule below to each method's predicted spectra on the fixed $100$-candidate set. VB gaps are detected with the same four-criterion rule as the in-distribution identification task of the main text (density below $0.9$~state/eV across a contiguous window at least $1$~eV wide whose high-energy edge lies in $[-1.5,-0.25]$~eV), applied identically to the DFT reference spectra, which define the ground-truth labels, and to every method's predicted spectra; under this rule the 100-material evaluation set contains $44$ genuine VB-gap materials ($25$ from the DeepDOSReasoner-screened half and $19$ from the Mat2Spec-screened half). The accompanying visualization (main-text Figure~3) shows representative flagged candidates alongside their DFT validations. The full per-method precision, recall, F1, and MCC on this set, together with the head-to-head interpretation, are reported in the main text (Table~4).

\begin{table*}[htbp]
\caption{eDOS-center ($\bar\varepsilon$) prediction accuracy on the held-out eDOS test set, comparing all methods. We report mean absolute error (MAE), root-mean-square error (RMSE), median absolute error (MedAE), and the 90th-percentile absolute error (P$_{90}|e|$), all in eV, together with the coefficient of determination $R^2$. These values are single-run point estimates rather than means over the five independent runs used for the main benchmark tables. Darkly shaded cells indicate the best result and lightly shaded cells the second best.}
\label{tab:edos_center}
\begin{center}
\begin{sc}
\begin{tabular}{l*{5}{c}}
\toprule
Method
& MAE (eV) $\downarrow$
& RMSE (eV) $\downarrow$
& MedAE (eV) $\downarrow$
& P$_{90}|e|$ (eV) $\downarrow$
& $R^2 \uparrow$ \\
\midrule

E3NN
& 0.198 & 0.337 & 0.112 & 0.462 & 0.862 \\

MEGNet
& 0.174 & 0.307 & 0.096 & 0.403 & 0.886 \\

GATGNN
& 0.193 & 0.334 & 0.110 & 0.438 & 0.864 \\

NequIP
& 0.189 & 0.331 & 0.106 & 0.434 & 0.867 \\

M3GNet
& 0.170 & 0.301 & 0.097 & 0.389 & 0.890 \\

MACE
& 0.268 & 0.418 & 0.179 & 0.567 & 0.787 \\

EquiformerV2
& 0.169 & 0.306 & \cellcolor{second}0.089 & 0.401 & 0.886 \\

\midrule

Mat2Spec
& 0.170 & 0.298 & 0.103 & 0.373 & 0.892 \\

DOSTransformer
& \cellcolor{second}0.166 & \cellcolor{second}0.281 & 0.101 & \cellcolor{second}0.370 & \cellcolor{second}0.904 \\

\midrule

DeepDOSReasoner
& \cellcolor{best}\textbf{0.138} & \cellcolor{best}\textbf{0.255} & \cellcolor{best}\textbf{0.083} & \cellcolor{best}\textbf{0.292} & \cellcolor{best}\textbf{0.921} \\

\bottomrule
\end{tabular}
\end{sc}
\end{center}
\end{table*}

\subsection{eDOS-Center Prediction Across the Benchmark}\label{app:edos_center}
The high-entropy-alloy case study (main text) uses the eDOS center $\bar\varepsilon$, the first energy moment of the eDOS, as a scalar descriptor of catalytic activity. To establish that DeepDOSReasoner predicts this descriptor reliably---and hence that its $\bar\varepsilon$ estimate for the high-entropy alloy can be trusted---we evaluate $\bar\varepsilon$ prediction on the held-out eDOS test set for every method. For each test material we compute $\bar\varepsilon$ from the predicted and the DFT-derived eDOS by the same numerical integral, and summarize the resulting error distribution with the mean absolute error (MAE), root-mean-square error (RMSE), median absolute error (MedAE), the 90th-percentile absolute error (P$_{90}|e|$), all in eV, and the coefficient of determination $R^2$ (\Cref{tab:edos_center}).

DeepDOSReasoner is the most accurate method on every metric (MAE $0.138$~eV, RMSE $0.255$, MedAE $0.083$, P$_{90}|e|$ $0.292$, $R^2=0.921$), ahead of the strongest crystal-to-sequence baseline (DOSTransformer, MAE $0.166$, $R^2=0.904$) and the strongest structure-encoder/interatomic-potential baselines (EquiformerV2 and M3GNet). The low MAE and high $R^2$ confirm that the model places spectral mass correctly across the full energy window, the property on which $\bar\varepsilon$ depends, and motivate its use as a catalytic-activity descriptor predictor in the HEA case study of the main text.

\subsection{Peak-Restricted MAE on van Hove Singularities}\label{app:vh}
Sharp peaks at critical points of the band structure (van Hove singularities) carry disproportionate physical weight: they encode the topology of the bands and amplify electronic responses such as superconducting transition temperatures, optical absorption strengths, and density-of-states-driven instabilities, and the peaks immediately adjacent to the VBM and CBM control near-edge optical processes and effective carrier behavior. They are also the spectral features most easily blurred by pointwise regression, so a model whose aggregate metrics are competitive can still under-resolve them. To probe peak fidelity directly, we extract the van Hove singularities adjacent to the VBM and CBM of each test material from the DFT ground truth (extraction procedure in \Cref{app:eval}) and report the mean absolute error of each method's predicted spectrum restricted to those peak bins; the summary appears in main-text Figure~5(d).

DeepDOSReasoner attains the lowest peak-restricted MAE on this evaluation, $15.95$, improving $4.3\%$ over the next-best method (Mat2Spec, $16.66$), $9.1\%$ over DOSTransformer ($17.55$), and $29.4\%$ over the median baseline. The ordering separates cleanly into three groups: the two crystal-to-sequence baselines (Mat2Spec, DOSTransformer) sit closest to DeepDOSReasoner; the structure encoders and interatomic potentials cluster in the $22$--$27$ range (M3GNet $22.52$, NequIP $22.55$, MACE $22.58$, E3NN $23.12$, MEGNet $23.88$, GATGNN $26.66$); and EquiformerV2, the strongest baseline on aggregate eDOS $R^2$ (main-text Table~1), becomes the \emph{worst} on peak MAE, at $33.47$. This last contrast is mechanistically informative: a method can achieve high aggregate $R^2$ by getting the broad envelope approximately right while compressing peak amplitudes, whereas the mass-conserving reasoning stage cannot lower its loss by thinning peaks --- under the continuity equation it can only transport mass \emph{onto} the peaks where DFT places it. The peak-MAE ranking is therefore a direct mechanism-level check on the constraint-preserving refinement, consistent with the WD and MSE rankings of main-text Table~1, both of which also reward correct peak placement. Source attention contributes the complementary axis: by grounding each energy bin in the atom-level embeddings whose orbital character drives the local critical-point structure, it supplies the spectral coordinates onto which the reasoning stage deposits mass.

\begin{table*}[htbp]
\caption{eDOS prediction performance under the stable-only protocol (training and test restricted to thermodynamically stable structures, energy above hull = 0). We report mean $\pm$ one standard deviation over five independent runs. Darkly shaded cells indicate the best results and lightly shaded cells indicate the second best.}
\label{tab:edos_stable}
\begin{center}
\begin{sc}
\begin{tabular}{lcccc}
\toprule
Model
& $R^2 \uparrow$
& MAE $\downarrow$
& MSE $\downarrow$
& WD $\downarrow$ \\
\midrule

E3NN
& 0.598\scriptsize{$\pm$0.002} & 3.131\scriptsize{$\pm$0.002} & 51.073\scriptsize{$\pm$0.256} & 0.276\scriptsize{$\pm$0.001} \\

MEGNet
& 0.631\scriptsize{$\pm$0.003} & 2.964\scriptsize{$\pm$0.001} & 46.874\scriptsize{$\pm$0.276} & \cellcolor{second}0.219\scriptsize{$\pm$0.003} \\

GATGNN
& 0.597\scriptsize{$\pm$0.005} & 3.212\scriptsize{$\pm$0.001} & 51.207\scriptsize{$\pm$0.398} & 0.279\scriptsize{$\pm$0.001} \\

NequIP
& 0.602\scriptsize{$\pm$0.005} & 3.164\scriptsize{$\pm$0.002} & 50.493\scriptsize{$\pm$0.334} & 0.263\scriptsize{$\pm$0.002} \\

M3GNet
& 0.633\scriptsize{$\pm$0.002} & 2.933\scriptsize{$\pm$0.004} & 46.528\scriptsize{$\pm$0.296} & 0.225\scriptsize{$\pm$0.003} \\

MACE
& 0.538\scriptsize{$\pm$0.004} & 3.526\scriptsize{$\pm$0.002} & 58.641\scriptsize{$\pm$0.331} & 0.337\scriptsize{$\pm$0.001} \\

EquiformerV2
& \cellcolor{second}0.634\scriptsize{$\pm$0.001} & \cellcolor{second}2.925\scriptsize{$\pm$0.003} & \cellcolor{second}46.445\scriptsize{$\pm$0.468} & 0.226\scriptsize{$\pm$0.001} \\

\midrule

Mat2Spec
& 0.613\scriptsize{$\pm$0.001} & 3.045\scriptsize{$\pm$0.001} & 49.129\scriptsize{$\pm$0.582} & 0.225\scriptsize{$\pm$0.001} \\

DOSTransformer
& 0.616\scriptsize{$\pm$0.001} & 3.022\scriptsize{$\pm$0.001} & 48.801\scriptsize{$\pm$0.435} & 0.224\scriptsize{$\pm$0.001} \\

\midrule

DeepDOSReasoner
& \cellcolor{best}\textbf{0.674\scriptsize{$\pm$0.001}} & \cellcolor{best}\textbf{2.809\scriptsize{$\pm$0.001}} & \cellcolor{best}\textbf{41.476\scriptsize{$\pm$0.016}} & \cellcolor{best}\textbf{0.197\scriptsize{$\pm$0.001}} \\

\bottomrule
\end{tabular}
\end{sc}
\end{center}
\end{table*}

\subsection{Data-Scaling Behavior}\label{app:scaling}
Sample efficiency is a practical concern in DOS prediction because the per-material cost of obtaining high-quality DFT eDOS supervision is large and scales unfavorably with system size, so reducing the training data needed to reach a target accuracy directly reduces the compute budget for training-data curation. We probe this by sub-sampling the full training set at three fractions ($25\%$, $50\%$, and $75\%$) and re-training DeepDOSReasoner and the two crystal-to-sequence baselines (Mat2Spec, DOSTransformer) from scratch on each subset, evaluating on the same held-out test set used in the main eDOS benchmark; the curves are shown in main-text Figure~5(a).

DeepDOSReasoner is consistently more sample-efficient than the baselines: at each training fraction it gives the best $R^2$, MAE, MSE, and WD among the methods plotted. All methods improve as the training set grows, but the separation between DeepDOSReasoner and the baseline envelope is already established at $25\%$ and does not close as the training set expands, indicating that the architecture is extracting useful structure--spectrum signal in the low-data regime rather than merely benefiting from data abundance. The separation is largest on MSE and WD, the two metrics that most directly reward sharp band-edge placement and correct distributional alignment, mirroring the full-data ranking and indicating that source-attentive decoding and mass-conserving refinement contribute to spectral geometry from the smallest training fraction onward. In practical terms, DeepDOSReasoner trained on $25\%$ of the data already lies inside the spread of baselines trained on the full set on several metrics, so the model can be deployed on smaller domain-specific spectrum benchmarks (e.g.\ a chalcogenide-focused eDOS subset, or the small phDOS benchmark) without sacrificing the architectural advantages established on the full benchmark.